\documentclass[modern, times]{aastex631}

\usepackage{CJK}
\usepackage{longtable}
\usepackage{makecell}
\usepackage{rotating}

\shorttitle{Optical component stars in X-ray binary systems}
\shortauthors{Zhang et al.}

\makeatletter
\let\frontmatter@title@above=\relax
\makeatother
\usepackage{lipsum}

\begin{document}
\begin{CJK}{UTF8}{gbsn}  

\title{Properties of the optical component stars in X-ray binary systems\footnote{An online version of the catalogs are publicly available at \url{https://astrophysics.cc/xray-binary-catalogs}} }

\correspondingauthor{Sheng-Bang Qian}
\email{qsb@ynao.ac.cn}
\author[0000-0003-2017-9151]{Jia Zhang (张嘉)}
\author{Sheng-Bang Qian (钱声帮)}
\affiliation{Yunnan Observatories, Chinese Academy of Sciences, Kunming 650216, China}
\affiliation{Department of Astronomy, Key Laboratory of Astroparticle Physics of Yunnan Province, Yunnan University, Kunming 650091, China}
\affiliation{Key Laboratory of the Structure and Evolution of Celestial Objects, Chinese Academy of Sciences, Kunming 650216, China}
\author{Guo-Bao Zhang (张国宝)}
\author{Xiao Zhou (周肖)}
\affiliation{Yunnan Observatories, Chinese Academy of Sciences, Kunming 650216, China}
\affiliation{Key Laboratory of the Structure and Evolution of Celestial Objects, Chinese Academy of Sciences, Kunming 650216, China}

\begin{abstract}
We collected a total of 4,058 X-ray binary stars, out of which 339 stars had three atmospheric parameters for optical companions from Gaia and LAMOST, while 264 stars had masses and radii of optical companions determined using stellar evolution models. We conducted a thorough discussion on the reliability of each parameter. The statistical analysis revealed a noticeable bimodal distribution in the mass, radius, and age of the optical components. Our findings led to the proposal of a new quantitative classification criterion for X-ray binary stars. In this classification, one type is categorized as high-mass, high-temperature, and young, while the other type is classified as low-mass, low-temperature, and old, corresponding to High-Mass X-Ray Binaries (HMXBs) and Low-Mass X-Ray Binaries (LMXBs), respectively. The dividing lines were established at 11,500 K, 1.7 $M_{\odot}$, and 0.14 Gyr. The classification results of the three parameters showed consistency with one another in 90\% of the cases and were in agreement with previous classifications in 80\% of the cases. We found that the parameters of the two types had well-defined boundaries and distinct patterns. Based on our findings, we suggest that temperature is the best parameter for classification. Therefore, we propose that X-ray binary stars should be classified into high-temperature X-ray binaries (HTXBs) and low-temperature X-ray binaries (LTXBs). We believe that this classification is more convenient in practice and aligns well with physics.

\end{abstract}


\section{Introduction} \label{sec:intro}

X-ray binaries are binary star systems in which one of the stars is a compact object, such as a neutron star or a black hole, and the other is a normal star. These systems are named as such because they release a significant amount of X-ray radiation, which is generated by material from the normal star being pulled onto the compact object and heated to extremely high temperatures \citep{2006csxs.book..623T}. X-ray binary stars are important objects for studying the physics of extreme environments, as well as for testing theories of gravity in the strong-field regime.

X-ray binaries can be divided into two main categories: high-mass X-ray binaries (HMXBs), in which the normal star is a high-mass star such as a Wolf-Rayet star or an O-type star, and low-mass X-ray binaries (LMXBs), in which the normal star is a low-mass star like a red giant or a main sequence star \citep{2006csxs.book..623T}.

There is also a category known as intermediate-mass X-ray binaries (IMXBs) as a transitional state between the two previously mentioned categories, as proposed by \citet{2003ApJ...597.1036P}.

In this classification, the mass refers to the mass of the optical companion, rather than the compact component. Regretfully, there is no unified quantitative standard for the dividing line between HMXBs and LMXBs. One solar mass was initially designated as the dividing line, but later there were also 1-10 $M_{\odot}$ \citep{2006csxs.book..623T}, 1-5 $M_{\odot}$ \citep{2011hea..book.....L}, or 8 $M_{\odot}$ \citep{2023A&A...671A.149F}.

The main issue with current X-ray binaries is that there is a lack of samples with reliable physical parameters. The majority of the compact components were not measured in mass, nor the reliable masses of the optical companions. \citep{2023A&A...671A.149F, 2023arXiv230316168A, 2023arXiv230316137N}.

Since the mass of the optical companion cannot be easily measured, X-ray binary stars are often roughly classified based on their X-ray observational features. The classification are sometimes based on the luminosity of the X-ray waveband, such as the luminosity of the X-ray band being between $10^{40-42}$ erg/s for HMXB and $10^{37-39}$ erg/s for LMXB. Additionally, factors such as hardness ratio (HR = (H-S)/(H+S)), type of time variability (pulses or bursts), infrared excess, and the ratio of X-ray to optical luminosity will also be taken into consideration.

Excitingly, the emergence of large-scale optical sky survey data (e.g., Gaia, LAMOST, and TESS) has provided a new opportunity for classification.

In this study, we collected a large sample of X-ray binary stars and used Gaia \citep{2022arXiv220605989B, 2022arXiv220800211G, 2016A&A...595A...1G, 2022arXiv220605992F, 2022arXiv220605864C} and LAMOST \citep{2022yCat.5156....0L} spectral data to calculate the physical parameters of optical companion star for over 200 systems. Further, based on these physical parameters, we obtained the parameter distribution and correlations of X-ray binary stars, and proposed a new classification criterion.

In Section 2, we will introduce the catalogs and their sources. Sections 3 to 5 focus on discussing the reliability of the parameters, while Sections 6 and 7 present the statistical results based on the catalogs, with a focus on the distinct differences between the two classes of X-ray binaries. The final section is the conclusion and discussion.

\section{Catalogs of X-ray Binary stars}

We provide a total of 6 catalogs, including a general catalog of all 4058 X-ray binary stars (Table \ref{tab:total}), cross-referenced catalog with General Catalog of Variable Stars GCVS \citep{2017ARep...61...80S} and The International Variable Star Index VSX \citep{2006SASS...25...47W} (Table \ref{tab:gv}), cross-referenced catalog with LAMOST DR9 (Table \ref{tab:lam}), cross-referenced catalog with Gaia DR3 physical parameters (Table \ref{tab:gaia_phy}) and astrometry parameters (Table \ref{tab:gaia_astrometry}), and the integrated catalog of physical parameters (Table \ref{tab:uni}). In addition to these 6 catalogs, there is also a table of number statistics (Table \ref{tab:statistic}).

\begin{deluxetable*}{llllllll}
\tablecaption{General catalog of X-ray binary systems} \label{tab:total}
\tabletypesize{\scriptsize}
\tablehead{
\colhead{\makecell[c]{Name\\from Simbad}} & \colhead{RA}  & \colhead{DEC}  & \colhead{RA$_{err}$}  & \colhead{DEC$_{err}$}  & \colhead{Type}  & \colhead{Reference of coordinate}  & \colhead{Reference of Type} \\
\colhead{ } & \colhead{(deg)}  & \colhead{(deg)}  & \colhead{(sec)}  & \colhead{(sec)}  & \colhead{ }  & \colhead{(Bibcode)}  & \colhead{(Bibcode)}
}
\startdata
CI Cam & 64.92556499855 & 55.99936294422 & 1e-12 & 1e-12 & HXB & 2020yCat.1350....0G & 2023A\&A...671A.149F \\
IGR J06074+2205 & 91.86088359229 & 22.09659996262 & 1e-12 & 1e-12 & HXB & 2020yCat.1350....0G & 2023A\&A...671A.149F \\
HZ Her & 254.45754588608997 & 35.342357376520006 & 1e-11 & 1e-12 & LXB & 2020yCat.1350....0G & 2023arXiv230316168A \\
HD 215227 & 340.73876315802704 & 44.721738639785 & 1e-11 & 1e-12 & HXB & 2020yCat.1350....0G & 2023A\&A...671A.149F \\
SWIFT J174510.8-262411 & 266.2952041666667 & -26.403499999999998 & 0.0001 & 1e-05 & LXB & 2013ApJS..209...14K & 2023arXiv230316168A \\
  \multicolumn{8}{c}{...} \\
  \multicolumn{8}{c}{A total of 4058 sources} \\
  \multicolumn{8}{c}{...} \\
SAX J1818.7+1424 & 274.6833333333333 & 14.403333333333334 & nan & nan & LXB & 2007A\&A...469..807L & 2023arXiv230316168A \\
MAXI J1848-015 & 282.19999999999993 & -1.4866666666666666 & nan & nan & LXB & 2001ApJS..134...77S & 2023arXiv230316168A \\
Swift J185003.2-005627 & 282.51373 & -0.9398 & 0.0007 & 0.0007 & LXB & 2020yCat.9058....0E & 2023arXiv230316168A \\
Swift J2037.2+4151 & 309.273335 & 41.834774 & 3e-05 & 3e-05 & LXB & 2003yCat.2246....0C & 2023arXiv230316168A \\
SAX J2224.9+5421 & 336.2068749999999 & 54.38613888888889 & 0.0006 & 0.0006 & LXB & 2014ApJ...787...67D & 2023arXiv230316168A \\
\enddata
\end{deluxetable*}

\begin{deluxetable*}{llllllcc}
\tablecaption{X-ray binaries collected by GCVS and VSX} \label{tab:gv}
\tablehead{
\colhead{\makecell[c]{Name\\from Simbad}} & \colhead{\makecell[c]{RA\\from\\VSX or GCVS}}  & \colhead{\makecell[c]{Dec\\from\\VSX or GCVS}}  &  \colhead{\makecell[c]{Type\\from\\VSX}}  & \colhead{\makecell[c]{Type\\from\\GCVS}} & \colhead{\makecell[c]{Period\\from\\VSX}} & \colhead{\makecell[c]{Period\\from\\GCVS}} \\
\colhead{ } & \colhead{(deg)}  & \colhead{(deg)}  & \colhead{ }  & \colhead{ }  & \colhead{(d)}  & \colhead{(d)}  
}
\startdata
CI Cam & 64.92558 & 55.99936 & HMXB/XN+PULS: & XNG & 19.407 &   \\
IGR J06074+2205 & 91.86087 & 22.09661 & HMXB & BE & --- &   \\
HZ Her & 254.45729 & 35.34242 & LMXB:/XPR+E & XPR & 34.875 &     1.700175 \\
HD 215227 & 340.73876 & 44.72174 & HMXB/BHXB: & --- & 60.37083 & --- \\
SWIFT J174510.8-262411 & 266.29508 & -26.40353 & LMXB/BHXB: & --- & 0.875 & --- \\
  \multicolumn{7}{c}{...} \\
  \multicolumn{7}{c}{A total of 219 sources} \\
 \multicolumn{7}{c}{...} \\
IGR J17329-2731 & 263.21112 & -27.50044 & ZAND: & --- & --- & --- \\
1RXH J173523.7-354013 & 263.84892 & -35.67128 & LMXB/XN & --- & --- & --- \\
IGR J16358-4726 & 248.97375 & -47.42772 & LMXB/XP & --- & --- & --- \\
IGR J17407-2808 & 265.17502 & -28.12345 & VAR & --- & --- & --- \\
SAX J1806.5-2215 & 271.63403 & -22.23817 & LMXB & --- & --- & --- \\
\enddata
\end{deluxetable*}

\begin{deluxetable*}{llllllccc}
\tablecaption{The optical components' parameters of X-ray binaries provided by LAMOST DR9} \label{tab:lam}
\tablehead{
\colhead{\makecell[c]{Name\\from Simbad}} & \colhead{\makecell[c]{RA\\from LAMOST}}  & \colhead{\makecell[c]{Dec\\from LAMOST}}  &  \colhead{$T_{eff}$}  & \colhead{logg} & \colhead{[M/H]} & \colhead{\makecell[c]{Mean\\Radial\\Velocity}} & \colhead{\makecell[c]{Spectral\\Type}} & \colhead{\makecell[c]{Number\\of\\Observations}}\\
\colhead{ } & \colhead{(deg)}  & \colhead{(deg)}  & \colhead{(K)}  & \colhead{(dex)}  & \colhead{(dex)} & \colhead{(km/s)} & \colhead{}  
}
\startdata
CI Cam & 64.925578 & 55.999361 & 3645$\pm$110 & 4.742$\pm$0.42 & -1.258$\pm$0.61 & --- & M & 1 \\
IGR J06074+2205 & 91.8608835 & 22.0965999 & --- & --- & --- & --- & --- & 5 \\
HZ Her & 254.4574208 & 35.3422972 & --- & --- & --- & --- & A & 1 \\
2MASS J21342037+4738002 & 323.584877 & 47.633387 & --- & --- & --- & --- & A & 1 \\
V418 Gem & 106.03612 & 26.419649 & --- & --- & --- & --- & \makecell[c]{Cataclysmic\\Variable} & 4 \\
HY Leo & 146.64361 & 13.849409 & --- & --- & --- & --- & \makecell[c]{Cataclysmic\\Variable} & 1 \\
2MASS J13293864+4713359 & 202.411027 & 47.226666 & 4857$\pm$2 & 3.664$\pm$0.01 & -0.491$\pm$0.01 & -29$\pm$1 & K & 2 \\
1ES 0851+39.2 & 133.5583628 & 39.0936827 & --- & --- & --- & --- & \makecell[c]{Cataclysmic\\Variable} & 3 \\
V647 Aur & 99.135666 & 35.595352 & --- & --- & --- & --- & \makecell[c]{Cataclysmic\\Variable} & 1 \\
$[$ZGV2011$]$ 13 & 10.6239 & 41.29928 & 4676$\pm$60 & 2.782$\pm$0.09 & -0.792$\pm$0.06 & -363$\pm$6 & K & 1 \\
$[$ZGV2011$]$ 18 & 10.81421 & 41.19027 & 5551$\pm$100 & 3.954$\pm$0.17 & -0.495$\pm$0.11 & -434$\pm$7 & G & 1 \\
$[$ZGV2011$]$ 20 & 10.60434 & 40.95489 & --- & --- & --- & --- & K & 1 \\
$[$ZGV2011$]$ 24 & 10.63793 & 41.05787 & --- & --- & --- & --- & G & 1 \\
$[$BGM2013$]$ BHC 4 & 10.60863 & 41.32072 & 4738$\pm$70 & 2.935$\pm$0.12 & -0.301$\pm$0.08 & -322$\pm$5 & G & 1 \\
3XMM J004212.1+411758 & 10.550629 & 41.299667 & --- & --- & --- & --- & DoubleStar & 1 \\
$[$ZGV2011$]$ 28 & 10.76375 & 41.35604 & --- & --- & --- & --- & DoubleStar & 1 \\
3XMM J004231.2+411938 & 10.63022 & 41.32748 & 4240$\pm$101 & 1.452$\pm$0.24 & -1.355$\pm$0.16 & -312$\pm$11 & G & 2 \\
$[$BGM2013$]$ BHC 28 & 10.76607 & 41.30136 & 5053$\pm$24 & 3.306$\pm$0.04 & -0.602$\pm$0.05 & -299$\pm$2 & G & 2 \\
$[$BGM2013$]$ BHC 25 & 10.74942 & 41.26829 & --- & --- & --- & --- & K & 1 \\
$[$BGM2013$]$ BHC 24 & 10.7485 & 41.32206 & 4615$\pm$66 & 2.922$\pm$0.10 & -0.415$\pm$0.07 & -183$\pm$7 & K & 1 \\
$[$BGM2013$]$ BHC 2 & 10.57773 & 41.23389 & 5671$\pm$13 & 3.949$\pm$0.15 & -0.732$\pm$0.06 & -251$\pm$19 & G & 2 \\
$[$BGM2013$]$ BHC 34 & 10.90534 & 41.24543 & --- & --- & --- & --- & G & 2 \\
$[$BGM2013$]$ BHC 31 & 10.79421 & 41.2476 & 4489$\pm$77 & 2.408$\pm$0.12 & -0.789$\pm$0.08 & -261$\pm$6 & G & 1 \\
$[$BGM2013$]$ BHC 1 & 10.56597 & 41.02069 & 5014$\pm$50 & 3.249$\pm$0.08 & -0.720$\pm$0.05 & -386$\pm$4 & K & 2 \\
2E 168 & 11.419018 & 42.135188 & 4323$\pm$34 & 4.542$\pm$0.05 & -0.162$\pm$0.03 & -0$\pm$5 & K & 1 \\
Feige 93 & 214.41762 & 13.030158 & --- & --- & --- & --- & \makecell[c]{White\\Dwarf} & 4 \\
GSC 03852-01275 & 210.624495 & 54.355237 & --- & --- & --- & --- & K & 1 \\
1RXS J134101.0+602556 & 205.2501 & 60.4367139 & --- & --- & --- & --- & \makecell[c]{White\\Dwarf} & 1 \\
TYC 3852-1069-1 & 211.0903 & 54.322882 & 5840$\pm$18 & 4.459$\pm$0.03 & -0.046$\pm$0.01 & +4$\pm$2 & G & 3 \\
PG 0205+134 & 32.0144729 & 13.6069902 & --- & --- & --- & --- & A & 3 \\
CXOM31 J004314.3+410721 & 10.80996 & 41.12253 & 5444$\pm$64 & 3.740$\pm$0.10 & -0.528$\pm$0.07 & -192$\pm$6 & G & 2 \\
Bol 158 & 10.80996 & 41.12253 & 5444$\pm$64 & 3.740$\pm$0.10 & -0.528$\pm$0.07 & -192$\pm$6 & G & 2 \\
$[$HPH2013$]$ 167 & 10.68472556 & 41.26891111 & 4497$\pm$43 & 2.590$\pm$0.06 & -0.178$\pm$0.03 & -314$\pm$6 & K & 1 \\
$[$R2005b$]$ 12 & 21.1454208 & 3.7915417 & --- & --- & --- & --- & --- & 1 \\
XMMM31 J004244.1+411607 & 10.68472556 & 41.26891111 & 4497$\pm$43 & 2.590$\pm$0.06 & -0.178$\pm$0.03 & -314$\pm$6 & K & 1 \\
SAX J0635.2+0533 & 98.826164 & 5.551743 & --- & --- & --- & --- & --- & 1 \\
\enddata
\end{deluxetable*}

\begin{longrotatetable} 
\begin{deluxetable*}{p{1.5cm}lllllllllc}
\tablecaption{The optical components' parameters of X-ray binaries provided by Gaia DR3: physical parameters} \label{tab:gaia_phy}
\tabletypesize{\tiny}
\tablehead{
\colhead{\makecell[c]{Name\\from Simbad}} & \colhead{\makecell[c]{RA\\from Gaia}} & \colhead{\makecell[c]{DEC\\from Gaia}} & \colhead{$T_{eff}$} & \colhead{logg} & \colhead{[M/H]} & \colhead{Mass} & \colhead{Radius} & \colhead{Luminosity} & \colhead{Age} & \colhead{\makecell[c]{Evolutionary\\stage}} \\ 
\colhead{ } & \colhead{(deg)} & \colhead{(deg)} & \colhead{(K)} & \colhead{(dex)} & \colhead{(dex)} & \colhead{($M_{\odot}$)} & \colhead{(R$_{\odot}$)} & \colhead{($L_{\odot}$)} & \colhead{(Gyr)} & \colhead{ }  
} 
\startdata
IGR J06074+2205 & 91.86088634197857 & 22.096597258920347 & $28984_{-70}^{+46}$ & $4.110_{-0.032}^{+0.03}$ & $+0.007_{-0.0054}^{+0.011}$ & --- & $5.359_{-0.2}^{+0.23}$ & --- & --- & --- \\
MAXI J0903-531 & 136.27840443491405 & -53.505413083748714 & $22971_{-327}^{+112}$ & $3.996_{-0.01}^{+0.017}$ & $-0.971_{-0.024}^{+0.09}$ & --- & $4.557_{-0.12}^{+0.056}$ & --- & --- & --- \\
MAXI J1820+070 & 275.0914127350555 & 7.185328968929734 & $5604_{-29}^{+35}$ & $4.694_{-0.022}^{+0.022}$ & $-4.064_{-0.062}^{+0.16}$ & --- & $0.5995_{-0.013}^{+0.0083}$ & --- & --- & --- \\
AAO+28 342 & 88.9793400429246 & 28.78510745617743 & $16927_{-305}^{+683}$ & $3.614_{-0.12}^{+0.37}$ & $+0.407_{-0.42}^{+0.14}$ & $6.272_{-0.12}^{+0.13}$ & $4.677_{-0.21}^{+0.25}$ & $1957_{-1.6e+02}^{+2e+02}$ & --- & 249 \\
NGC 6649 9 & 278.3679185343803 & -10.40243295086028 & $35005_{-9}^{+21}$ & $3.835_{-0.056}^{+0.022}$ & $+0.006_{-0.004}^{+0.0099}$ & --- & $10.24_{-0.36}^{+1}$ & --- & --- & --- \\
V1341 Cyg & 326.17146666896133 & 38.321405939264146 & $14210_{-105}^{+54}$ & $4.174_{-0.029}^{+0.053}$ & $-1.023_{-0.078}^{+0.063}$ & $3.781_{-0.078}^{+0.066}$ & $2.388_{-0.15}^{+0.083}$ & $206.9_{-25}^{+12}$ & --- & 133 \\
PSR J1023+0038 & 155.94870458335123 & 0.6446470547528212 & $5966_{-4}^{+12}$ & $4.654_{-0.016}^{+0.015}$ & $-4.119_{-0.022}^{+0.038}$ & --- & $0.6593_{-0.0082}^{+0.0082}$ & --- & --- & --- \\
AX J1755.7-2818 & 268.9284641514467 & -28.30260153482759 & $3290_{-11}^{+14}$ & $3.970_{-0.023}^{+0.037}$ & $-1.327_{-0.021}^{+0.024}$ & --- & $0.5195_{-0.02}^{+0.014}$ & --- & --- & --- \\
V2116 Oph & 263.0089546460945 & -24.74560029077066 & $4275_{-32}^{+21}$ & $1.227_{-0.023}^{+0.03}$ & $-0.836_{-0.076}^{+0.12}$ & --- & $28.92_{-0.97}^{+0.8}$ & --- & --- & --- \\
NP Ser & 274.0058035972191 & -14.036598833320399 & $4788_{-39}^{+37}$ & $4.637_{-0.015}^{+0.015}$ & $-0.377_{-0.059}^{+0.064}$ & --- & $1.006_{-0.1}^{+0.1}$ & $0.4958_{-0.098}^{+0.099}$ & --- & 391 \\
 \multicolumn{11}{c}{...} \\
  \multicolumn{11}{c}{A total of 329 sources} \\
 \multicolumn{11}{c}{...} \\
Cir X-1 & 230.1701764075732 & -57.166742166539116 & $4296_{-19}^{+17}$ & $2.029_{-0.011}^{+0.028}$ & $-0.970_{-0.053}^{+0.065}$ & --- & $11.47_{-0.38}^{+0.14}$ & --- & --- & --- \\
PSR J2032+4127 & 308.0546693559062 & 41.4567587500388 & $29996_{-126}^{+129}$ & $4.037_{-0.0091}^{+0.01}$ & $+0.264_{-0.034}^{+0.025}$ & --- & $6.257_{-0.083}^{+0.097}$ & --- & --- & --- \\
SRGE J204319.0+443820 & 310.8277570377428 & 44.638891277204316 & $7543_{-77}^{+76}$ & $3.172_{-0.039}^{+0.033}$ & $-0.085_{-0.0076}^{+0.0045}$ & --- & $7.187_{-0.27}^{+0.32}$ & --- & --- & --- \\
1FGL J1417.7-4407 & 214.37735925120347 & -44.04932767896711 & $5026_{-22}^{+21}$ & $3.650_{-0.034}^{+0.035}$ & $+0.049_{-0.03}^{+0.036}$ & $1.142_{-0.052}^{+0.056}$ & $2.6_{-0.13}^{+0.13}$ & $3.875_{-0.35}^{+0.34}$ & $7.074_{-1.1}^{+0.99}$ & 552 \\
3FGL J1544.6-1125 & 236.1641203188938 & -11.468022194700493 & $4488_{-42}^{+285}$ & $4.900_{-0.04}^{+0.022}$ & $-1.676_{-0.57}^{+0.17}$ & --- & $0.405_{-0.022}^{+0.026}$ & --- & --- & --- \\
IGR J17329-2731 & 263.2111131733354 & -27.500438107295558 & $3515_{-2}^{+4}$ & $0.422_{-0.0033}^{+0.0075}$ & $-0.130_{-0.0001}^{+0.0001}$ & --- & $125.4_{-1}^{+0.44}$ & --- & --- & --- \\
4FGL J0427.8-6704 & 66.95685947442904 & -67.07640585049464 & $5962_{-2}^{+6}$ & $4.693_{-0.0068}^{+0.005}$ & $-4.130_{-0.016}^{+0.026}$ & $0.9159_{-0.05}^{+0.046}$ & $0.7527_{-0.11}^{+0.16}$ & $0.4463_{-0.12}^{+0.19}$ & $2.797_{-2.6}^{+4.9}$ & 204 \\
2MASS J16233414-2631336 & 245.8921168911989 & -26.525988121153727 & $5921_{-90}^{+78}$ & $3.686_{-0.057}^{+0.054}$ & $-2.187_{-0.28}^{+0.13}$ & $1.132_{-0.065}^{+0.059}$ & $1.77_{-0.15}^{+0.18}$ & $3.288_{-0.59}^{+0.75}$ & $6.552_{-1.2}^{+1.5}$ & 418 \\
PSR J1723-2837 & 260.8465812119027 & -28.63266533424399 & $5828_{-41}^{+46}$ & $4.489_{-0.027}^{+0.022}$ & $+0.079_{-0.05}^{+0.044}$ & $0.9962_{-0.043}^{+0.049}$ & $1.082_{-0.045}^{+0.038}$ & $1.217_{-0.088}^{+0.072}$ & $6.814_{-2.1}^{+1.9}$ & 316 \\
AX J1735.8-3207 & 263.9428385638315 & -32.119401005517396 & $7166_{-6}^{+5}$ & $4.587_{-0.0012}^{+0.0009}$ & $-3.381_{-0.011}^{+0.016}$ & --- & $0.8357_{-0.0011}^{+0.0012}$ & --- & --- & --- \\
\enddata
\end{deluxetable*}
\end{longrotatetable}

\begin{longrotatetable}
\begin{deluxetable*}{p{1.5cm}lllllcclcc}
\tablecaption{The optical components' parameters of X-ray binaries provided by Gaia DR3: astrometry parameters}  \label{tab:gaia_astrometry}
\tabletypesize{\tiny}
\tablehead{
\colhead{\makecell[c]{Name\\from Simbad}} & \colhead{\makecell[c]{RA\\from Gaia}} & \colhead{\makecell[c]{DEC\\from Gaia}} & \colhead{distance} & \colhead{\makecell[c]{Galactic\\longitude\\l}} & \colhead{\makecell[c]{Galactic\\latitude \\b}} & \colhead{vsini} & \colhead{$[$Alpha/Fe$]$} & \colhead{Gmag} & \colhead{\makecell[c]{Spectral\\type}} & \colhead{ \makecell[c]{Gaia\\source id}} \\ 
\colhead{ } & \colhead{(deg)} & \colhead{(deg)} & \colhead{(pc)} & \colhead{(deg)} & \colhead{(deg)} & \colhead{(km/s)} & \colhead{(dex)} & \colhead{ } & \colhead{ }
} 
\startdata
IGR J06074+2205 & 91.86088634197857 & 22.096597258920347 & $3724_{-1.4e+02}^{+1.6e+02}$ & 188.38532157936345 & 0.81375073815519 & --- & --- & 12.166335 & B & 3423526544838563328 \\
MAXI J0903-531 & 136.27840443491405 & -53.505413083748714 & $5546_{-1.6e+02}^{+70}$ & 273.07612343985846 & -4.303878231912306 & $54_{-30}^{+30}$ & --- & 13.145346 & B & 5311384333263075840 \\
MAXI J1820+070 & 275.0914127350555 & 7.185328968929734 & $1820_{-51}^{+34}$ & 35.85354166356241 & 10.159160736016908 & --- & --- & 17.386671 & O & 4477902563164690816 \\
AAO+28 342 & 88.9793400429246 & 28.78510745617743 & $2299_{-9.3e+02}^{+3.9e+02}$ & 181.2839667823892 & 1.8595308169574565 & --- & --- & 10.010054 & B & 3431561565357225088 \\
NGC 6649 9 & 278.3679185343803 & -10.40243295086028 & $2033_{-72}^{+2.1e+02}$ & 21.63824683138313 & -0.7896417890712586 & --- & --- & 10.961636 & O & 4155023796790984064 \\
V1341 Cyg & 326.17146666896133 & 38.321405939264146 & $3990_{-2.7e+02}^{+1.6e+02}$ & 87.32818255374067 & -11.316361330752494 & --- & --- & 14.701829 & F & 1952859683185470208 \\
PSR J1023+0038 & 155.94870458335123 & 0.6446470547528212 & $1418_{-18}^{+19}$ & 243.4895915647394 & 45.78224492606224 & --- & --- & 16.232056 & O & 3831382647922429952 \\
AX J1755.7-2818 & 268.9284641514467 & -28.30260153482759 & $705.6_{-27}^{+20}$ & 1.6786929226550131 & -1.5782105523758543 & --- & --- & 18.685217 & -- & 4063384450408345216 \\
V2116 Oph & 263.0089546460945 & -24.74560029077066 & $1402_{-48}^{+52}$ & 1.9370210642332468 & 4.794968837444518 & --- & --- & 15.517966 & M & 4110236324513030656 \\
NP Ser & 274.0058035972191 & -14.036598833320399 & $893_{-24}^{+28}$ & 16.43191649427052 & 1.2774774537036035 & --- & --- & 17.013025 & K & 4146621775597340544 \\
\multicolumn{11}{c}{...} \\
\multicolumn{11}{c}{A total of 329 sources} \\
\multicolumn{11}{c}{...} \\
Cir X-1 & 230.1701764075732 & -57.166742166539116 & $1411_{-37}^{+23}$ & 322.1184786573702 & 0.03771540102283994 & --- & --- & 17.775951 & -- & 5883218164517055488 \\
PSR J2032+4127 & 308.0546693559062 & 41.4567587500388 & $1609_{-24}^{+28}$ & 80.2238146279451 & 1.0278993456654848 & --- & --- & 11.277184 & B & 2067835682818358400 \\
SRGE J204319.0+443820 & 310.8277570377428 & 44.638891277204316 & $1925_{-68}^{+85}$ & 83.98370959231536 & 1.3407233597302817 & --- & --- & 16.56358 & M & 2070085317968809216 \\
1FGL J1417.7-4407 & 214.37735925120347 & -44.04932767896711 & $2540_{-1.1e+02}^{+88}$ & 318.86122490455944 & 16.144039568152028 & --- & --- & 15.771324 & K & 6096705840454620800 \\
3FGL J1544.6-1125 & 236.1641203188938 & -11.468022194700493 & $1395_{-99}^{+1.1e+02}$ & 356.17059058559914 & 32.961168517773444 & --- & --- & 18.603163 & -- & 6268529198286308224 \\
IGR J17329-2731 & 263.2111131733354 & -27.500438107295558 & $3918_{-14}^{+8.5}$ & 359.71698950280415 & 3.1464832658437185 & --- & --- & 16.052113 & M & 4061336747511224704 \\
4FGL J0427.8-6704 & 66.95685947442904 & -67.07640585049464 & $2197_{-13}^{+11}$ & 279.13809497569895 & -38.55409363178523 & --- & --- & 17.726036 & -- & 4656677385699742208 \\
2MASS J16233414-2631336 & 245.8921168911989 & -26.525988121153727 & $2189_{-1.1e+02}^{+1.2e+02}$ & 350.9698412995653 & 15.975087455996512 & --- & --- & 15.997249 & unknown & 6045465884181756800 \\
PSR J1723-2837 & 260.8465812119027 & -28.63266533424399 & $806.1_{-28}^{+33}$ & 357.6161833862773 & 4.260059100541004 & --- & --- & 15.540544 & F & 4059795674516044800 \\
AX J1735.8-3207 & 263.9428385638315 & -32.119401005517396 & $1406_{-3.5}^{+3.6}$ & 356.17687590356815 & 0.11034387356033036 & --- & --- & 14.982109 & O & 4055019091071763712 \\
\enddata
\end{deluxetable*}
\end{longrotatetable}

\begin{longrotatetable}
\begin{deluxetable*}{p{1.5cm}lllllllll@{\extracolsep{\fill}}c@{\extracolsep{\fill}}l}
\tablecaption{The optical components' parameters of X-ray binaries integrated  by this paper}  \label{tab:uni}
\tabletypesize{\tiny}
\tablehead{
\colhead{\makecell[c]{Name\\from Simbad}} & \colhead{RA} & \colhead{DEC} & \colhead{$T_{eff}$} & \colhead{logg} & \colhead{[M/H]} & \colhead{Mass} & \colhead{Radius} & \colhead{Luminosity} & \colhead{Age} & \colhead{\makecell[c]{Evolutionary\\stage}} & \colhead{Period}  \\ 
\colhead{ } & \colhead{(deg)} & \colhead{(deg)} & \colhead{(K)} & \colhead{(cgs)} & \colhead{(dex)} & \colhead{($M_{\odot}$)} & \colhead{(R$_{\odot}$)} & \colhead{($L_{\odot}$)} & \colhead{(Gyr)} & \colhead{ } & \colhead{(d)} 
} 
\startdata
CI Cam & 64.92556499855 & 55.99936294422 & $3645_{-110}^{+110}$ & $4.742_{-0.42}^{+0.42}$ & $-1.258_{-0.61}^{+0.61}$ & $0.239_{-0.079}^{+0.17}$ & $0.2465_{-0.073}^{+0.14}$ & $0.00975_{-0.0055}^{+0.017}$ & $11.48_{-11}^{+1.7}$ & $1_{-0}^{+0}$ & 19.407 \\
IGR J06074+2205 & 91.86088359229 & 22.09659996262 & $28984_{-70}^{+46}$ & $4.110_{-0.032}^{+0.03}$ & $+0.007_{-0.0054}^{+0.011}$ & $13.47_{-0.14}^{+0.4}$ & $5.359_{-0.2}^{+0.23}$ & $1.786e+04_{-1.5e+03}^{+3.3e+03}$ & $0.005248_{-0.00098}^{+0.0014}$ & $1_{-0}^{+0}$ & --- \\
MAXI J0903-531 & 136.27841934719544 & -53.50542318726667 & $22971_{-327}^{+112}$ & $3.996_{-0.01}^{+0.017}$ & $-0.971_{-0.024}^{+0.09}$ & $7.497_{-0.21}^{+0.097}$ & $4.557_{-0.12}^{+0.056}$ & $5200_{-4.7e+02}^{+2.1e+02}$ & $0.03389_{-0.00077}^{+0.0016}$ & $1_{-0}^{+0}$ & 0.79114 \\
MAXI J1820+070 & 275.09142659019994 & 7.1853569054 & $5604_{-29}^{+35}$ & $4.694_{-0.022}^{+0.022}$ & $-4.064_{-0.062}^{+0.16}$ & --- & $0.5995_{-0.013}^{+0.0083}$ & --- & --- & --- & 0.1417 \\
AAO+28 342 & 88.9793368304 & 28.78511718449 & $16927_{-305}^{+683}$ & $3.614_{-0.12}^{+0.37}$ & $+0.407_{-0.42}^{+0.14}$ & $6.272_{-0.12}^{+0.13}$ & $4.677_{-0.21}^{+0.25}$ & $1957_{-1.6e+02}^{+2e+02}$ & $0.03981_{-0.014}^{+0.029}$ & $1_{-0}^{+1}$ & --- \\
NGC 6649 9 & 278.36791847053 & -10.402432490840003 & $35005_{-9}^{+21}$ & $3.835_{-0.056}^{+0.022}$ & $+0.006_{-0.004}^{+0.0099}$ & --- & $10.24_{-0.36}^{+1}$ & --- & --- & --- & --- \\
V1341 Cyg & 326.17147681053 & 38.32140738083 & $14210_{-105}^{+54}$ & $4.174_{-0.029}^{+0.053}$ & $-1.023_{-0.078}^{+0.063}$ & $3.781_{-0.078}^{+0.066}$ & $2.388_{-0.15}^{+0.083}$ & $206.9_{-25}^{+12}$ & $0.2399_{-0.0055}^{+0.023}$ & $1_{-0}^{+0}$ & 9.8435 \\
PSR J1023+0038 & 155.94868409363 & 0.64472385931 & $5966_{-4}^{+12}$ & $4.654_{-0.016}^{+0.015}$ & $-4.119_{-0.022}^{+0.038}$ & --- & $0.6593_{-0.0082}^{+0.0082}$ & --- & --- & --- & 0.198094 \\
AX J1755.7-2818 & 268.92847781693996 & -28.302572668440003 & $3290_{-11}^{+14}$ & $3.970_{-0.023}^{+0.037}$ & $-1.327_{-0.021}^{+0.024}$ & --- & $0.5195_{-0.02}^{+0.014}$ & --- & --- & --- & --- \\
V2116 Oph & 263.00897184809 & -24.745591435860003 & $4275_{-32}^{+21}$ & $1.227_{-0.023}^{+0.03}$ & $-0.836_{-0.076}^{+0.12}$ & $1.089_{-0.35}^{+0.88}$ & $28.92_{-0.97}^{+0.8}$ & $529.7_{-1.6e+02}^{+4.9e+02}$ & $4.787_{-3.7}^{+8.4}$ & $3_{-0}^{+5}$ & 1160.8 \\
\multicolumn{12}{c}{...} \\
  \multicolumn{12}{c}{A total of 339 sources} \\
  \multicolumn{12}{c}{...} \\
Cir X-1 & 230.17017640757 & -57.16674216654 & $4296_{-19}^{+17}$ & $2.029_{-0.011}^{+0.028}$ & $-0.970_{-0.053}^{+0.065}$ & --- & $11.47_{-0.38}^{+0.14}$ & --- & --- & --- & 0.03563 \\
PSR J2032+4127 & 308.05466935591 & 41.45675875004 & $29996_{-126}^{+129}$ & $4.037_{-0.0091}^{+0.01}$ & $+0.264_{-0.034}^{+0.025}$ & --- & $6.257_{-0.083}^{+0.097}$ & --- & --- & --- & 17000.0 \\
SRGE J204319.0+443820 & 310.82775703774 & 44.6388912772 & $7543_{-77}^{+76}$ & $3.172_{-0.039}^{+0.033}$ & $-0.085_{-0.0076}^{+0.0045}$ & $2.719_{-0.15}^{+0.19}$ & $7.187_{-0.27}^{+0.32}$ & $145.5_{-27}^{+47}$ & $0.4787_{-0.081}^{+0.071}$ & $2_{-0}^{+0}$ & --- \\
1FGL J1417.7-4407 & 214.3773592512 & -44.04932767897 & $5026_{-22}^{+21}$ & $3.650_{-0.034}^{+0.035}$ & $+0.049_{-0.03}^{+0.036}$ & $1.142_{-0.052}^{+0.056}$ & $2.6_{-0.13}^{+0.13}$ & $3.875_{-0.35}^{+0.34}$ & $7.074_{-1.1}^{+0.99}$ & $3_{-1}^{+0}$ & 5.373672 \\
3FGL J1544.6-1125 & 236.16412031889 & -11.4680221947 & $4488_{-42}^{+285}$ & $4.900_{-0.04}^{+0.022}$ & $-1.676_{-0.57}^{+0.17}$ & $0.45_{-0.023}^{+0.063}$ & $0.405_{-0.022}^{+0.026}$ & $0.05598_{-0.0065}^{+0.032}$ & $7.587_{-6.8}^{+5.6}$ & $1_{-0}^{+0}$ & --- \\
IGR J17329-2731 & 263.21111317334 & -27.5004381073 & $3515_{-2}^{+4}$ & $0.422_{-0.0033}^{+0.0075}$ & $-0.130_{-0.0001}^{+0.0001}$ & --- & $125.4_{-1}^{+0.44}$ & --- & --- & --- & --- \\
4FGL J0427.8-6704 & 66.95685947443 & -67.07640585049 & $5962_{-2}^{+6}$ & $4.693_{-0.0068}^{+0.005}$ & $-4.130_{-0.016}^{+0.026}$ & $0.9159_{-0.05}^{+0.046}$ & $0.7527_{-0.11}^{+0.16}$ & $0.4463_{-0.12}^{+0.19}$ & $2.797_{-2.6}^{+4.9}$ & --- & --- \\
2MASS J16233414-2631336 & 245.8921168912 & -26.52598812115 & $5921_{-90}^{+78}$ & $3.686_{-0.057}^{+0.054}$ & $-2.187_{-0.28}^{+0.13}$ & $1.132_{-0.065}^{+0.059}$ & $1.77_{-0.15}^{+0.18}$ & $3.288_{-0.59}^{+0.75}$ & $6.552_{-1.2}^{+1.5}$ & $2_{-0}^{+0}$ & --- \\
PSR J1723-2837 & 260.8465812119 & -28.63266533424 & $5828_{-41}^{+46}$ & $4.489_{-0.027}^{+0.022}$ & $+0.079_{-0.05}^{+0.044}$ & $0.9962_{-0.043}^{+0.049}$ & $1.082_{-0.045}^{+0.038}$ & $1.217_{-0.088}^{+0.072}$ & $6.814_{-2.1}^{+1.9}$ & $1_{-0}^{+0}$ & --- \\
AX J1735.8-3207 & 263.94283856383 & -32.11940100552 & $7166_{-6}^{+5}$ & $4.587_{-0.0012}^{+0.0009}$ & $-3.381_{-0.011}^{+0.016}$ & --- & $0.8357_{-0.0011}^{+0.0012}$ & --- & --- & --- & --- \\
\enddata
\end{deluxetable*}
\end{longrotatetable}

\begin{deluxetable*}{ll|ll|ll|ll}
\tablecaption{X-ray Binary Stars Statistical Table}  \label{tab:statistic}
\tablehead{
\colhead{X-ray Type} & \colhead{Number}  & \colhead{From LAMOST} & \colhead{Number}  & \colhead{From Gaia} & \colhead{Number} & \colhead{Intergrated by this paper} & \colhead{Number}
}
\startdata
\textbf{Total X-ray Binaries}               &   4058  &  \textbf{Observed by LAMOST (DR9)}      &    36  &  \textbf{Observed by Gaia (DR3)}     &   828   &  \textbf{Integrated Catalog}    &   344  \\  
\textbf{confirmed}                              &   3725  &       teff                                                     &    17  &     teff                                     &   329   &   teff                                        &   344  \\ 
 confirmed HMXB                              &   1501  &        [M/H]                                                    &    17  &    logg                                    &   329   &   logg                                      &   344  \\ 
 confirmed LMXB                               &   819   &        logg                                                   &    17  &    [M/H]                                      &   324   &   [M/H]                                       &   339  \\ 
 confirmed Unclassified                     &   1405  &                                                                  &         &     mass                                 &    59    &     mass                                  &   264  \\ 
\textbf{candidate}                              &   333    &                                                                  &         &     radius                                &   320   &     radius                                 &   338  \\ 
 HMXB candidate                              &   107    &                                                                  &         &     age                                    &    41   &     age                                     &   258  \\ 
 LMXB candidate                              &    32     &                                                                  &          &     pmra                                  &  603    &    Luminosity                              &   279        \\                    
 Unclassified candidate                    &   194    &                                                                  &          &      pmdec                               &  603    &                                               &            \\             
\textbf{With precise coordinates}       &   1413  &                                                                  &          &      distance                            &  320   &                                               &            \\        
\textbf{Included by GCVS or VSX}   &   219    &                                                                  &          &       rv                                      &  30      &                                               &            \\         
\hline
\enddata
\end{deluxetable*}

We retrieved 4058 X-ray binary stars from Simbad \citep{2000A&AS..143....9W} and three recent X-ray binary catalogs \citep{2023A&A...671A.149F, 2023arXiv230316168A, 2023arXiv230316137N}, including 3725 confirmed and 333 candidates. We eliminated sources with low coordinates precision, leaving 1413 X-ray binary stars with high-precision coordinates, whose errors in right ascension and declination are less than 0.4 arc seconds. Only high-precision coordinates can be used for further cross-identification with other sky surveys. Among the X-ray binary stars with high-precision coordinates, 1186 are confirmed and 227 are candidates.

Through cross-referencing with LAMOST, 36 sources were observed by LAMOST, 17 of which were provided with atmospheric parameters, and 16 with radial velocities. Two sources 2E 168 and TYC 3852-1069-1 have atmospheric parameters provided by both Gaia and LAMOST, and by comparing the results, the temperatures are very close to each other, but the differences in logg and [M/H] are 0.6 dex and 0.6 dex, respectively. For source TYC 3852-1069-1, Gaia provided a mass of 0.943 $M_{\odot}$, and our calculation based on LAMOST's parameters is 0.974 $M_{\odot}$, which is very close. 

We also cross-referenced the X-ray binaries with the GCVS and VSX variable star catalogs and found that 219 sources have been included as variable stars. Through cross-referencing with Gaia DR3, we found that more than 300 sources have temperature and other parameters provided by Gaia.

The information in Table \ref{tab:total} includes the star name, coordinates (RA and DEC) and their errors (RA$_{err}$ and DEC$_{err}$). The errors from Simbad are not included in the original information, they are calculated based on the coordinate values and the precision information. If the errors in RA and DEC are less than 0.00011 degrees, that is less than 0.4 arc seconds, it is considered to have high precision errors.

Table \ref{tab:total} also provides the X-ray type of each target, which is divided into confirmed but unclassified X-ray binary XB*, confirmed high-mass X-ray Binary HXB, confirmed low-mass X-ray Binary LXB, unclassified X-ray binary candidate XB?, high-mass X-ray Binary candidate HX?, and low-mass X-ray Binary candidate LX?.

The last two columns of Table \ref{tab:total}  provide the literature reference bibcode for the coordinates and types.

Table \ref{tab:gv} presents the cross-referenced results of X-ray binaries with GCVS and VSX catalogs, with a total of 219 common targets, representing 5\% of all X-ray binaries. It should be noted that we only cross-referenced 1413 X-ray binaries, because the coordinates of other X-ray binaries are too low in precision to be used for comparison, and even if cross-referenced successfully, it cannot be considered as the same target. All the tables in this paper are comparison results with 1413 binaries.

GCVS and VSX provide the variable star type of each target, which is too complex to be discussed in detail in this paper, please refer to the official instructions at \url{http://cdsarc.u-strasbg.fr/ftp/cats/B/gcvs/vartype.txt} and \url{https://www.aavso.org/vsx/index.php?view=about.vartypes}. In addition to the variable star type, we consider the period as another important parameter. The last two columns are the periods of the variable stars. The periods come from different literature, and generally refer to the light variation period of the source. Here, it can be understood as the orbital period of the X-ray binary system, as well as the light variation period.

Table \ref{tab:lam} is the cross-result of X-ray binaries with LAMOST DR9. The first column is still the star name from Simbad, then the coordinates from LAMOST. The most important information is the atmospheric parameters given by LAMOST: effective temperature $T_{eff}$, surface gravity logg, metallicity [M/H]. These three names are commonly used in the following tables with the same meanings. The metallicity given in the official star catalog of LAMOST is [Fe/H]. Although physically, [Fe/H] is not equal to [M/H], it is generally believed that the conversion coefficient between them is very close to 1. In order to unify with the metallicity from Gaia, we use the name [M/H] here, although its actual situation is [Fe/H]. The last three columns are the average radial velocity, spectral type, and number of observations. Spectral types are all single letters from MK spectral types.

Table \ref{tab:gaia_phy} is the cross-result of X-ray binaries with Gaia DR3 and is the main data source of this paper. Gaia not only provides high-precision coordinates and three atmospheric parameters, but also provides mass, radius, luminosity, age, and evolutionary stage for a target. Gaia provides multiple sets of parameters from various models for a target, so we need to make a selection among them. For the astrophysical parameters, the available models include The Final Luminosity Age Mass Estimator (FLAME), The General Stellar Parametrizer from Photometry (GSP-Phot), The General Stellar Parametrizer from Spectroscopy (GSP-Spec), The Extended Stellar Parametrizer for Hot Stars (ESP-HS), and The Extended Stellar Parametrizer for Ultra Cool Dwarfs (ESP-UCD). We select the parameters in this order and prioritize the results from the earlier models.

For Mass, Age, and Luminosity, only FLAME provides them, and for radius, although GSP-Phot also provides it, in order to consider the self-consistency between parameters, we prefer to use FLAME's radius. For the three atmospheric parameters, there is no best model in most cases. We had roughly compared them and tend to use GSP-Phot's results. This model uses low-resolution BP/RP spectra, although its resolution is not as high as the RVS spectra that GSP-Spec is based on, but its spectral range is wider, and it also takes into account the apparent G magnitude and parallax, so compared to GSP-Spec, it can get more self-consistent parameters.

For ESP-HS and ESP-UCD, they are aimed at hot stars and cool stars, respectively, but after our comparison, we did not find their superiority. So their priority is the lowest.

The final selection results show that most atmospheric parameters are selected from GSP-Phot, not only because of its high priority, but also because it provides much more parameters than other models. It can be said that most of Gaia's atmospheric parameters are provided by the GSP-Phot model.

Table \ref{tab:gaia_astrometry} is also the result of cross-matching X-ray binaries with Gaia DR3, and its parameters include not only the star names and coordinates, but also distance, the projected rotational velocity (vsini), the abundance of alpha elements ([Alpha/Fe]), the g-band magnitude (Gmag), the spectral type, and the Gaia source id. According to some literature, many X-ray binaries come from outside the Milky Way, but according to Gaia's distance measurements, except for the Large and Small Magellanic Cloud, all the sources are within the Milky Way. Therefore, the distance parameter may not be reliable. l and b are the Galactic longitude and latitude. The projected rotational velocity (vsini) refers to the rotation speed of the star at the equator, and i is the inclination angle of the equatorial plane of the orbit.

Table \ref{tab:uni} is a table of physical parameters that integrates all collected information, including the star names and coordinates, three atmospheric parameters, mass, radius, luminosity, age, evolutionary stage and period. The three atmospheric parameters come from the official tables of LAMOST or Gaia. A small portion of the physical parameters, such as mass, come from the original Gaia table, but most of them come from our calculations based on the atmospheric parameters. 22\% of the mass is provided by Gaia, and the remaining 78\% is from our calculation. If Gaia provides physical parameters, then we use Gaia's parameters; if Gaia does not provide parameters, we perform calculations.

The evolutionary stage is determined by our calculations, with different numbers representing different evolutionary phases. 1 represents the main sequence, 2 represents the subgiant branch, or the Hertzsprung gap for intermediate and massive stars, 3 represents the red giant branch, or the rapid stage of red giant for intermediate and massive stars. For a complete description, please refer to the website \url{http://stev.oapd.inaf.it/cmd_3.7/faq.html}. The period is derived from VSX or GCVS.

The last Table \ref{tab:statistic} is a stars count of the previous six tables. From the first two columns of the table, it can be seen that there are 1413 X-ray binary stars with precise coordinates, which is only 1413/4058=35\% of the total, this is because the majority of detections are based on X-ray wavelength data, which has low spatial resolution compared to optics, so it often cannot provide precise coordinates. Of these 1413, only 219 have been included in GCVS or VSX as variable stars, a ratio of 219/1413=15\%. This demonstrates a lack of optical observation for X-ray binary stars, as well as a lack of detailed physical parameters.

The third and fourth columns of Table \ref{tab:statistic} show the number of X-ray binary stars that have been cross-matched with the LAMOST database. Although LAMOST is the second largest database of stellar spectra (just overtaken by Gaia DR3), providing over 10 million stellar spectra, we only cross-matched 36 X-ray binary stars, of which only 17 provided atmospheric parameters. This small proportion suggests that X-ray binary stars are generally dim and difficult to observe.

The fifth and sixth columns of Table \ref{tab:statistic} show the number of X-ray binary stars that have been cross-matched with the Gaia database. There were 828 X-ray binaries observed by Gaia, of which more than 300 provided atmospheric parameters with a proportion of 300/1413=21\%. These data formed the basis of this study. Although Gaia only provided 59 masses and 41 ages, we calculate over 200 physical parameters based on their atmospheric parameters. We, consequently, developed a new quantitative classification criteria based on these physical parameters.

The last two columns of Table \ref{tab:statistic} show the number of parameters that we obtained by compiling all of the surveys. We provide over 300 atmospheric parameters and over 200 physical parameters. These parameters constitute the largest and also the most parameter-rich X-ray binary star catalog to date. The understanding of X-ray binary stars has been largely expanded in terms of observation.

\subsection{The physical parameters of the optical components} 

In order to obtain as many physical parameters as possible, we calculated physical parameters based on atmospheric parameters. It should be emphasized that we calculated the parameters of the optical component stars in X-ray binary systems, not the compact components. The input parameters we used were atmospheric parameters. Although these atmospheric parameters are the results of observing the entire X-ray binary system, the optical luminosity of the compact component is much lower than its companion, and can be almost ignored in the optical wavebands, so the observed atmospheric parameters almost reflect the surface atmospheric conditions of the optical component. Therefore, these atmospheric parameters can be used to calculate the physical parameters of the optical components.

This method is basically the isochrone interpolation method, which is a classic method that has been used for a long time. It is fundamentally based on atmospheric parameters and estimates parameters such as mass by comparing to a library of stellar database. It has been successfully applied and described by \citet{2019ApJS..244...43Z}. Here, we will introduce it briefly.

We already have a stellar parameter database that covers all possible stars, containing complete parameters for all ages, initial masses, and metallicities. Within this comprehensive database, we look for stellar samples with atmospheric parameters closest to our target, usually at least several dozens can be found. Then, the masses and radii of these dozens of stars becomes our estimated values for the target. We take the mass and radius of the sample that has the closest atmospheric parameters as the central value, and the range of parameters for all samples as the error value.

We used the stellar database of PARSEC (PAdova and TRieste Stellar Evolution Code; \citealp{2012MNRAS.427..127B, 2014MNRAS.444.2525C, 2015MNRAS.452.1068C, 2014MNRAS.445.4287T, 2017ApJ...835...77M,2019MNRAS.485.5666P,2020MNRAS.498.3283P}), version CMD 3.6, to calculate the absolute parameters of the optical star in X-ray binary systems. The Reimers mass loss coefficient on the RGB stage was set to $\eta_{Reimers}=0.2$, and the two-part-power law initial mass function (IMF; \citealp{2001MNRAS.322..231K, 2002Sci...295...82K, 2013pss5.book..115K}) was used. We downloaded the isochrone tables with the metallicity [M/H] ranging from -2.19 to +0.7 with a step of 0.02, and the log(age/yr) from 6.6 to 10.13 with a step of 0.01. We collected approximately 22 million sets of stellar parameters to interpolate the atmosphere parameters of our targets.

The FLAME model of Gaia provides parameters such as mass, radius, and age, which are essentially the same method as ours, using also the PARSEC database. The difference is that in addition to the atmospheric parameters, Gaia also uses extinction $A_{G}$, magnitude, and an estimate of the distance. With these three parameters, the radius can be directly obtained without relying on the stellar evolutionary database. Gaia provides a lot of radii, and we also provide radii that only rely on atmospheric parameters. Later, we will see that the radii given by the two methods are highly consistent.

\begin{figure*}
\gridline{\fig{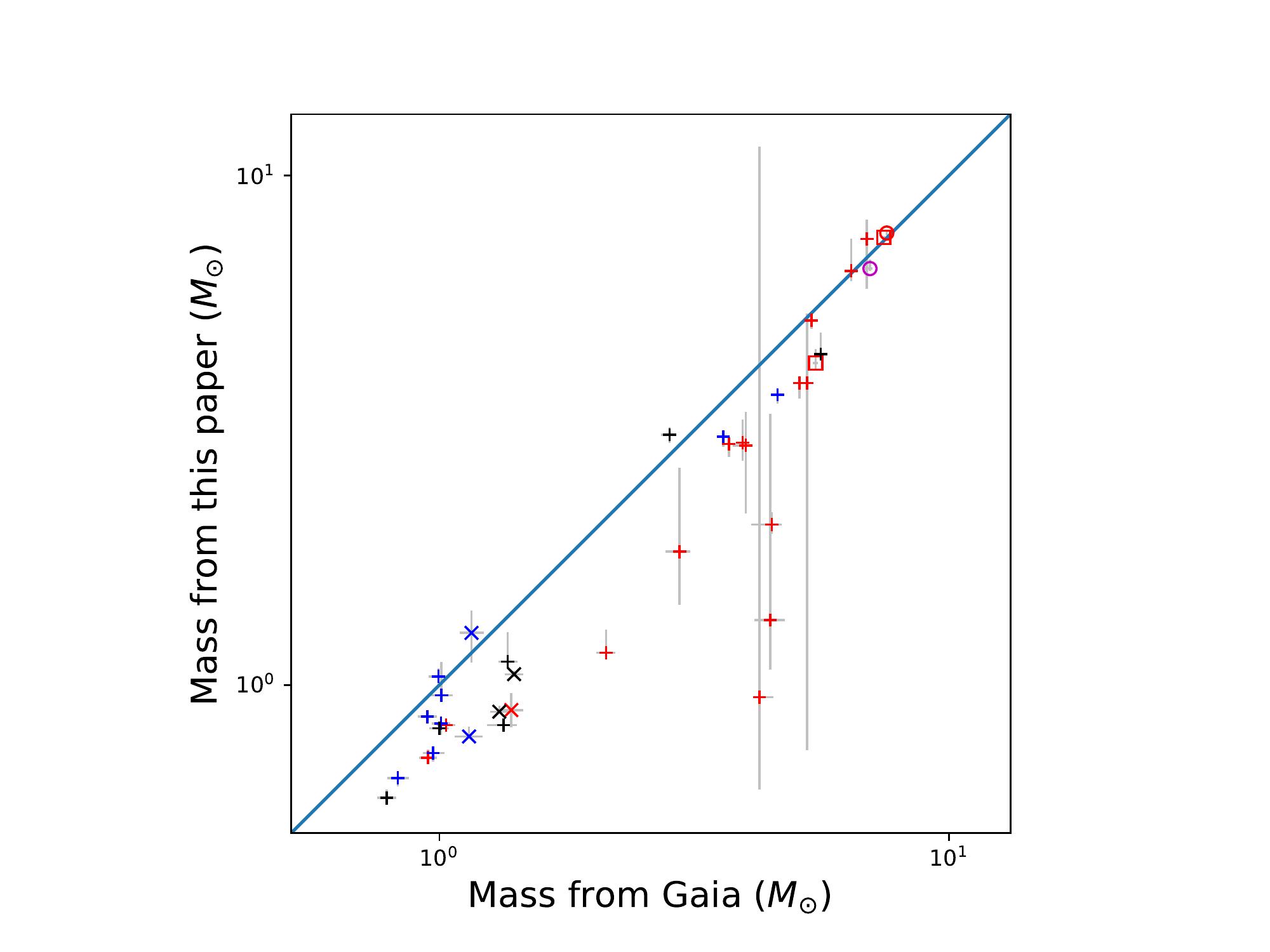}{0.5\textwidth}{(1)}
              \fig{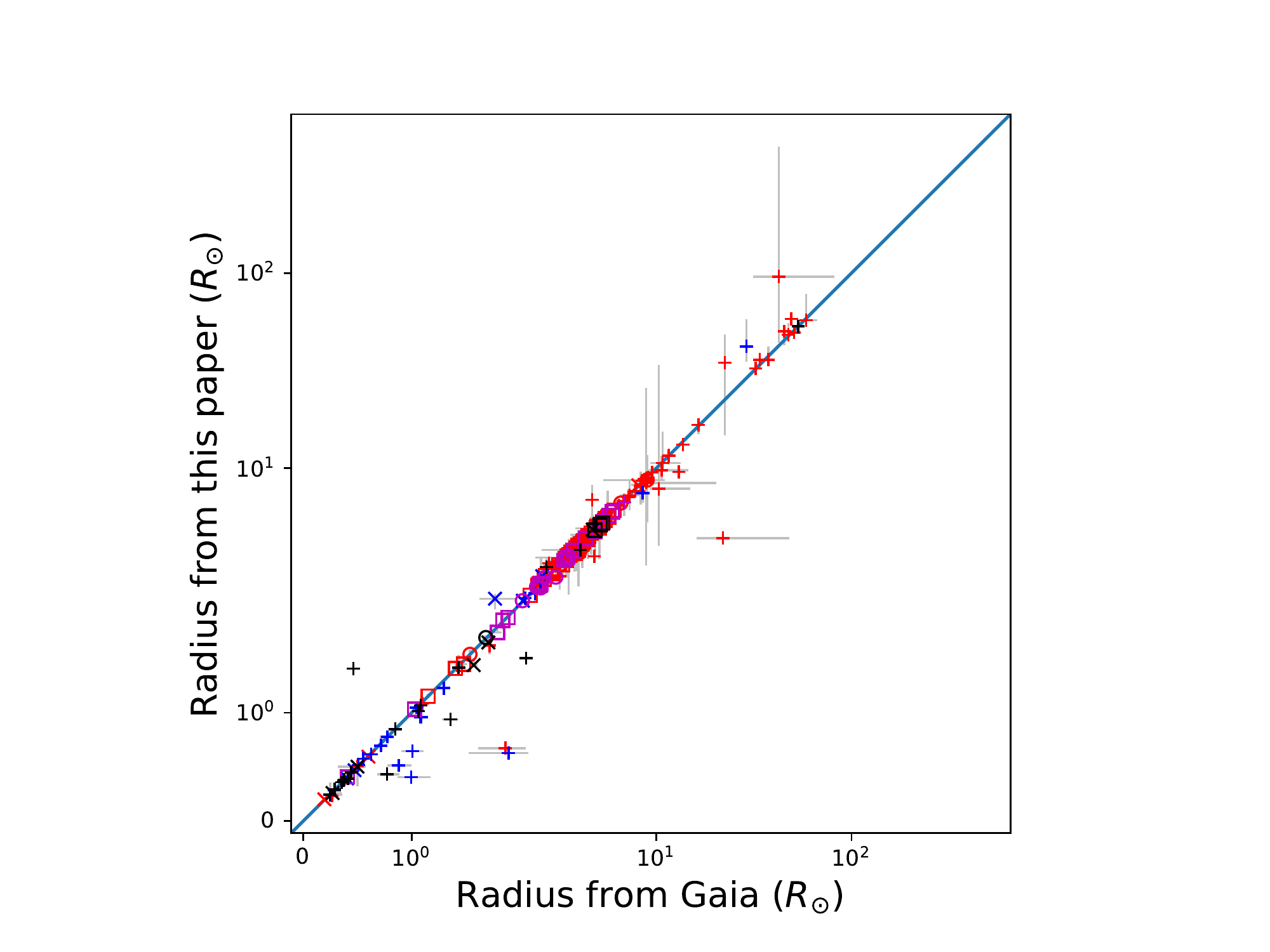}{0.5\textwidth}{(2)}
              }
\gridline{\fig{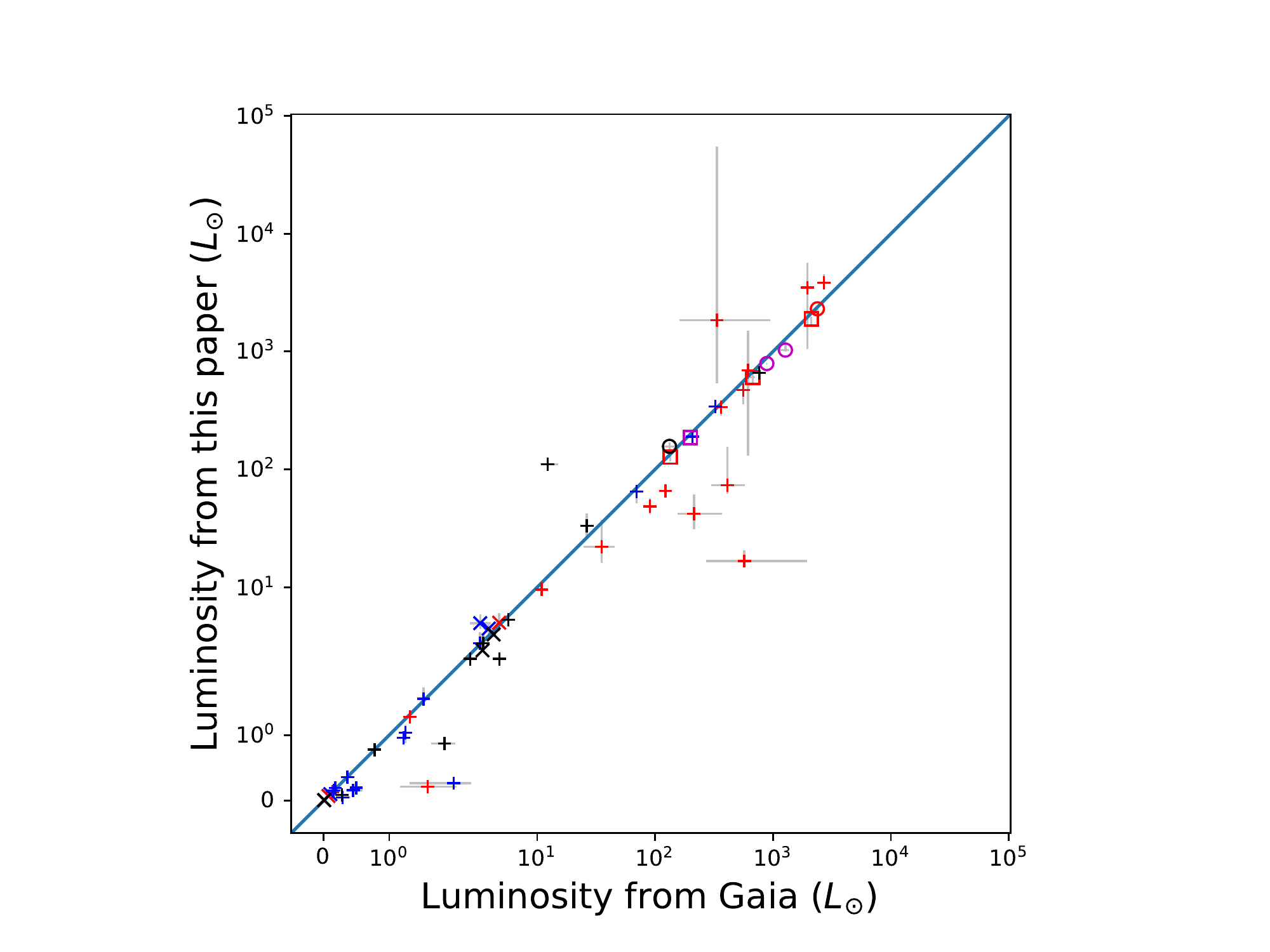}{0.5\textwidth}{(3)}
              \fig{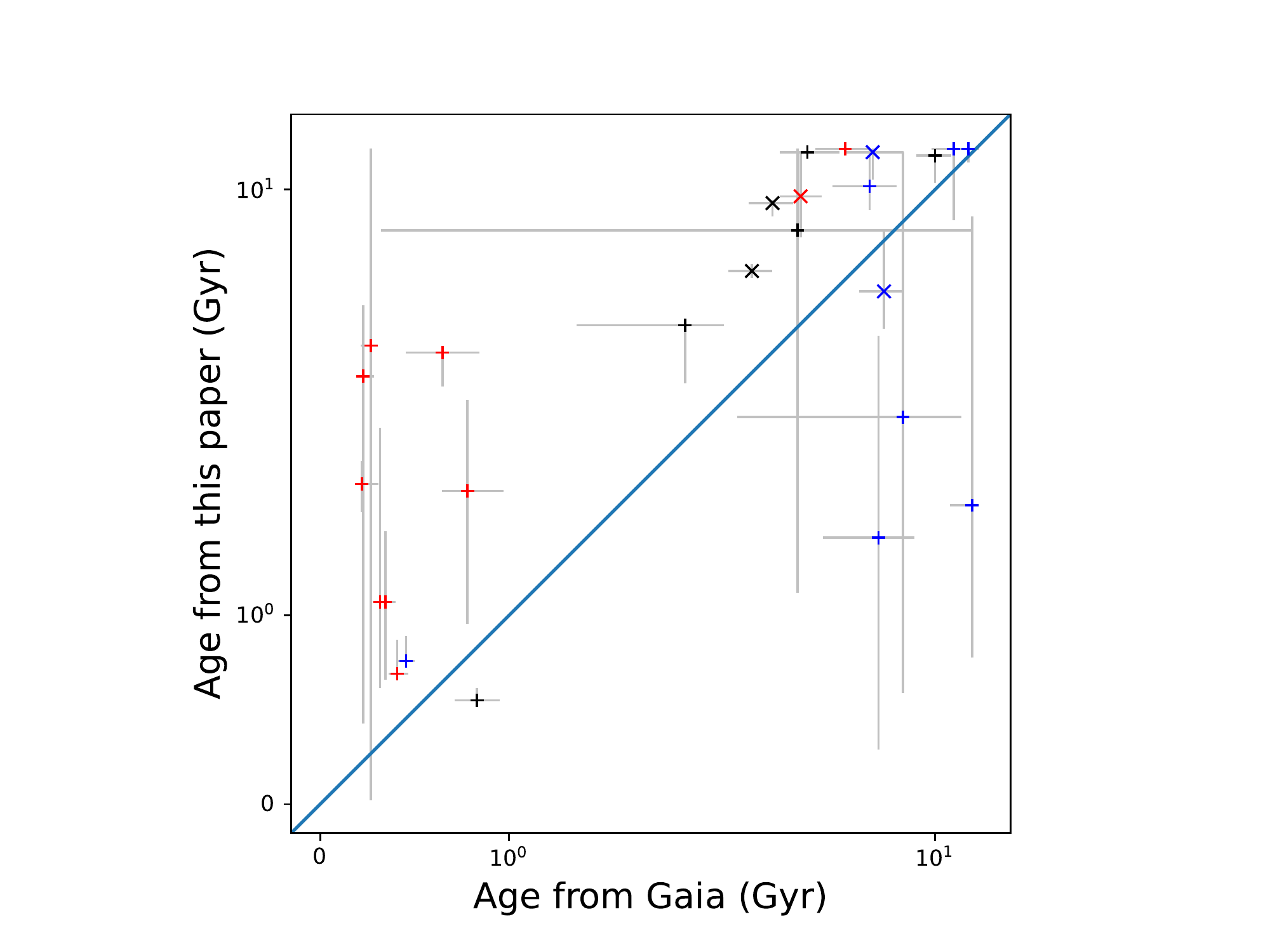}{0.5\textwidth}{(4)}
              }
\caption{Comparison of mass, radius, luminosity and age between Gaia and our calculation.  \label{fig:Comparison_gaia_us}}
\end{figure*}

In order to check the reliability of our results, we compared our results with those of Gaia, as shown in Figure \ref{fig:Comparison_gaia_us}. It can be seen that our calculated masses are generally lower than Gaia's results, with a mean deviation of 23\%, and six targets have very large errors.

For the comparison of radii, the consistency is surprising. Within a range of three orders of magnitude, there is no overall bias between them, and their errors are similar to each other.

For the comparison of luminosity, the two results are generally consistent. For the comparison of age, there is a huge difference, with a relative deviation from 0.05 to 12 times, and a mean value of 2.4 times. At the same time, the age errors are very large, some exceeding one order of magnitude. Despite this, there is still a weak positive relationship between them.

In conclusion, compared to Gaia, our radii and luminosity parameters are generally reliable, the mass is generally lower by 23\%, and the age is very unreliable, with a maximum difference of 12 times.

\subsection{Coordinate errors and the cross matching radius} 

Although this work is to study X-ray binaries using data from the optical wavelength, the initial discovery of X-ray binaries is from X-ray wavelength observations. 

In X-ray observations, the targets cannot be located with accurate coordinates like in optical observations. A typical example is 4U 2129+12 and CXOU J212958.1+121002, in the M15 star cluster, which are separated by 2.7 arc seconds \citep{2001ApJ...561L.101W}, and cannot be distinguished by ASM/RXTE. This is because X-ray sources are rare and faint in images, not to mention the need to subtract the effects of cosmic X-ray background sources. For faint sources, multiple observations need to be combined to detect them. 

To obtain accurate coordinates for X-ray targets, an absolute coordinate correction is needed and cross-correlation with known optical star catalogs (2MASS) is performed. If successful, the coordinates may be obtained accurately.

Relative to X-ray band, optical observation can obtain a large number of dense sources. Therefore, the first problem when crossing the X-ray stars with optical stars is the precision of the coordinates of X-ray sources. Although we have collected 4058 X-ray binary stars, the vast majority of targets do not have an accurate coordinate. Our requirement is that the errors of RA and DEC cannot exceed one ten thousandth of a degree, that is, 0.36 arc seconds. According to this requirement, only 1413 sources meet the requirements, with a proportion of 1413/4058=35\%. So we can only use these 1413 targets to cross with other optical survey catalogs. Among them, 831 targets are crossed to the optical counterpart, with a proportion of 831/1413=59\%.

The cross matching radius we used is 2 arc seconds, which is the lower limit commonly used in matching different star catalogs. A smaller radius may cause a problem of low matching success rate, but it has the advantage of higher accuracy. This makes the targets we match have a high probability of being the real optical counterparts of X-ray targets.

According to Gaia official recommendations, the matching radius is 1.8 arc seconds, where "1" is used to account for potential astrometric uncertainties between catalogs, and 0.8 arc seconds is for the possible deviation of the target caused by their proper motion from 2000 to 2016, calculated at 50 mas/yr. The proper motion of stars exceeding 50 mas/yr in the entire sky target is less than 0.2\%. Among the stars in this paper, 603 stars have proper motion data, and only four stars (2MASS J14023374+5422331, 2XMM J140234.4+542117, GSC 03852-01275, V822 Cen) have proper motion exceeding 50 mas/yr, accounting for 0.7\%.

For the LAMOST, GCVS and VSX tables, because the density of stars in them is significantly lower than that of Gaia, it is even less likely to cause problems with incorrect matches, that is, the radius of 2 arc seconds is also small enough. From past experience, false positive results are very rare and will not affect our statistical conclusions.

If the target is in a dense star field, then even within a radius of 2 arc seconds, there may be more than one optical target, making it difficult to determine which one is the corresponding one. The usual solution is to require a higher coordinate accuracy and then use a smaller radius to match.

However, due to observational constraints, the accuracy of the target cannot be infinitely high. The matching radius cannot be too small either, as that will certainly strengthen the reliability, but the number of matched sources will greatly decrease.

In practice, if there are multiple targets within the matching radius, we choose the closest and brightest one as the corresponding target. The deviation value of each target multiplied by its magnitude, the target with the smallest value is considered to be the most likely corresponding target.

\section{Discussion on the methods of obtaining the physical parameters of X-ray binaries}

This paper provides absolute parameters of optical companions in X-ray binary systems in bulk, based on spectral observations from Gaia and LAMOST and using stellar evolutionary models. We will discuss two cases: one that depends on stellar evolutionary models and another that does not.

\subsection{Independent of stellar evolutionary models}\label{sec:independent_model}

In the case that can not depend on stellar evolutionary models, it is not allowed to estimate mass and other parameters using atmospheric parameters or the spectral type. In this case, two conditions need to be met to measure the masses: first, the radial velocity curves (or the time variations due to the light time travel  effect) of both component stars need to be measured, and second, the inclination of the binary orbit needs to be determined.

For X-ray binaries, achieving the first condition is difficult. It is easy to measure the radial velocity of the optical companion star, but it is challenging to measure the radial velocity of the compact star because it has low luminosity (such as white dwarfs and neutron stars) or no luminosity (such as black holes), making it difficult to observe spectral lines directly. An indirect method is to measure the rotation velocity of the accretion disk around the compact star. For example,  \citet{2019Natur.575..618L, 2020ApJ...900...42L} measured the orbital velocity of the accretion disk around a stellar-mass black hole (with a mass of 12-50 $M_{\odot}$, which is still in intense debate) using four large telescopes (LAMOST, GTC, Keck, and the 3.5 m telescope at the Calar Alto Observatory). The semi-amplitude of the orbital velocity was found to be 11.2$\pm$0.3 km/s.

Although it is difficult to measure the radial velocity of the compact star directly, there is an indirect and feasible method: measuring the light time travel effect of the compact star, similar to measuring the mass of a third body using the eclipsing signal of a binary system. If the compact star is a neutron star, its pulse (spin) period can be measured in the X-ray band as it varies during the orbital period (\citealp{2015A&A...577A.130F} and the references therein). This method can be used to determine the projected semi-major axis $a_{x}\sin{i}$ and orbital eccentricity (e) of the compact star, and then calculate the radial velocity semi-amplitude of the compact star. This indirect method only applies to pulsars and requires high precision and continuous X-ray observations.

The second condition is the measurement of the orbital inclination $\sin{i}$, which is rarely measured reliably in practice. Currently, the methods used to measure $\sin{i}$ include analyzing the ellipsoidal modulation of the light curve and using approximate formulas (see \citealp{1984ARA&A..22..537J}, or review papers \citealp{2007A&A...473..523V} and \citealp{2015A&A...577A.130F})

The cases where the inclination can be reliably measured are usually the close eclipsing binaries, where one or both of the stars fill their Roche lobes and exhibit significant eclipse variations and ellipsoidal distortions. The inclination can be determined by analyzing the light curve. For X-ray binaries, it is difficult to observe eclipses between the two component stars, and even if eclipses occur, the resulting light variations are usually quite small. In addition, compact stars cannot exhibit ellipsoidal modulations, and although the optical companion of LMXBs may show ellipsoidal variations, the amplitude is typically small and highly dispersed. Furthermore, both eclipsing and ellipsoidal modulations are heavily affected by accretion processes. Even if high-quality light curves of some X-ray binaries are obtained and carefully corrected for the effects of accretion disks, pulsations, and magnetic activity, the small amplitude usually results in a large uncertainty in the analysis results.

\citet{2022NatAs...6.1203Y} obtain the orbital inclination of a non-accreting neutron star binary system through the analysis of its optical light curve. It should be noted that the physical parameters of the companion star were treated as known input parameters during the analysis. If not, the analysis will be highly uncertain. This is because the small amplitude of the light curve makes atmospheric conditions of the companion star a major factor affecting the results. The mass and semi-major axis of the companion star are important constraints, otherwise severe parameter degeneracy may occur.

In fact, the analysis of the light curve can simultaneously obtain both the orbital inclination and the mass ratio, so the radial velocity curve is not necessarily required. However, for X-ray binaries with small amplitude ellipsoidal modulations, the results are often unreliable. Parameters that differ significantly can give almost identical fitting curves. Considering that the observed light curve usually has a large dispersion, this method has unpredictable uncertainties.

The inclination can also be calculated using an approximate formula, which is derived from \citet{1984ARA&A..22..537J}. Inclination can be derived from four parameters: $R_{L}/a$ (the ratio between the Roche lobe radius and the separation of the two components), the Roche lobe filling factor $R_{star}/R_{L}$ (the ratio of the companion radius to the Roche lobe radius), the semi-eclipse angle $\theta_{E}$ fitted from light curve, and $\Omega$ (the ratio of the rotational frequency of the optical companion to the orbital frequency of the system).

This method requires an estimate or assumption of $R_{star}/R_{L}$ and $\Omega$. For example, assuming that the accretion is Roche lobe overflow, the companion star should be close to filling its Roche lobe, so both $R_{star}/R_{L}$ and $\Omega$ should be close to 1.

If $R_{star}/R_{L}$ is unknown, we cannot apply this method. The problem is if the star does not fill its Roche lobe, it is impossible to know $R_{star}/R_{L}$ in advance, because that means we already know the mass ratio, semi-major axis and radius.

\subsection{Dependent on stellar evolutionary models}

Since the above dynamical approach is fraught with difficulties, dependence on stellar evolutionary models has become an inevitable way. In the era of surveys, utilizing stellar evolutionary models can allow us to obtain a large number of relatively reliable parameters .

Relying on stellar models means inferring other stellar parameters from a partial set of observed stellar parameters. In this paper, we estimate the mass, radius, and age of stars using three atmospheric parameters. It is also feasible to estimate  
by using two (temperature and surface gravity) or one (temperature or spectral type) parameter, or by using temperature and luminosity \citep{2019ApJ...872...43H}.

We believe that the number and precision of input parameters determine the reliability of the output estimated parameters. The results obtained from three input parameters are better than those from two input parameters, and high-precision are better than low-precision.

Obtaining atmospheric parameters for stars is much easier compared to measuring radial velocity amplitudes and light curves. Extensive spectral surveys have already provided massive amounts of atmospheric parameters for stars. Over 300 million and 7 million stars with atmosphere parameters were released by Gaia DR3 and LAMOST DR9.

Before Gaia DR3, spectroscopic observations on X-ray binaries were quite rare, mainly due to their faintness and imprecise coordinates, which resulted in a lack of basic parameters for X-ray binaries as a whole. Nowadays, with the help of Gaia DR3, we can obtain basic parameters of X-ray binaries on a massive scale for the first time.

Previously, to obtain the mass of an optical companion of an X-ray binary, most studies relied on the companion's spectral type and compared it with empirical relationships to estimate the mass (see \citealp{2023A&A...671A.149F} and the references therein). This comparison process often involves assuming a main-sequence star, although it is not explicitly stated. A small number of studies have obtained temperature and surface gravity through spectral observations and assumed solar metallicity, and have used stellar models to obtain more reliable results than just using spectral types \citep{2015MNRAS.453..976F, 2010ApJ...724..306A}.

This paper's method relies on the stellar evolutionary model, and we will discuss its reliability in detail in next section.

\section{Discussion on the reliability of the parameters presented in this paper}

\subsection{Mass}

The masses are based on three atmospheric parameters ($T_{eff}$, logg, and [M/H]) and the PARSEC database. The reliability depends on three aspects: the accuracy of the three atmospheric parameters, the accuracy of the PARSEC model, and the applicability of the single-star model to the component star of X-ray binaries.

\subsubsection{The accuracy of the three atmospheric parameters}

The atmospheric parameters mainly come from Gaia observations with multiple models. We collected the results from all models and selected them according to a set priority. Overall, most parameters come from the GSP-Phot model. According to the official website of Gaia, the atmospheric parameters given by GSP-Phot are compared with the results of APOGEE DR16 \citep{2020AJ....160..120J}, GALAH DR3 \citep{2021MNRAS.506..150B}, LAMOST DR4 \citep{2011RAA....11..924W, 2014IAUS..306..340W} and RAVE DR6 \citep{2020AJ....160...83S}, with mean absolute differences of 150-418 K, 0.1-0.4 dex, and 0.24-0.3 dex for $T_{eff}$, logg, and [M/H], respectively. This deviation transferred to the stellar mass will cause a deviation of 10\% to 40\%.

\subsubsection{The accuracy of the PARSEC model}

The reliability of the stellar model is crucial. PARSEC is a very mature isochrone database developed for more than 10 years. It is designed to simulate the entire range of stellar evolution, from the birth of a star in a molecular cloud to its final stages as a white dwarf, neutron star, or black hole.

The accuracy of the PARSEC has been tested and validated against stars with known properties. For example, one study by \citet{2015ApJ...812...96G} compared PARSEC results to "a sample of 59 benchmark evolved stars with model-independent masses (from binary systems or asteroseismology) obtained from the literature." They found that "The average fractional difference in the mass interval $\sim$0.7-4.5 $M_{\odot}$ is consistent with zero (-1.30 $\pm$ 2.42\%)".

We compared the results of a B-type star with a temperature of about 14,000 K and logg around 3.5 under different evolutionary models. Under the same atmospheric parameters, the mass given by the PARSEC model has a difference of 3-18\% from other three models \citep{2020Natur.580E..11A, 2020A&A...634L...7S, 2021A&A...649A.167L}.

Compared with errors caused by other factors, the errors caused by the model itself are relatively minor.

\subsubsection{The applicability of the single-star model to the component star of X-ray binaries}

All isochrone databases are based on a single star model. However, the evolutionary history of X-ray binary stars is different from that of single stars, as their components have undergone significant mass loss or accretion. Our method uses atmospheric parameters to estimate its mass, which requires that the correspondence between the surface atmosphere of the star and its mass conform to the single star model. Whether the optical star of X-ray binary systems can meet this requirement requires examining whether their surface atmospheric composition is close to that of a single star. If close, then our method is feasible. If not close, then the method is not feasible.

The so-called surface atmospheric composition is close to that of a single star, means that the products of nuclear reaction in the core cannot affect the atmospheric composition. If the hydrogen shell of the optical star is completely lost, so that the helium produced by nuclear reaction has been exposed on the surface, or the abnormal imprints of CNO cycle (Increase in N abundance, decrease in C and O abundance, \citealp{2020A&A...633L...5I}) is found on its surface, then this situation is that the nuclear reaction products affect the atmospheric composition of the surface.

For the case that nuclear reaction products do not affect the composition of the surface atmosphere, it is feasible to use surface parameters to estimate the mass, regardless of whether the star has experienced material loss or gain. \citet{2020ApJ...895..136C, 2021ApJ...920...76C} calculated the parameters of pulsating star of two binary system (KIC 10736223 and OO Dra) using single star and binary star (considering matter accretion) models, respectively. These two pulsating stars have accumulated 2 and 1.7 times their initial mass during the evolution of binary stars, respectively, but the calculation results of the single star model and binary model are very close. The mass deviation calculated by the two models is only 2.5\% and 1\%. The deviation of the radius is also very small, at 0.7\% and 0.3\%, respectively. Although accretion significantly increases the mass of a star, its surface atmospheric composition has not changed too much. So the current internal and external structures still conform to the single star model, and their mass and radius can still be accurately calculated based on the single star model.

In the case of nuclear reaction affects the composition of the surface atmosphere, it is not feasible to use surface parameters to estimate the mass. If the hydrogen envelope is severely stripped, and so the star becomes He-rich or the uncommon CNO-processed material appears on the surface, the atmosphere of such a stripped star will differ greatly from a normal single star. In this case, using single-star models to estimate the mass based on atmospheric parameters would lead to a serious overestimation of the true mass. By comparing the stripped star masses provided by \citet{2020A&A...633L...5I} and \citet{2021A&A...649A.167L}, we found that the masses predicted by the PARSEC model are overestimated 5 to 7 times.

Therefore, the situation that can cause incorrect results is striped He-rich star, and we need to correct or eliminate this situation. However, this article chooses to ignore it because this situation is very rare. These stars have a short lifetime and are very rarely in observation. According to \citet{2021MNRAS.502.3436E}, the time spent in the stripped star phase is on the order of few $\times$ 10$^{5}$ yr, leading to a low probability of detection. It may take observing around 3000 targets to find one such rare object \citep{2020MNRAS.493L..22E}.

In this paper, we provide mass for over 250 X-ray binaries. Based on the above probability, the number of stars overestimated in mass due to the inapplicability of single-star models is less than 0.1. Thus, we consider this event to be of low probability and disregard it.

In summary on mass, taking into account the above three influencing factors, the final mass error mainly comes from the error of atmospheric parameters themselves. For over 99\% of the targets (excluding the less than 1\% of stripped He-rich stars), we suggest the typical relative errors of the masses are 10\% to 60\%.

\subsection{Radius}

As mentioned earlier, radius can also be estimated based on atmospheric parameters and using a single star model as mass. Furthermore, we will explain that the reliability of the radius can also be verified using independent methods.

We compared the radii obtained from Gaia with those calculated from atmospheric parameters, as shown in the figure \ref{fig:Comparison_gaia_us}. The standard deviation of the relative difference is 19.6\%. Gaia's method for determining radius does not rely on stellar models or assumptions. Gaia calculates luminosity from distance, apparent magnitude, and extinction, and then obtains radius from luminosity and temperature. This method is independent of stellar models, and therefore the mutual corroboration of the two methods can demonstrate the reliability of the results. Based on this, we suggest the typical relative errors of the radii are around 20\%.

\subsection{Age}

The method of using a single star model can calculate mass and radius, but cannot calculate age. Because age reflects the historical process rather than the current status. A component star that has lost a large amount of mass can be identical to an older single star of the same mass, but their formation history is completely different.

For the optical companion stars in X-ray binary systems that have experienced lots of mass loss, the single star model may severely overestimate their age, which may be several times larger than the true value.

Even without considering the overestimation caused by the model itself, the errors in atmospheric parameters can transmit huge errors on age. From the following figure \ref{fig:relation}, it will be seen that the error in age often spans an order of magnitude. The overestimation brought by the model may have been submerged in the huge error bars.

The age presented in this paper is in large uncertainty and severely overestimated. Therefore, the age parameters can only be used for semi-quantitative discussions and cannot be used for detailed analysis.

\section{Comparison of parameters between this paper and the literature}

Recently, \citet{2023A&A...671A.149F} and \citet{2023arXiv230316168A} reported catalogs of X-ray binaries, comprising 152 HMXBs and 348 LMXBs. Among them, 90 and 87 systems included the mass of the companion star. Using atmospheric parameters from Gaia and LAMOST, we computed masses for 37 and 16 of these stars. These 37 + 16 = 53 targets can be compared with previous results. Unfortunately, our results do not match the previous ones at all.

To explore the reasons for the conflicts and the reliability of the data as much as possible, We need to check each and every target in detail. Among the 53 targets that can be compared, more than 30 companion masses were estimated based on spectral types. Estimation based solely on spectral type should not be more reliable than results based on three atmospheric parameters combined with mature evolutionary models. Therefore, this kind of results cannot be used to examine the reliability of ours, and so we do not use them for comparison.

After excluding targets whose masses were estimated based on spectral type and targets of no mass information could be found in the original literature, we listed 12 X-ray binaries in Table \ref{tab:comp_lit_and_us} to compare with our results.

We further divided these 12 targets into four groups divided by horizontal lines in the Table \ref{tab:comp_lit_and_us}. The companion masses of the four targets in the first group were calculated based on the temperature $T_{eff}$, surface gravity logg (or luminosity), and a stellar model. It can be seen that our results are generally higher, with a mass ratio ranging from 1 to 2.3 in the sixth column of the Table. The difference mainly comes from the temperature difference. The temperatures of the first three targets are around 30,000 K, while Gaia's measurement are around 40,000 K, resulting in a larger mass. For the fourth target, V934 Her, its temperatures are all around 3,700 K, so the derived masses are also very close. If temperatures from Gaia are more accurate, the masses in this paper should be more reliable.

\begin{deluxetable*}{lllllllll}
\tablecaption{Comparison of optical companion mass between literature and this paper} \label{tab:comp_lit_and_us}
\tablewidth{0pt}
\tablehead{
\colhead{\makecell[l]{Name}} & \colhead{\makecell[l]{Mass from\\Literature ($M_{\odot}$)}}  & \colhead{\makecell[l]{Literature}}  &  \colhead{\makecell[l]{Description about the source of mass}}  & \colhead{\makecell[l]{Mass ($M_{\odot}$)}} & \colhead{\makecell[l]{$\frac{\textrm{Mass from this paper}}{\textrm{Mass from Literature}}$}}  &   \colhead{\makecell[l]{Temperature (K)}} & \colhead{\makecell[l]{logg}} & \colhead{\makecell[l]{$[$M/H$]$}} 
}
\startdata
1H 1555-552 & 19.4$\pm$5 & 2015MNRAS.453..976F & \makecell[l]{ \hspace{0.2cm}Based on $T_{eff}$ (28000$\pm$1500 K) and logg (3.75$\pm$0.25)\\ with the stellar model of PARSEC} & $29.4_{-0.7}^{+2}$ & 1.5 & $41287_{-175}^{+75}$ & $4.07_{-0.02}^{+0.01}$ & $-1.00_{-0.003}^{+0.01}$ \\
HD 259440 & 13.2-19.0 & 2010ApJ...724..306A & \makecell[l]{ \hspace{0.2cm}Based on $T_{eff}$ ($\approx$ 30,000 K) and logg ($\approx$ 4.0)\\ with evolutionary tracks of Schaller et al. (1992)} & $30.3_{-2}^{+1}$ & 1.6-2.3 & $41272_{-214}^{+85}$ & $4.06_{-0.03}^{+0.02}$ & $-0.99_{-0.004}^{+0.01}$ \\
X Per & 15.5 & 1997MNRAS.286..549L & \makecell[l]{ \hspace{0.2cm}Based on $T_{eff}$ (29500$\pm$1500 K) and logg (4.0$\pm$0.2)\\ with evolutionary tracks calculated by Claret (1995)} & $29.4_{-0.8}^{+0}$ & 1.9 & $41187_{-120}^{+78}$ & $4.09_{-0.008}^{+0.008}$ & $-1.00_{-0.003}^{+0.009}$ \\
V934 Her & $1.6^{+0.1}_{-0.2}$ & 2019ApJ...872...43H & \makecell[l]{ \hspace{0.2cm}Based on $T_{eff}$ (3650 K) and Luminosity (1200 $L_{\odot}$)\\ with models of Charbonnel et al. (1996) and Escorza et al. (2017)} & $1.58_{-0.04}^{+0.08}$ & 1 & $3707_{-2}^{+1}$ & $0.47_{-0.01}^{+0.01}$ & $-0.87_{-0}^{+0.02}$ \\
\hline
V662 Cas & 16$\pm$2 & 2017ApJ...844...16H & \makecell[l]{ \hspace{0.2cm}Mass function from Grundstrom et al. (2007)\\ inclination from light curve fitting \\ assumed neutron star mass of 1.4 $M_{\odot}$} & $25.1_{-2}^{+0}$ & 1.6 & $18566_{-64}^{+84}$ & $2.45_{-0.008}^{+0.01}$ & $+0.21_{-0.004}^{+0.02}$ \\
V2116 Oph & $<$1.22 & 2006ApJ...641..479H & \makecell[l]{ \hspace{0.2cm}Mass function and assumed neutron star mass of 1.35 $M_{\odot}$\\ result in a maximum mass of 1.22 $M_{\odot}$} & $1.09_{-0.3}^{+0.9}$ & $>$0.9 & $4275_{-32}^{+21}$ & $1.23_{-0.02}^{+0.03}$ & $-0.84_{-0.08}^{+0.1}$ \\
PSR J1723-2837 & 0.42$_{-0.02}^{+0.28}$ & 2013ApJ...776..20C & \makecell[l]{ \hspace{0.2cm}Mass ratio from double radial velocity curves \\ assumed neutron star mass of 1.4-2.0 $M_{\odot}$} & $0.996_{-0.04}^{+0.05}$ & 2.4 & $5828_{-41}^{+46}$ & $4.49_{-0.03}^{+0.02}$ & $+0.08_{-0.05}^{+0.04}$ \\
\hline
V818 Sco & 0.4 & 2021MNRAS.508.1389C & \makecell[l]{ \hspace{0.2cm}Mass ratio and inclination from light curve fitting\\ assumed neutron star mass 1.4 $M_{\odot}$} & $3.99_{-0.1}^{+0}$ & 10 & $16202_{-27}^{+12}$ & $4.40_{-0.002}^{+0.005}$ & $-0.25_{-0.04}^{+0.03}$ \\
1FGL J1417.7-4407 & 0.35 & 2015ApJ...804L..12S & \makecell[l]{ \hspace{0.2cm}Mass ratio and inclination from light curve fitting\\single radial velocity curve} & $1.14_{-0.05}^{+0.06}$ & 3.3 & $5026_{-22}^{+21}$ & $3.65_{-0.03}^{+0.04}$ & $+0.05_{-0.03}^{+0.04}$ \\
4FGL J0427.8-6704 & 0.65$\pm$0.08 & 2016ApJ...831...89S & \makecell[l]{ \hspace{0.2cm}Inclination from light curve fitting\\ mass ratio from double radial velocity curves} & $0.916_{-0.05}^{+0.05}$ & 1.4 & $5962_{-2}^{+6}$ & $4.69_{-0.007}^{+0.005}$ & $-4.13_{-0.02}^{+0.03}$ \\
\hline
\makecell[l]{V779 Cen \\ (Cen X-3)} & $20.2^{+1.8}_{-1.5}$ & 2007A\&A...473..523V & \makecell[l]{ \hspace{0.2cm}Mass ratio from double radial velocity amplitudes \\ inclination by the approach of Rappaport \& Joss (1983)\\ semi-eclipse angle $\theta_{e}$ from the light curve\\ adopting a Roche-lobe overflow systems\\ so the $\beta$ (the ratio of the companion radius to that of its Roche lobe)\\ is in the range 0.9 to 1.0} & $10.3_{-0.4}^{+0.6}$ & 0.6 & $19330_{-234}^{+228}$ & $3.33_{-0.02}^{+0.03}$ & $+0.20_{-0.03}^{+0.03}$ \\
\makecell[l]{GP Vel \\ (Vela X-1)} & 26$\pm$1 & 2015A\&A...577A.130F & \makecell[l]{ \hspace{0.2cm}Same method as V779 Cen(Cen X-3)} & $20_{-3}^{+1}$ & 0.8 & $18327_{-260}^{+352}$ & $2.63_{-0.04}^{+0.04}$ & $-0.73_{-0.02}^{+0.009}$ \\
\enddata
\end{deluxetable*}

The three targets in the second group are all based on the assumption of neutron star mass. The masses of this paper are generally larger by 1 to 2.4 times. As the true masses of the neutron stars are not clear, these three sources are not suitable for checking our results. It is worth noting that for PSR J1723-2837, the authors obtained double radial velocity curves, so the mass ratio of the two components should be reliable. Based on the assumption of a neutron star mass of 1.4 - 2.0 $M_{\odot}$, the mass of the companion star was estimated to be in the range of 0.4 - 0.7 $M_{\odot}$. Its upper limit 0.7 $M_{\odot}$ is close to our result of 1 $M_{\odot}$. If the mass of the neutron star is assumed to be 2.6 $M_{\odot}$ (like 4U 1700-37, \citealp{2007A&A...473..523V}),  then the mass of the companion star is 1 $M_{\odot}$.

The next three targets in the third group were calculated by using the light curves to fit the mass ratios or inclinations. The reliability of the light curve fitting depends heavily on the amplitude of the light curve. The smaller the amplitude, the less reliable the results, even if the theoretical curve perfectly fits the data. Checking the targets in detail, the light curve amplitudes of V818 Sco and 1FGL J1417.7-4407 are only 0.15 mag, while 4FGL J0427.8-6704 has an amplitude of 1 - 2 mag and shows obvious eclipsing structure. Therefore, the fitting results of 4FGL J0427.8-6704 should be very reliable, and its mass ratio from the radial velocity curves is also very reliable. We believe that the masses given by \citet{2016ApJ...831...89S} are the most reliable masses of X-ray binaries discovered by us. The companion star mass in this paper is 1.4 times of theirs, i.e. larger by 40\%. This is consistent with the conclusion of the previous section that the typical error of mass is 10\% - 60\%.

It is worth mentioning that the neutron star mass given by \citet{2016ApJ...831...89S} is 1.86 $\pm$ 0.1 $M_{\odot}$, higher than the common estimate of 1 to 1.4 $M_{\odot}$ for neutron stars.

For V818 Sco and 1FGL J1417.7-4407, the masses in this paper are 2.3 and 9 times larger than previous studies. Due to the small light curve amplitudes, the fitting results bear unpredictable uncertainties, so they are also not suitable for comparison.

The last two targets belong to a group where the inclinations were not obtained by fitting light curves, but calculated by an approximation formula from \citet{1984ARA&A..22..537J} described above in Section \ref{sec:independent_model} Independent of stellar evolutionary models. The input parameters $R_{star}/R_{L}$ and $\Omega$ in the formula were set to near 1 based on the assumption of Roche lobe overflow, as the usual understanding on LMXBs.  However, their resulting companion masses (20.2 and 26 $M_{\odot}$) were too large to be like LMXBs. If the optical companion does not fill its Roche lobe, the $R_{star}/R_{L}$ should be lower, and the inclination will be lower, then the mass will be smaller. Comparing these two targets, the masses in this paper are 40\% and 20\% smaller.

In conclusion, the target with the most reliable companion mass is considered to be 4FGL J0427.8-6704 \citep{2016ApJ...831...89S}, and our estimated mass are 40\% larger, consistent with the conclusion of a 10\% - 60\% relative error. The second most reliable targets in our opinion are PSR J1723-2837, V779 Cen, and GP Vel, our estimated mass are 0.6-2.4 times of the those from the literature. Unless more data become available, we can only come to these limited comparative results.

\section{Parameters distribution and relationship} \label{sec:dist_rela}

\begin{figure*}
\gridline{\fig{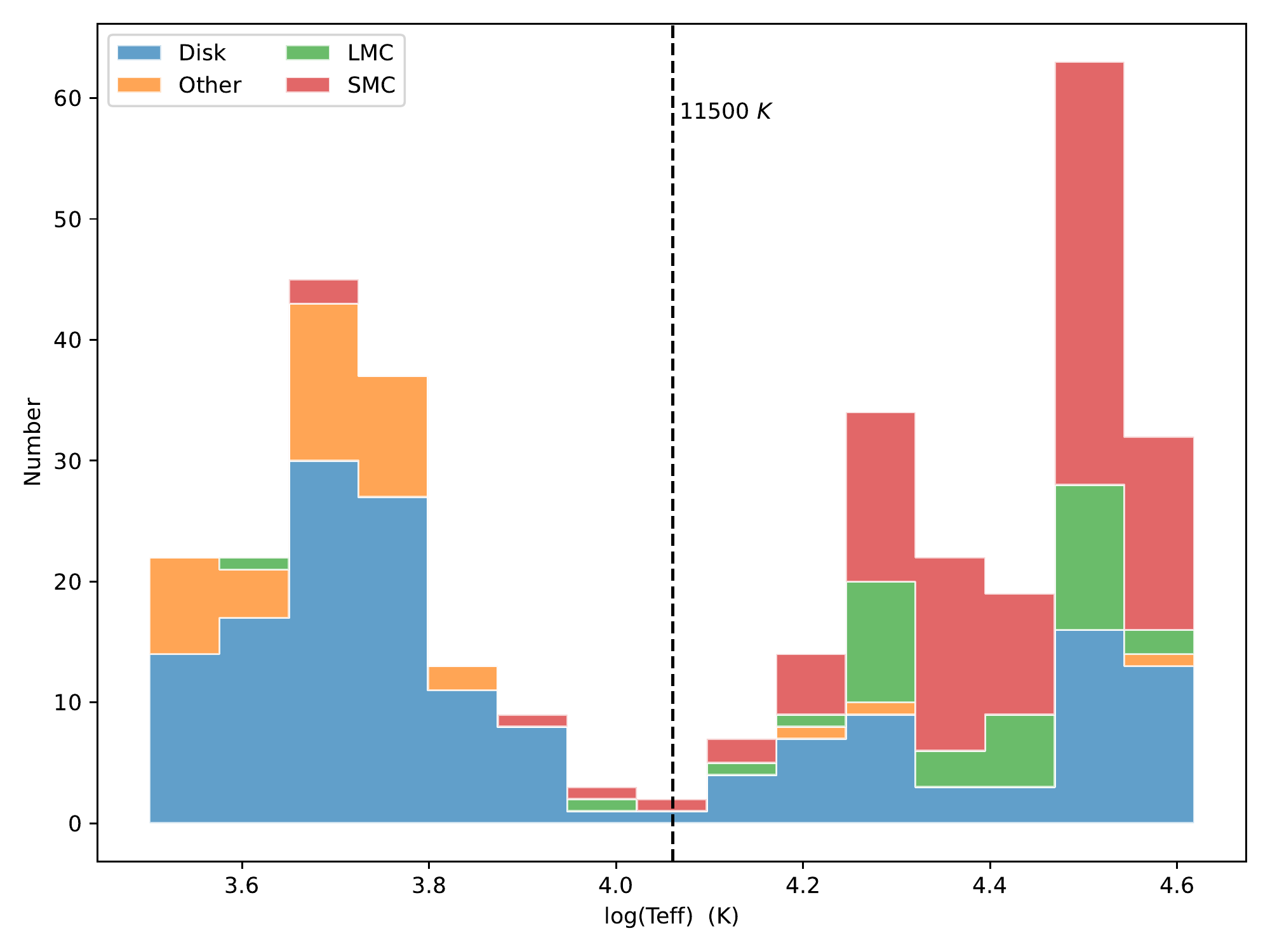}{0.3\textwidth}{(1)}
              \fig{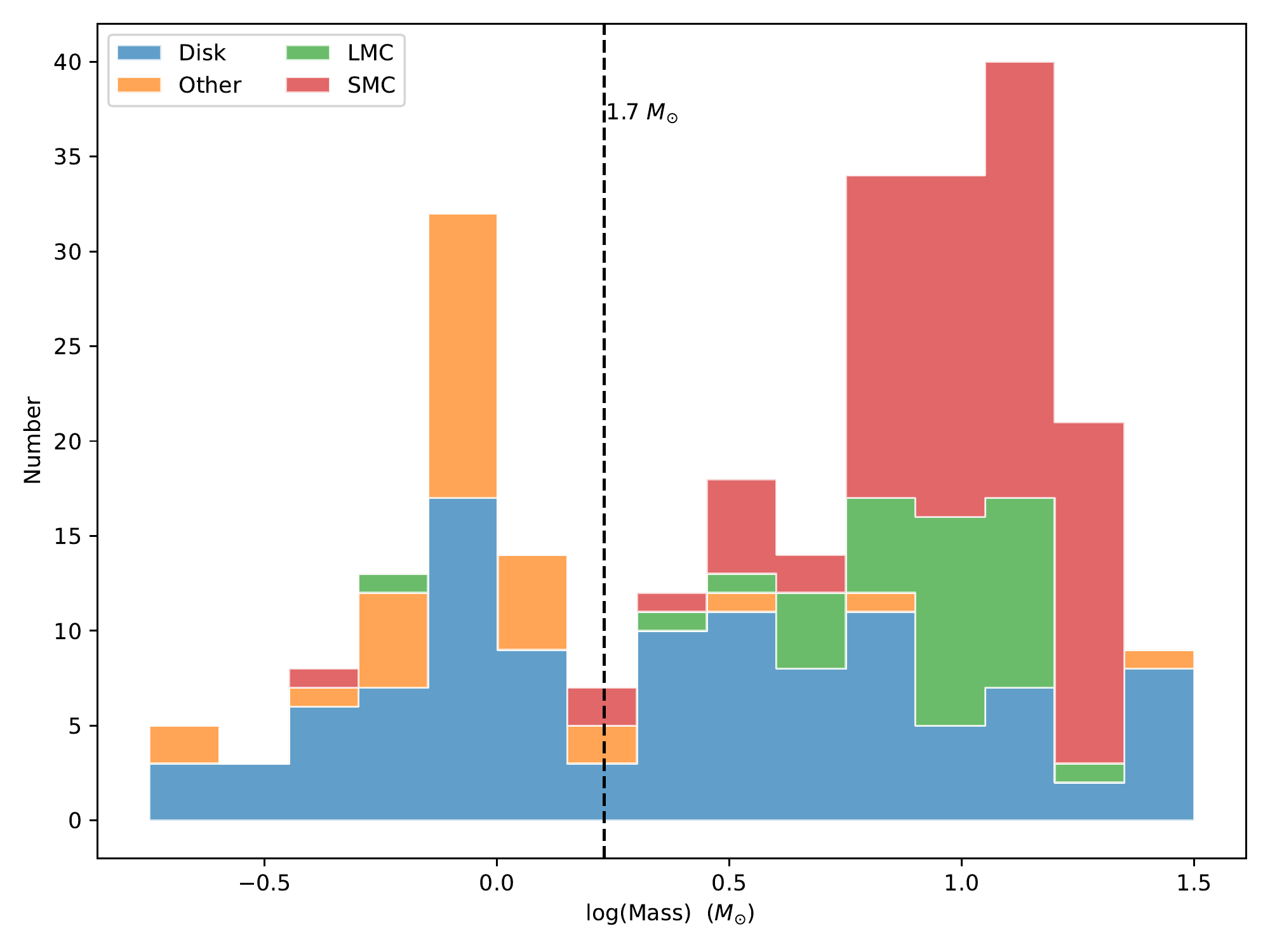}{0.3\textwidth}{(2)}
              \fig{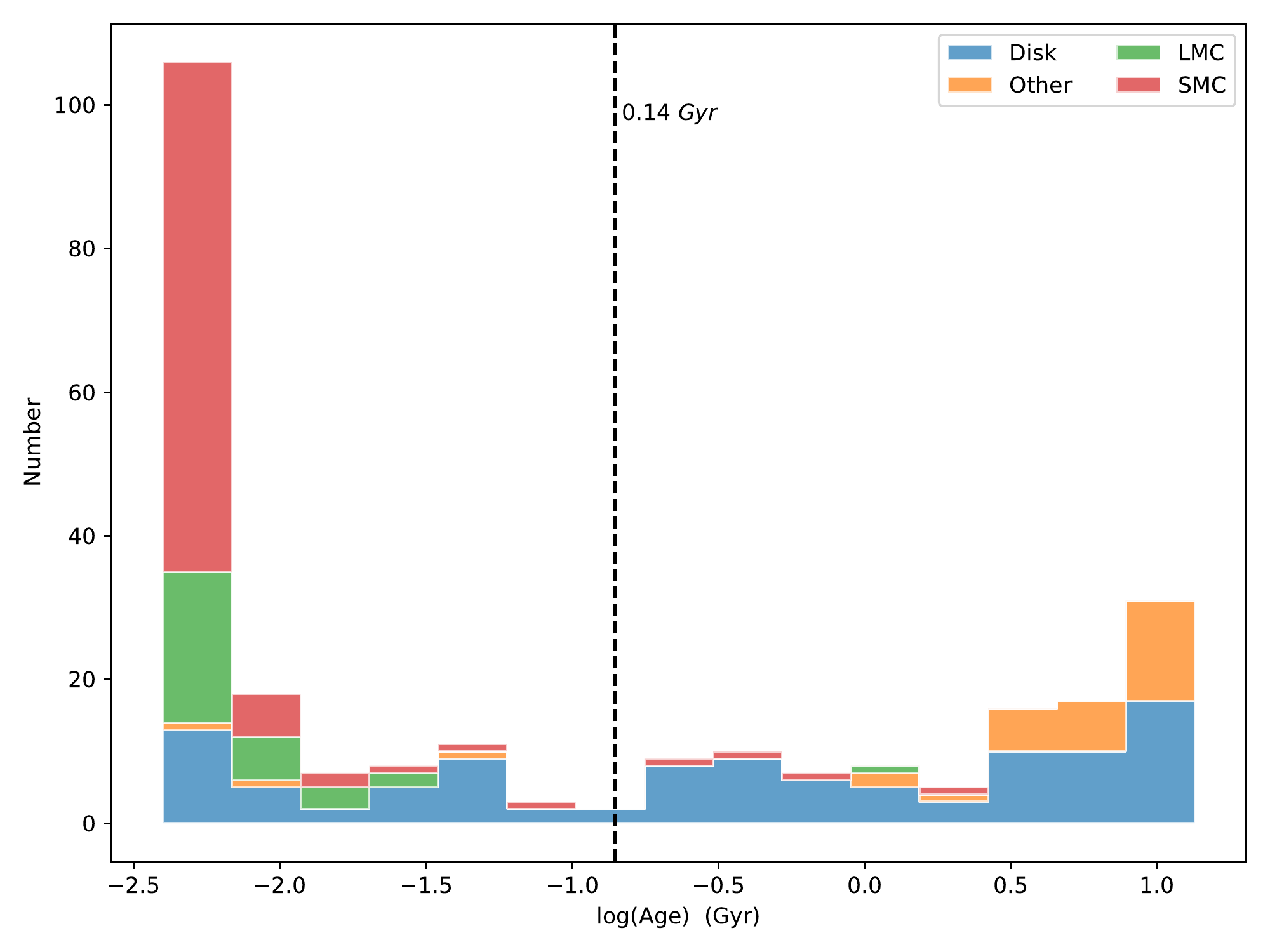}{0.3\textwidth}{(3)}
              }
\gridline{\fig{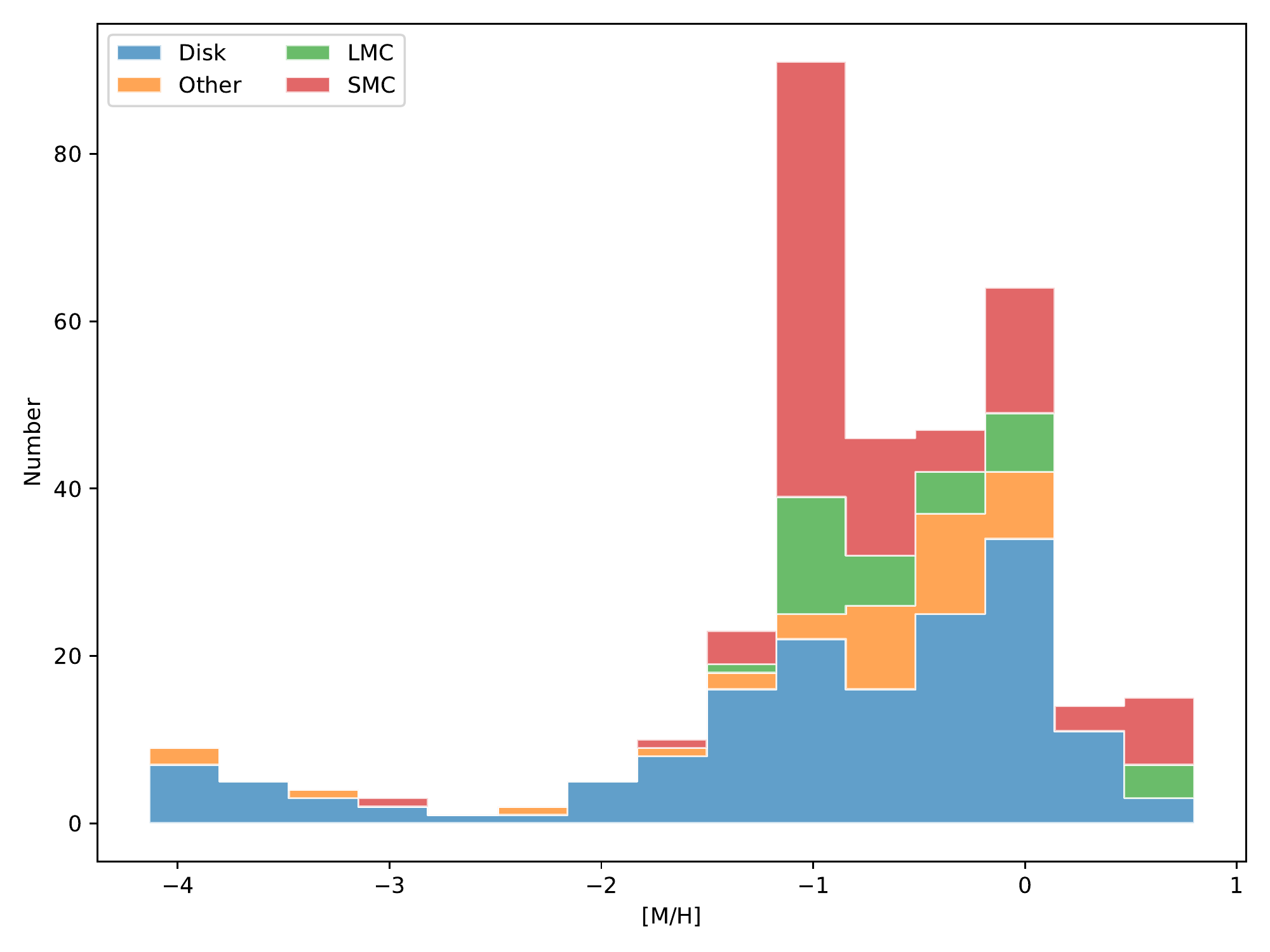}{0.3\textwidth}{(4)}
              \fig{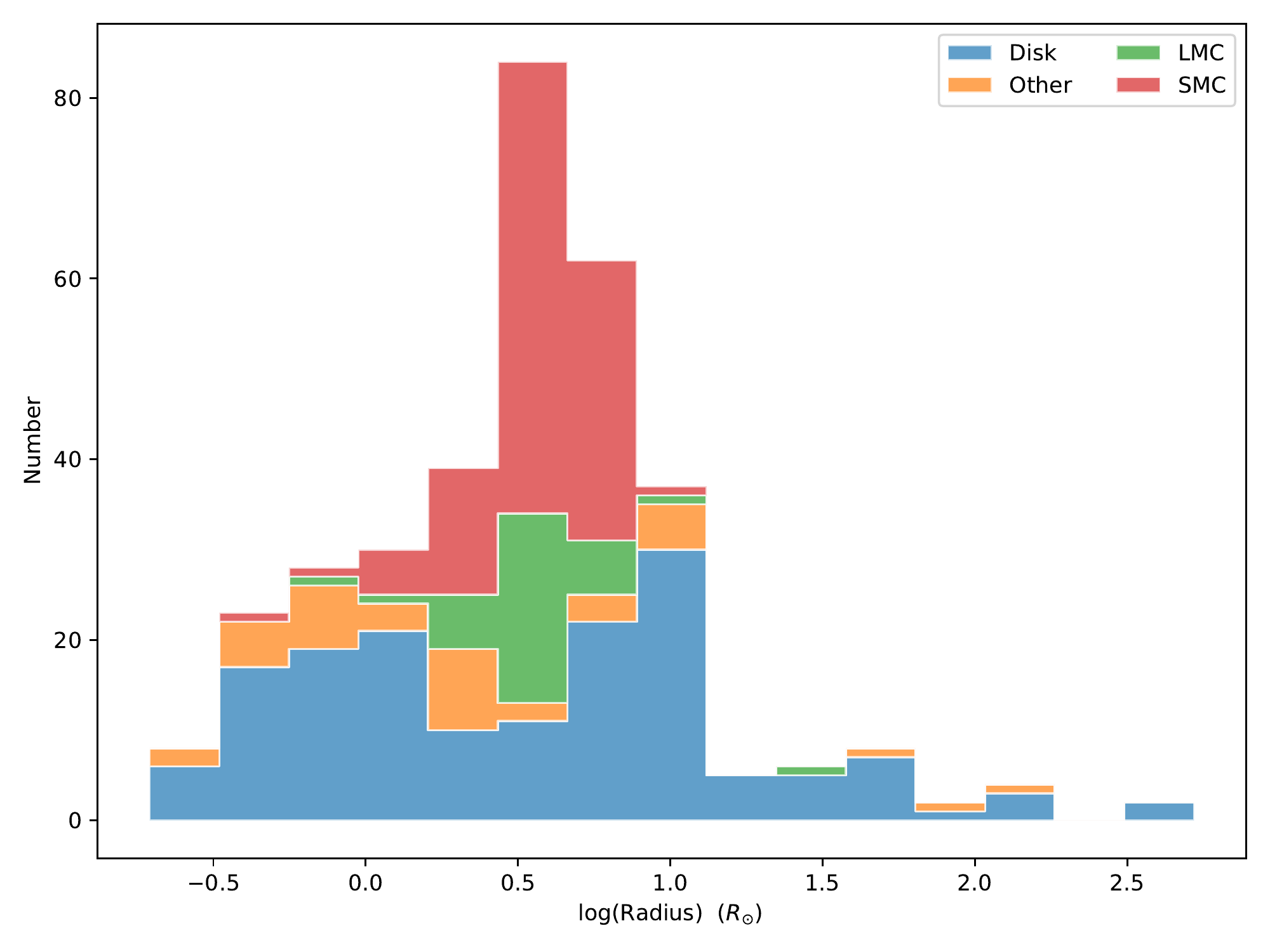}{0.3\textwidth}{(5)}
              \fig{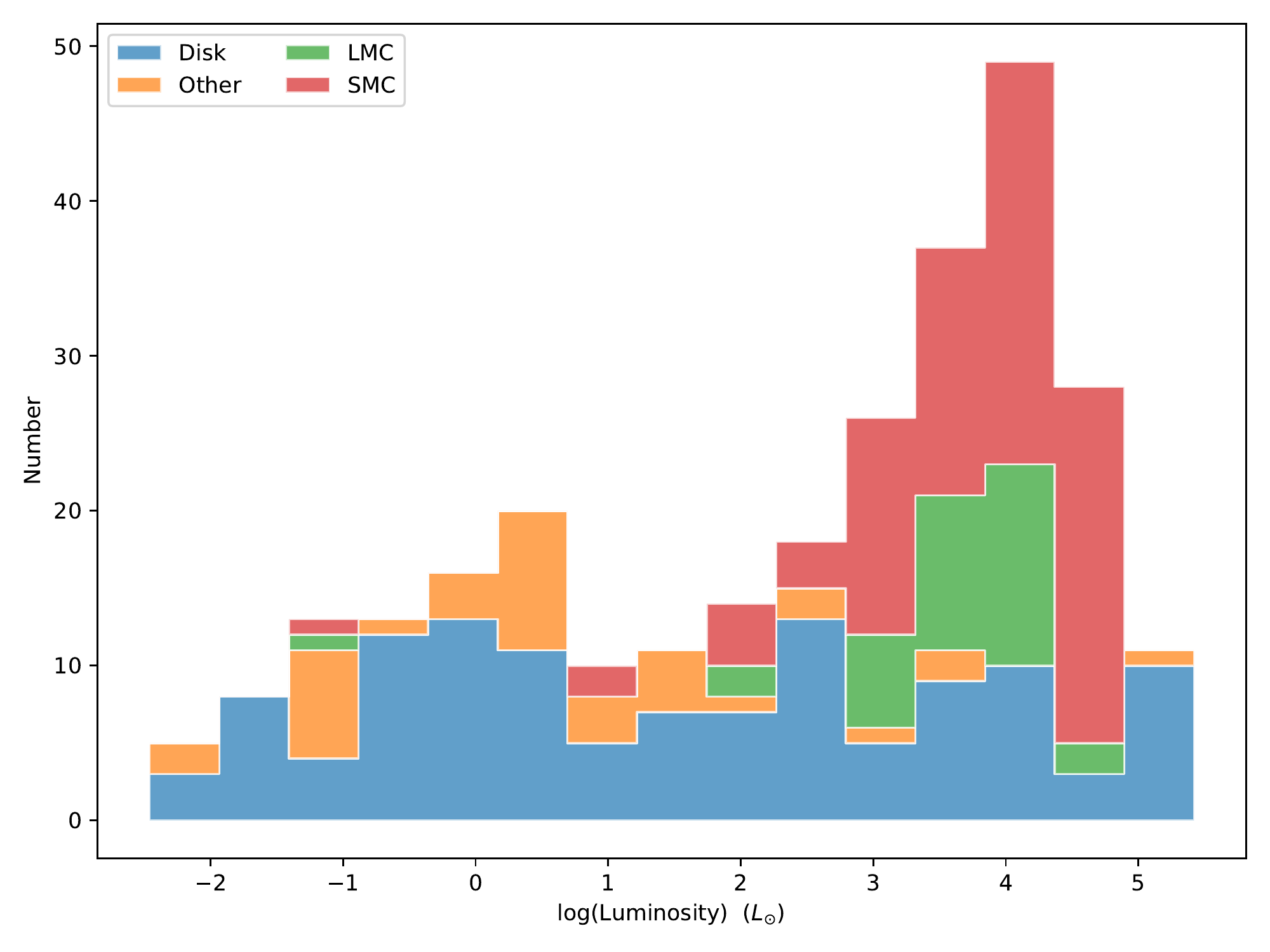}{0.3\textwidth}{(6)}
              } 
\gridline{\fig{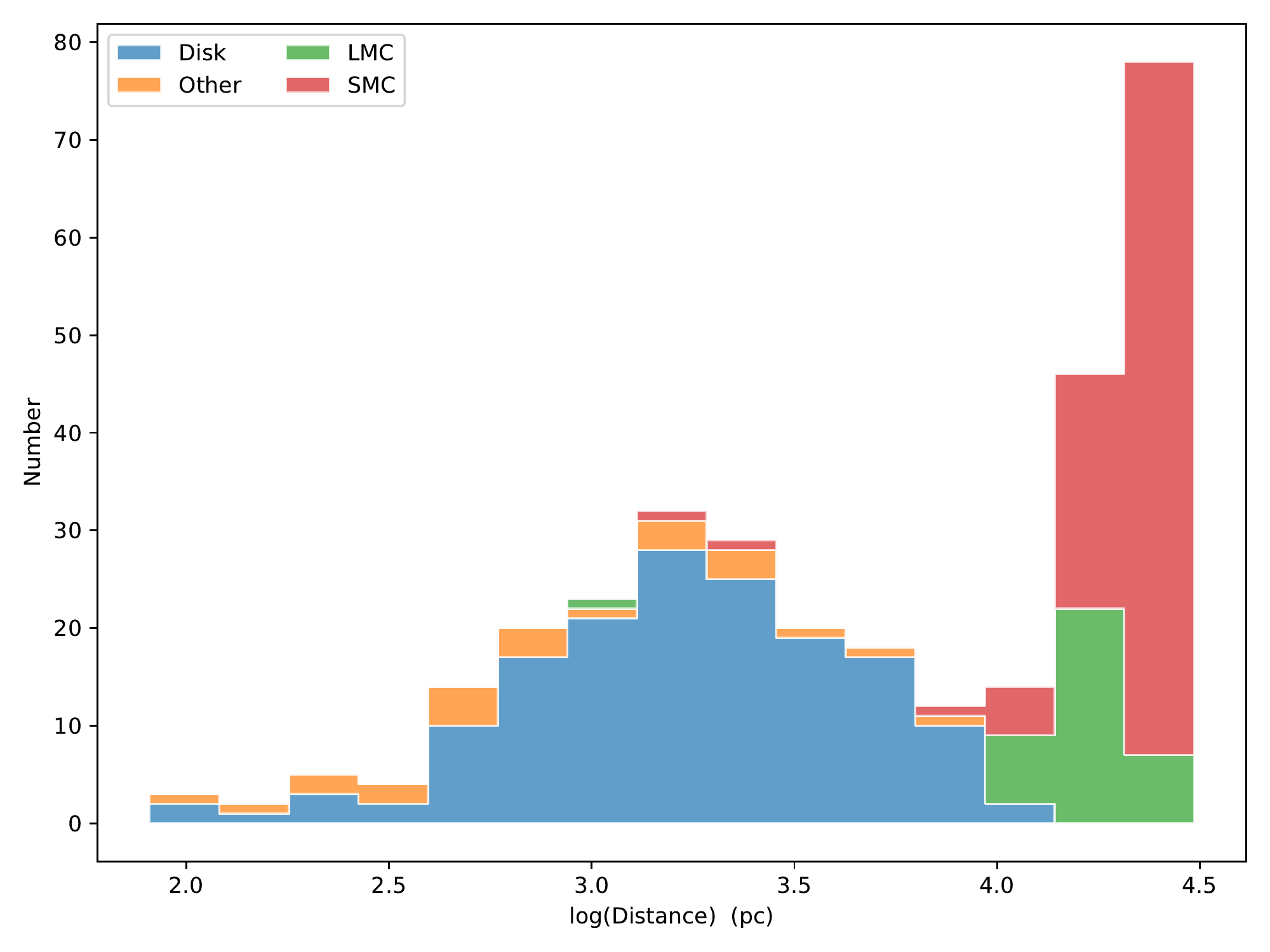}{0.3\textwidth}{(7)}
              \fig{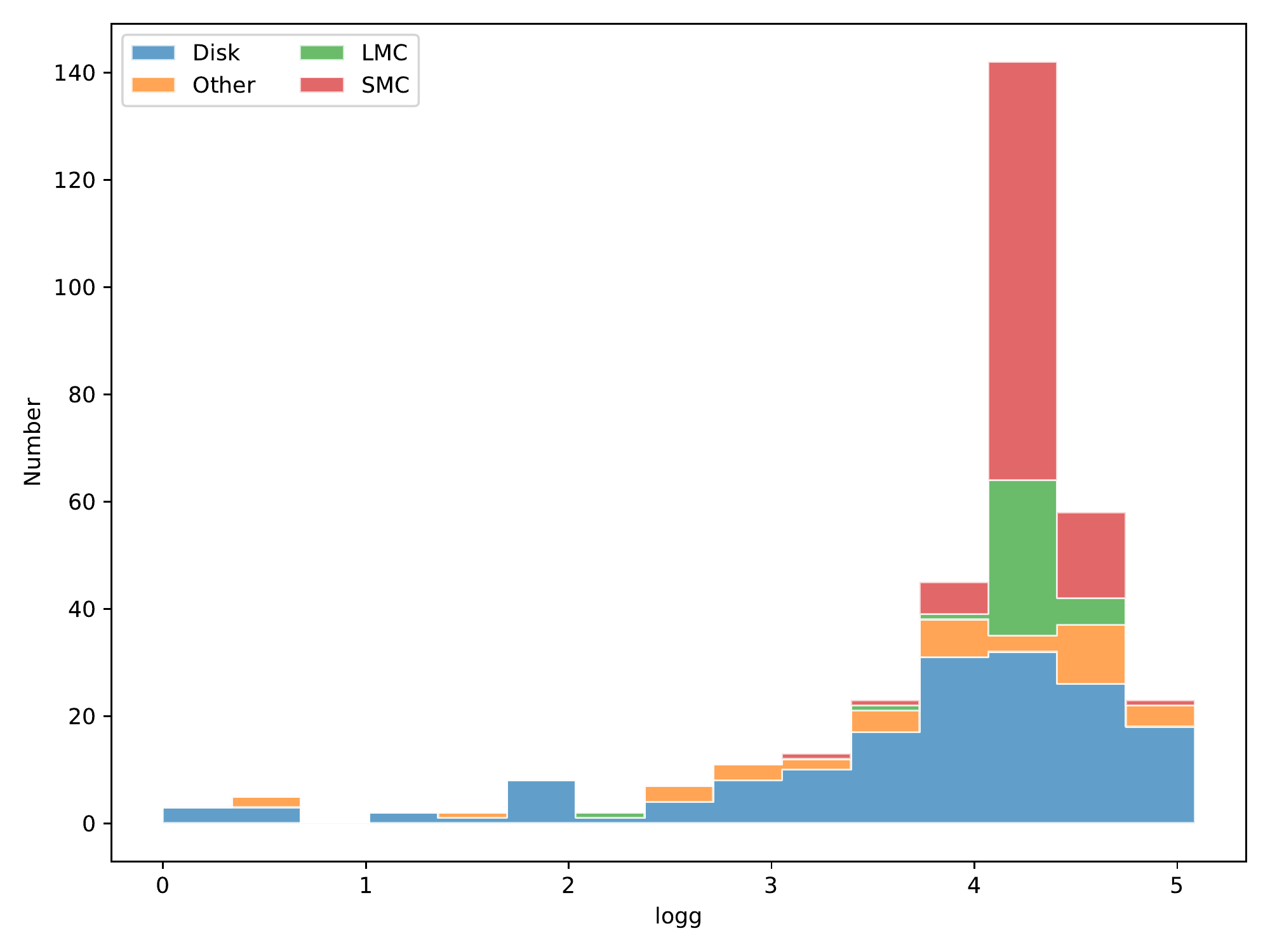}{0.3\textwidth}{(8)}
              \fig{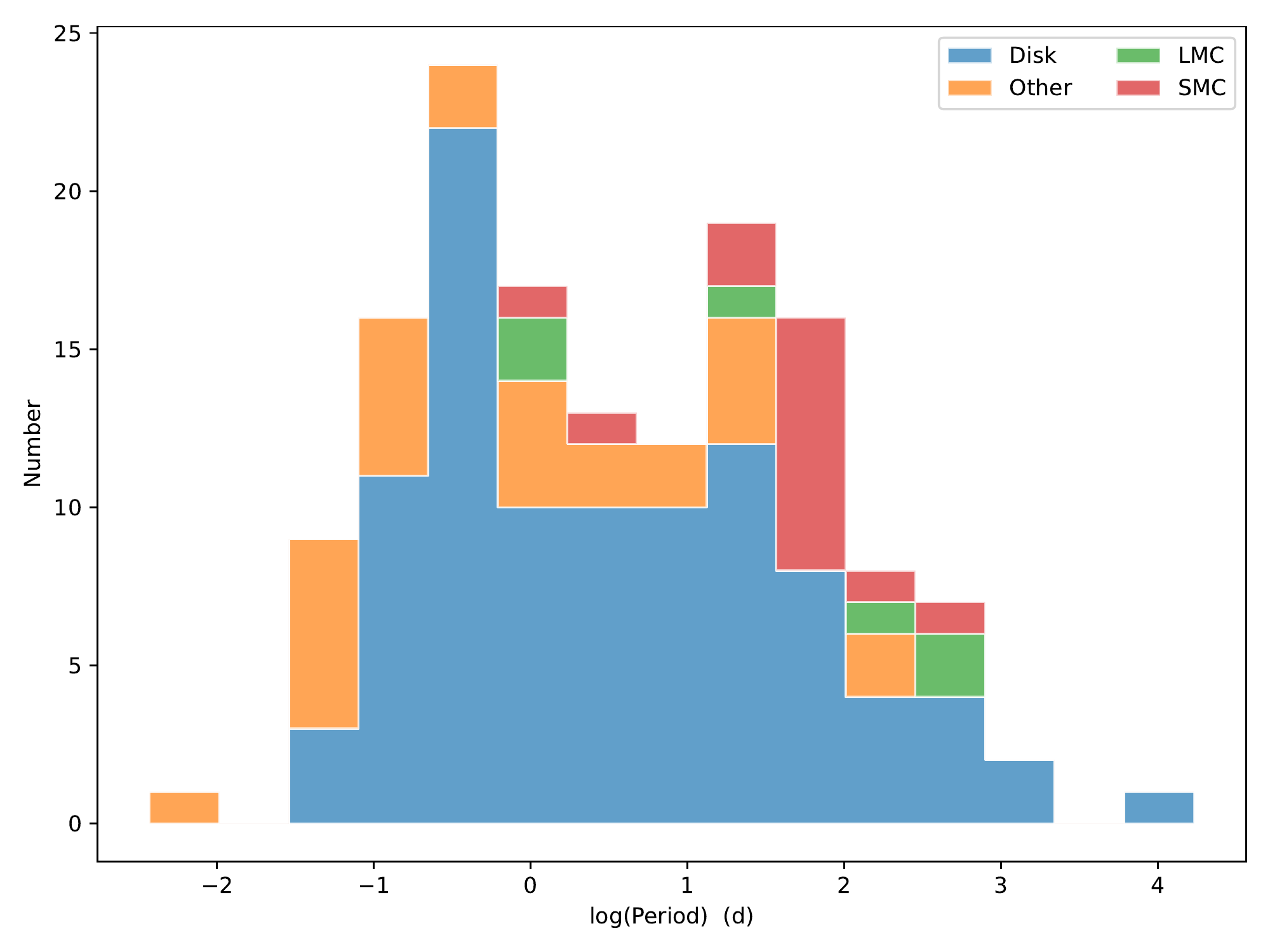}{0.3\textwidth}{(9)}
             }
\caption{Parameters distribution for the optical components of X-ray binary systems. Different colors represent different coordinate regions. Green represents Large Magellanic Clouds, red represents Small Magellanic Clouds, blue represents the direction of the Galactic disk (-15 $<$ b $<$ 15), and orange represents other directions. \label{fig:distribution}}
\end{figure*}

Figure \ref{fig:distribution} shows the distribution of various parameters of the optical stellar components in X-ray binary systems, with the exception of the last parameter being the orbital period of the binary.

Panels 1-3 show the logarithmic distribution of temperature, mass, and age, respectively. It can be seen that they all have a lowest point, where the distribution can be divided into two parts. 

We marked the lowest point with a dotted line, which are 11,500 K, 1.7 $M_{\odot}$, and 0.14 Gyr, respectively. These three positions are the new classification criteria for X-ray binaries proposed in this paper. Based on these three parameters, we suggest dividing X-ray binary stars into two categories. One is high-temperature, high-mass, and young X-ray binary stars, and the other is low-temperature, low-mass, and old X-ray binary stars.

To give only one classification result, all three parameters must be highly consistent with each other, meaning that a high-temperature X-ray binary star must also be high-mass and young, and a low-temperature X-ray binary star must also be low-mass and old.

\begin{figure*}
\gridline{\fig{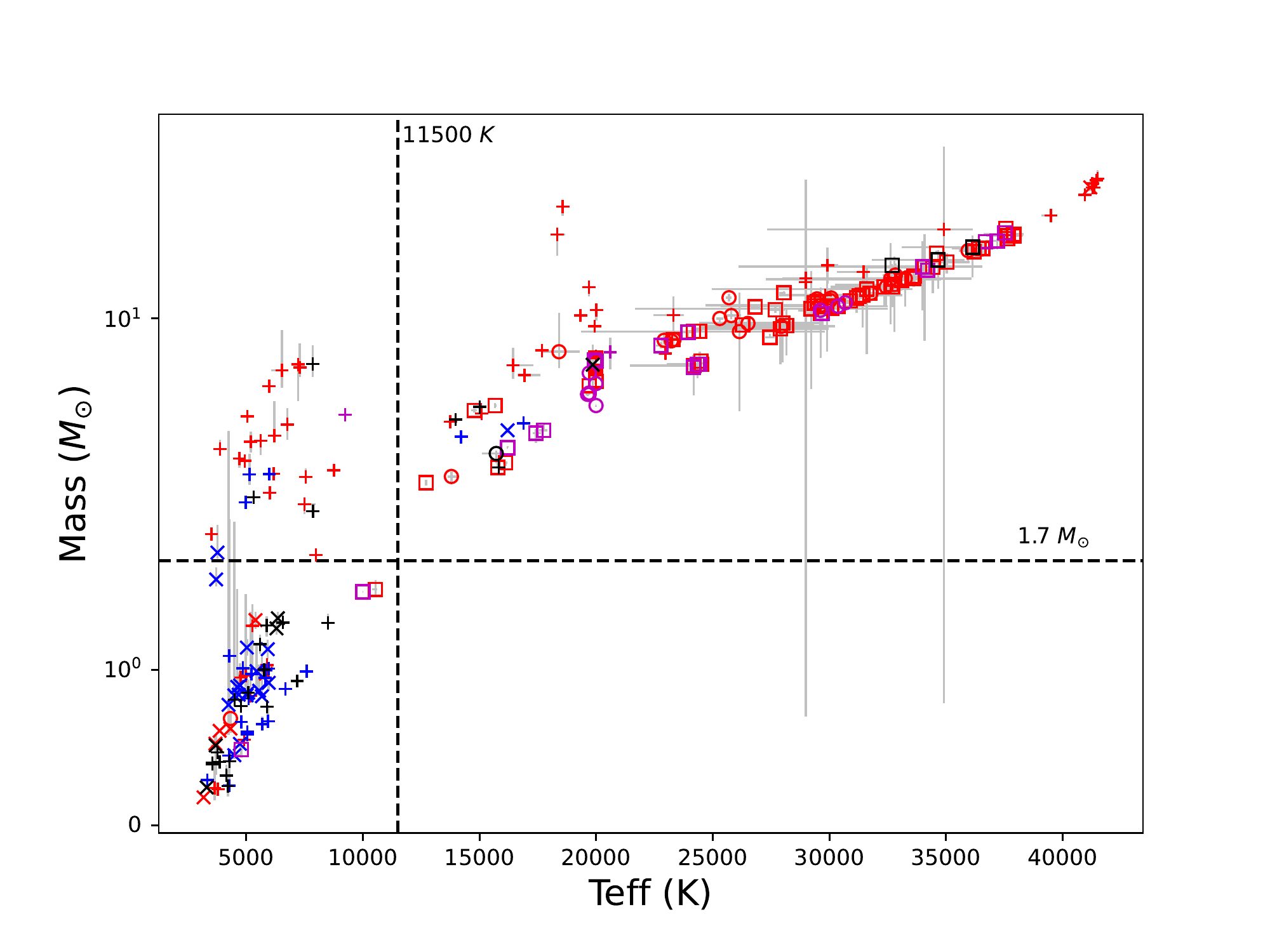}{0.3\textwidth}{(1)}
              \fig{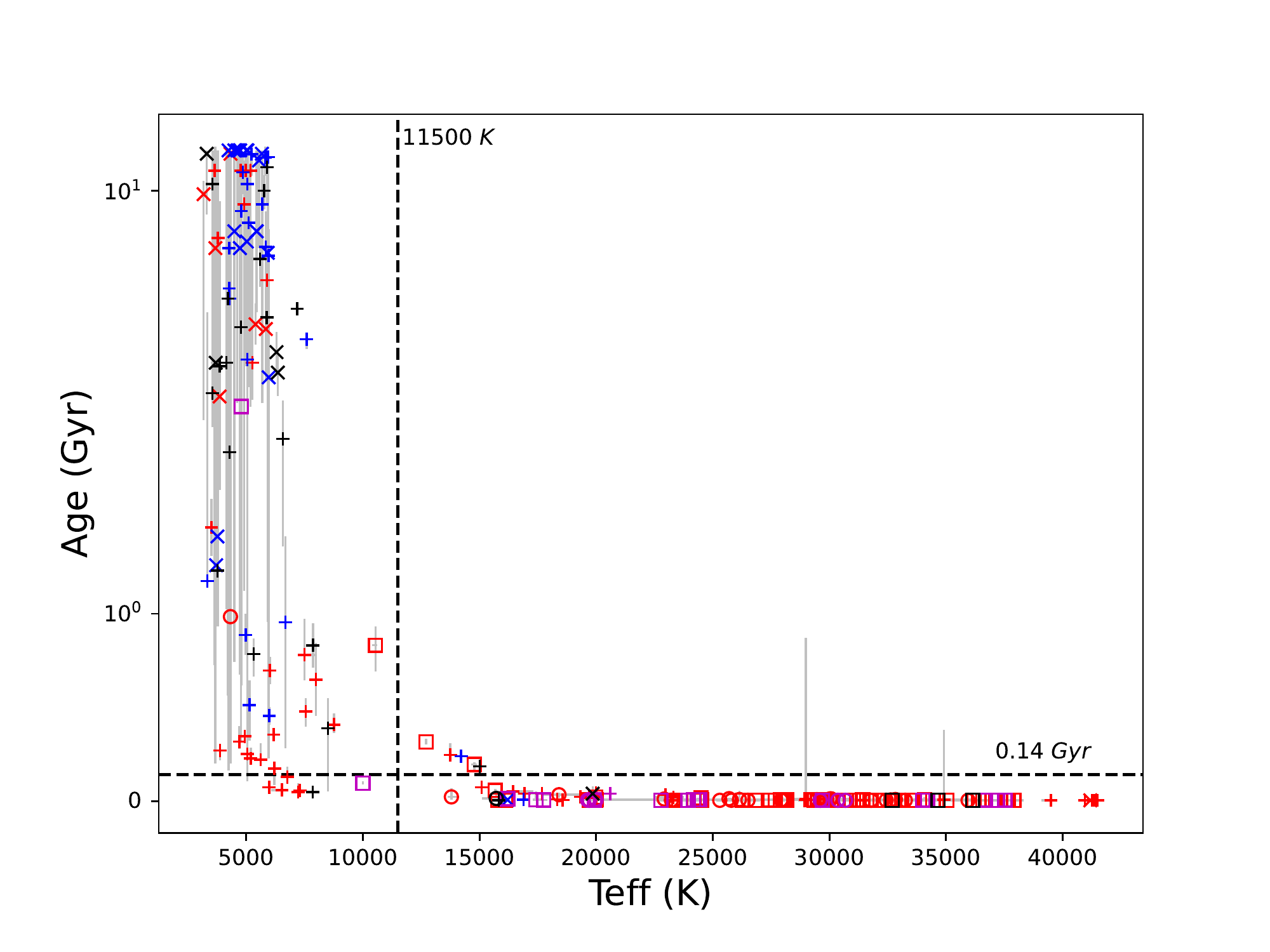}{0.3\textwidth}{(2)}
              \fig{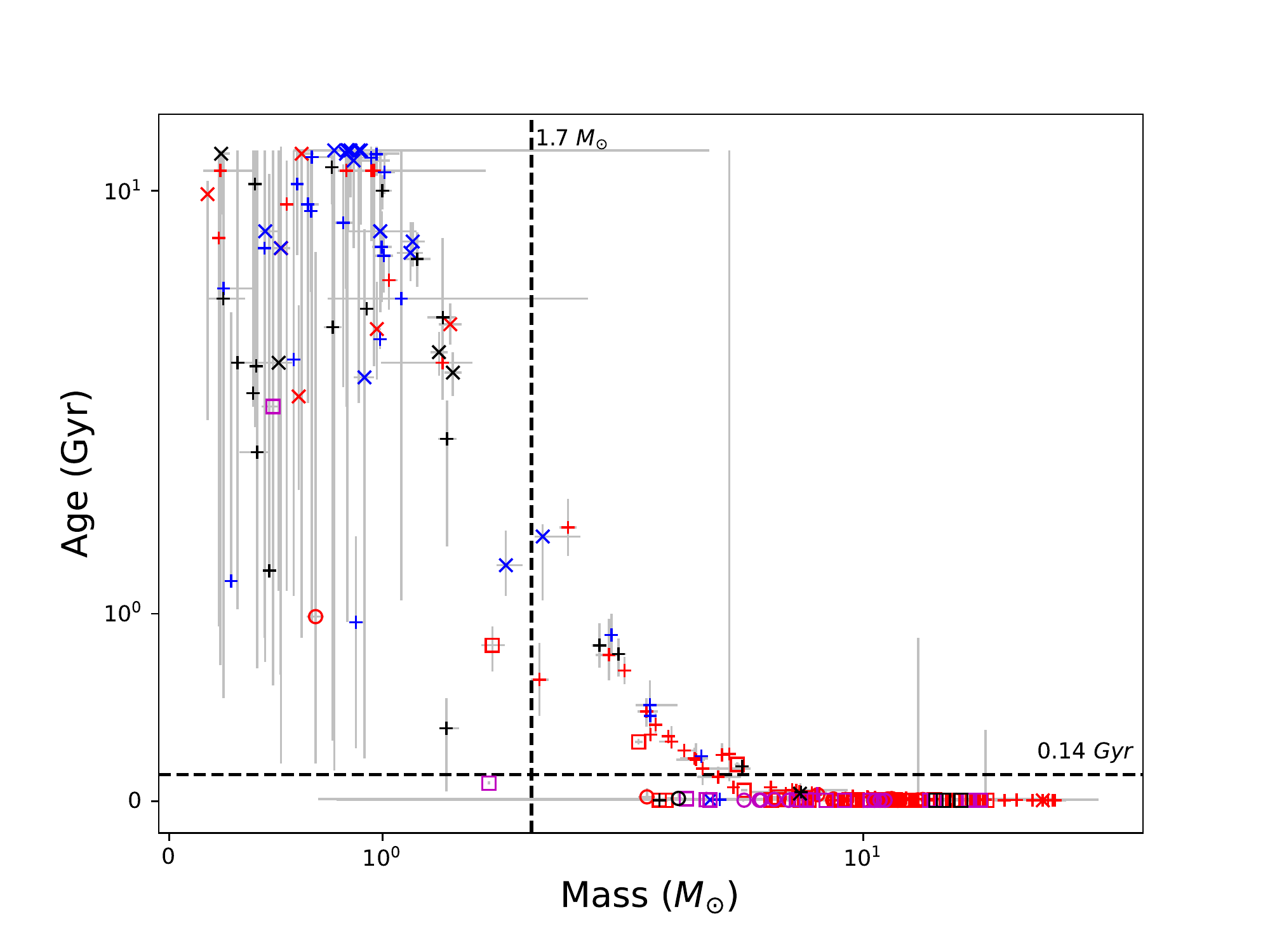}{0.3\textwidth}{(3)}
              }
\vspace{-0.6cm}
\gridline{\fig{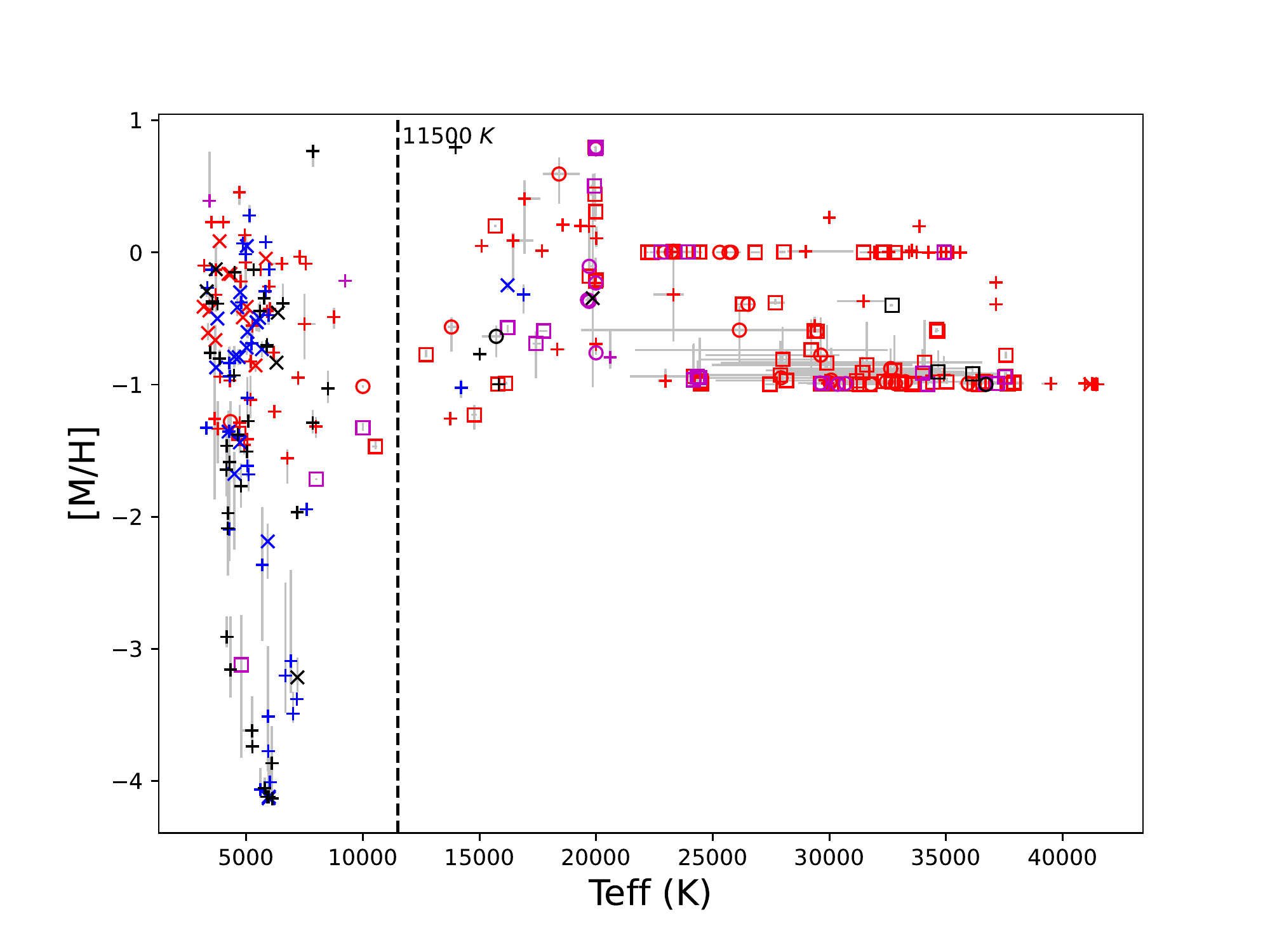}{0.3\textwidth}{(4)}
              \fig{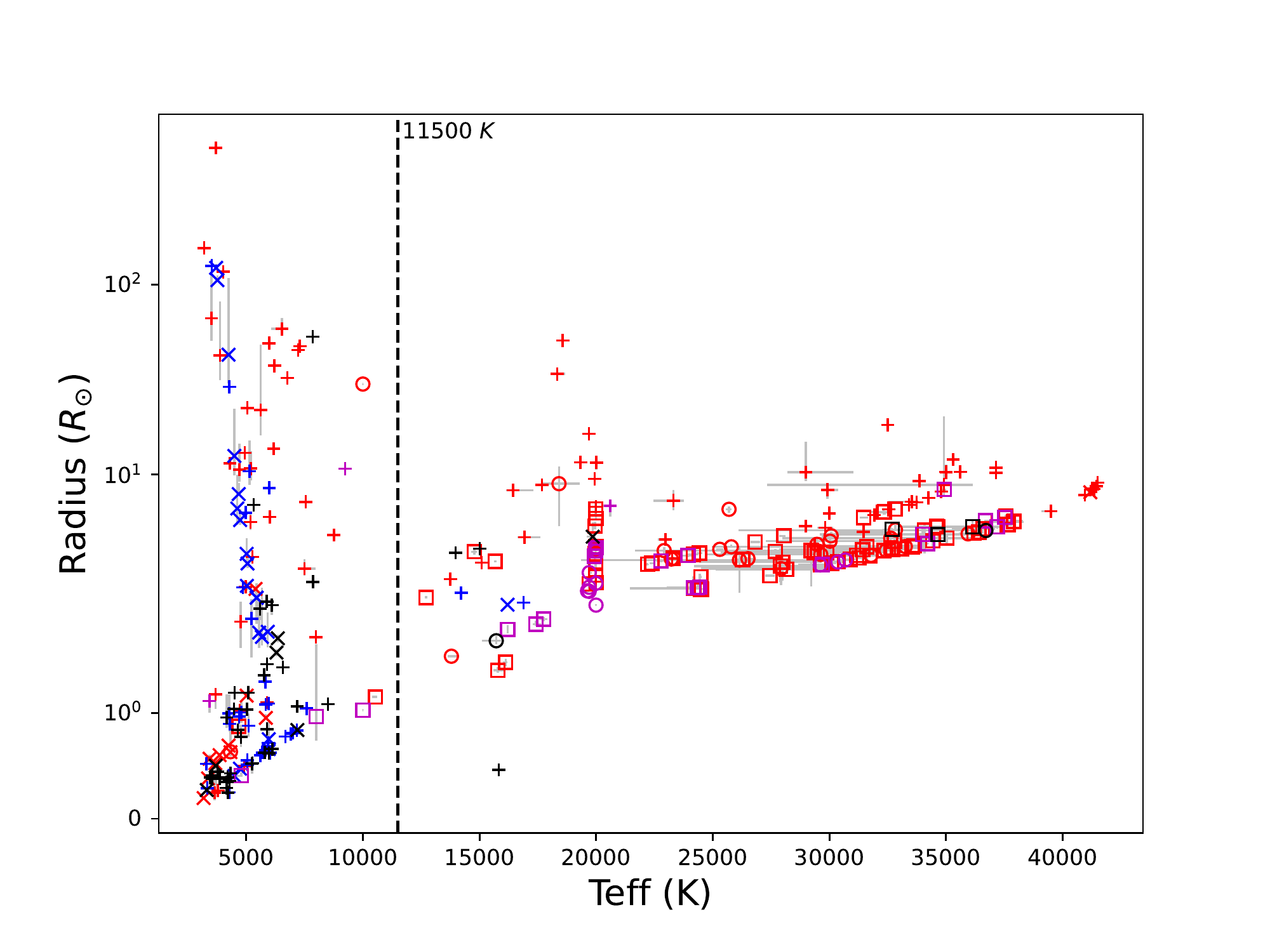}{0.3\textwidth}{(5)}
              \fig{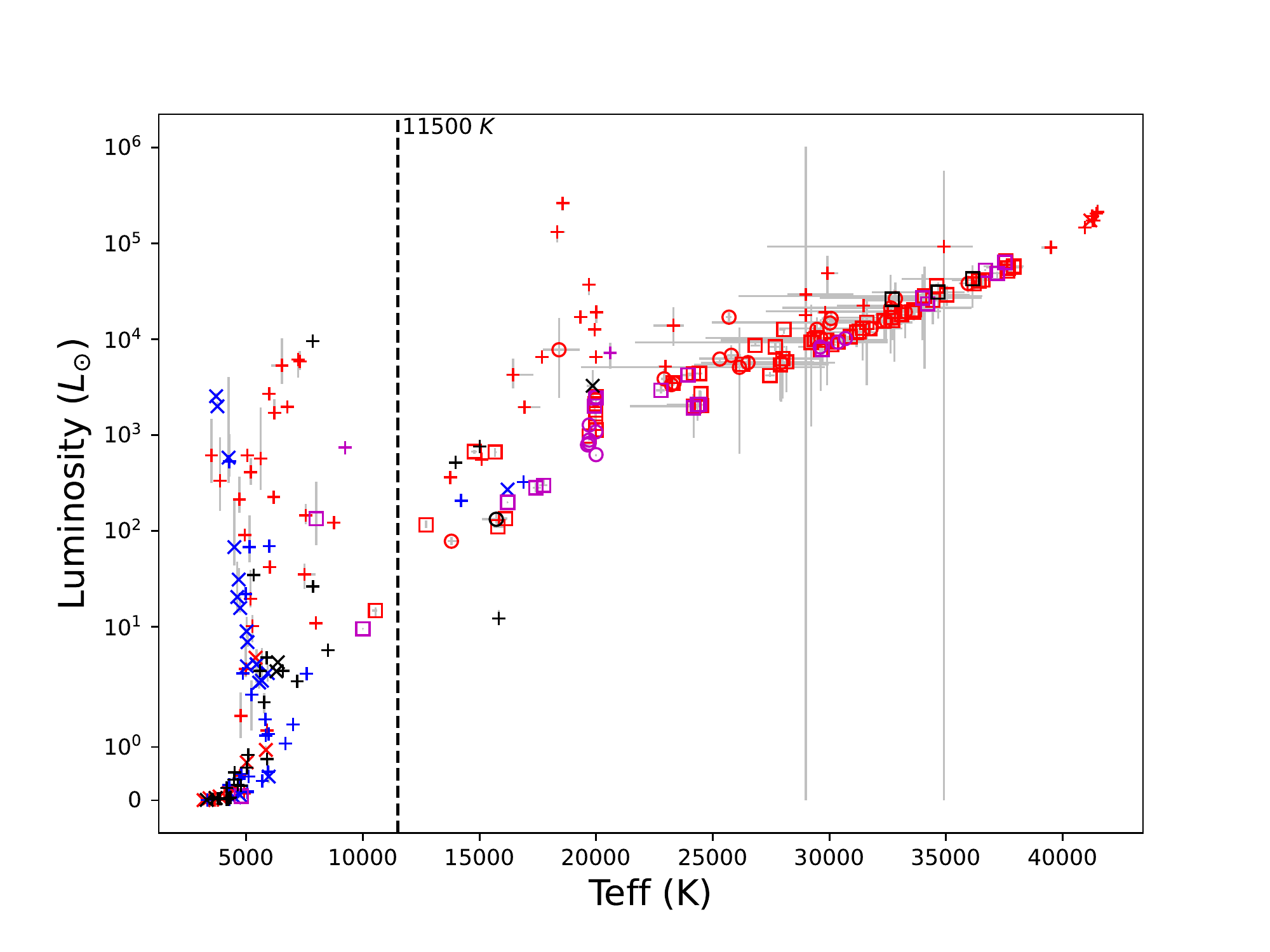}{0.3\textwidth}{(6)}
              }
\vspace{-0.6cm}
\gridline{\fig{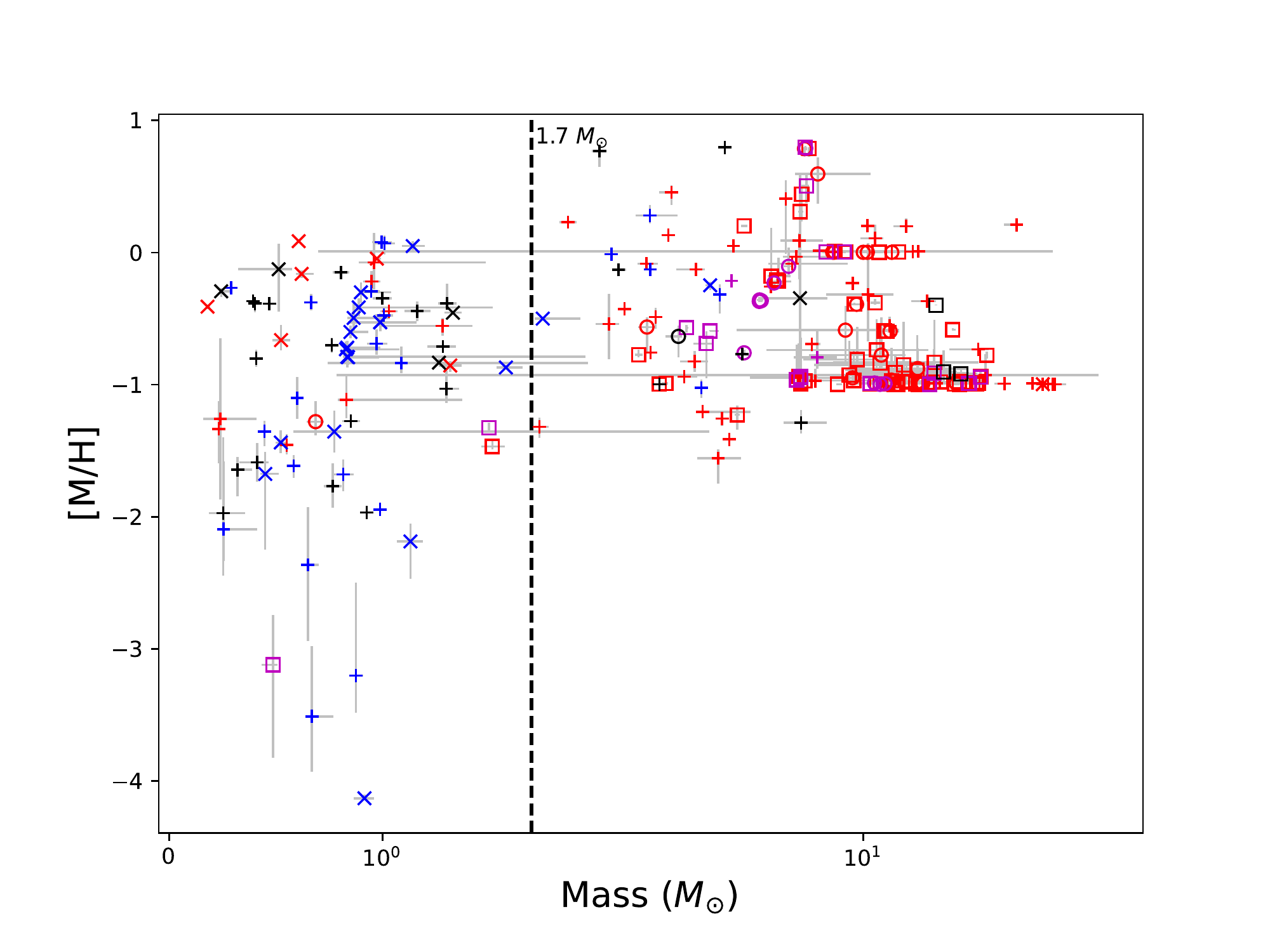}{0.3\textwidth}{(7)}
              \fig{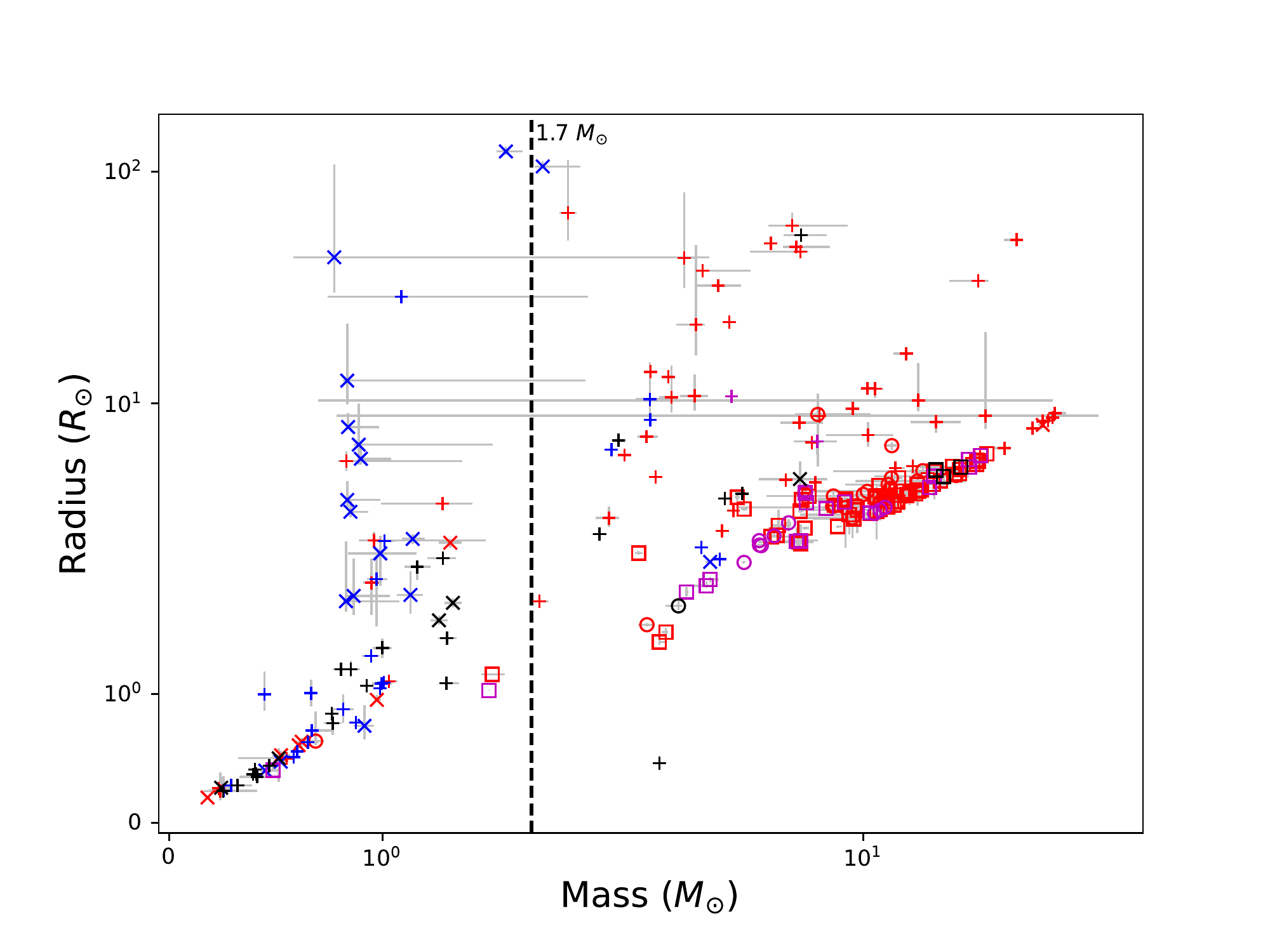}{0.3\textwidth}{(8)}
              \fig{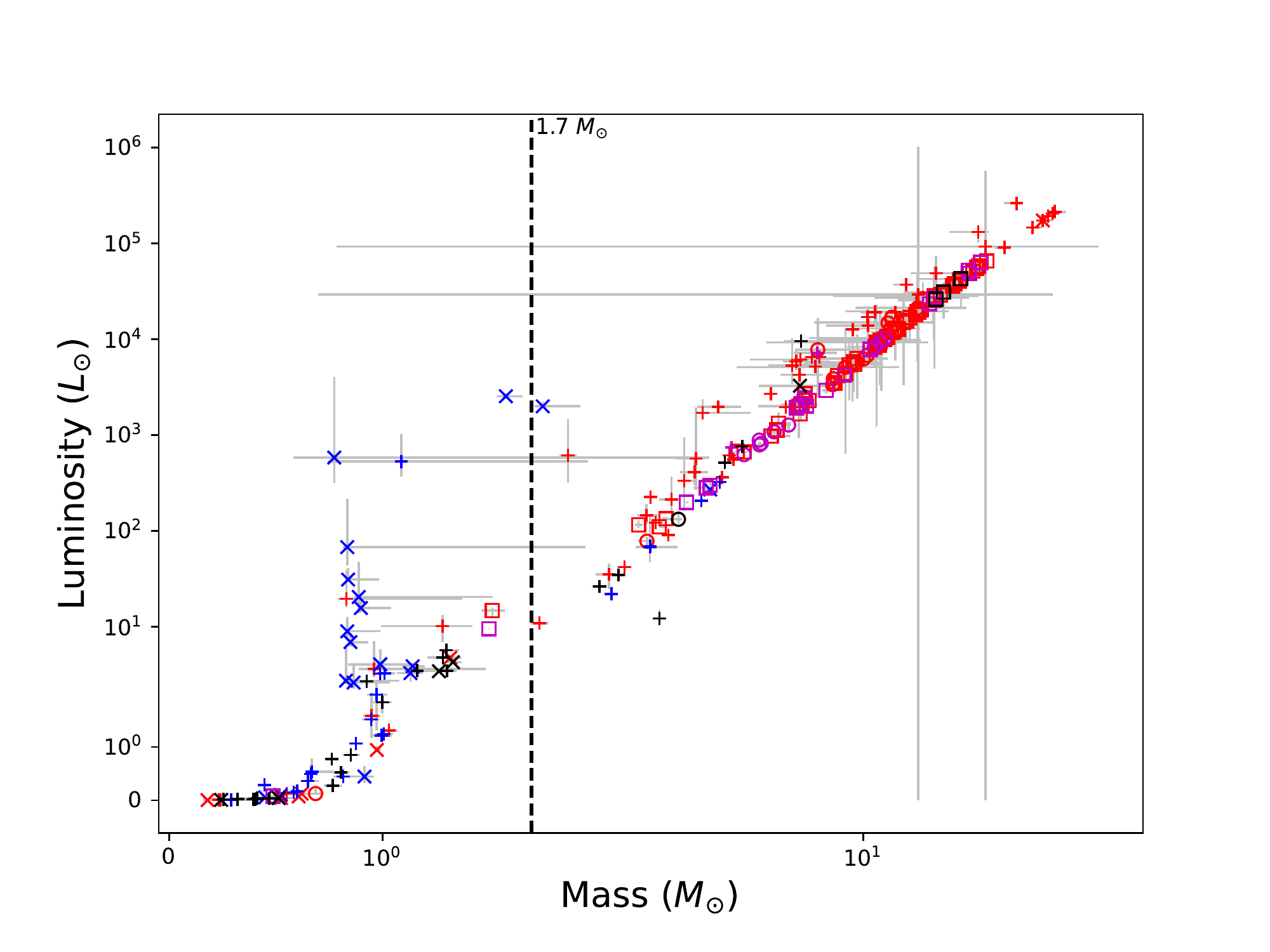}{0.3\textwidth}{(9)}
             }
\vspace{-0.6cm}
\gridline{\fig{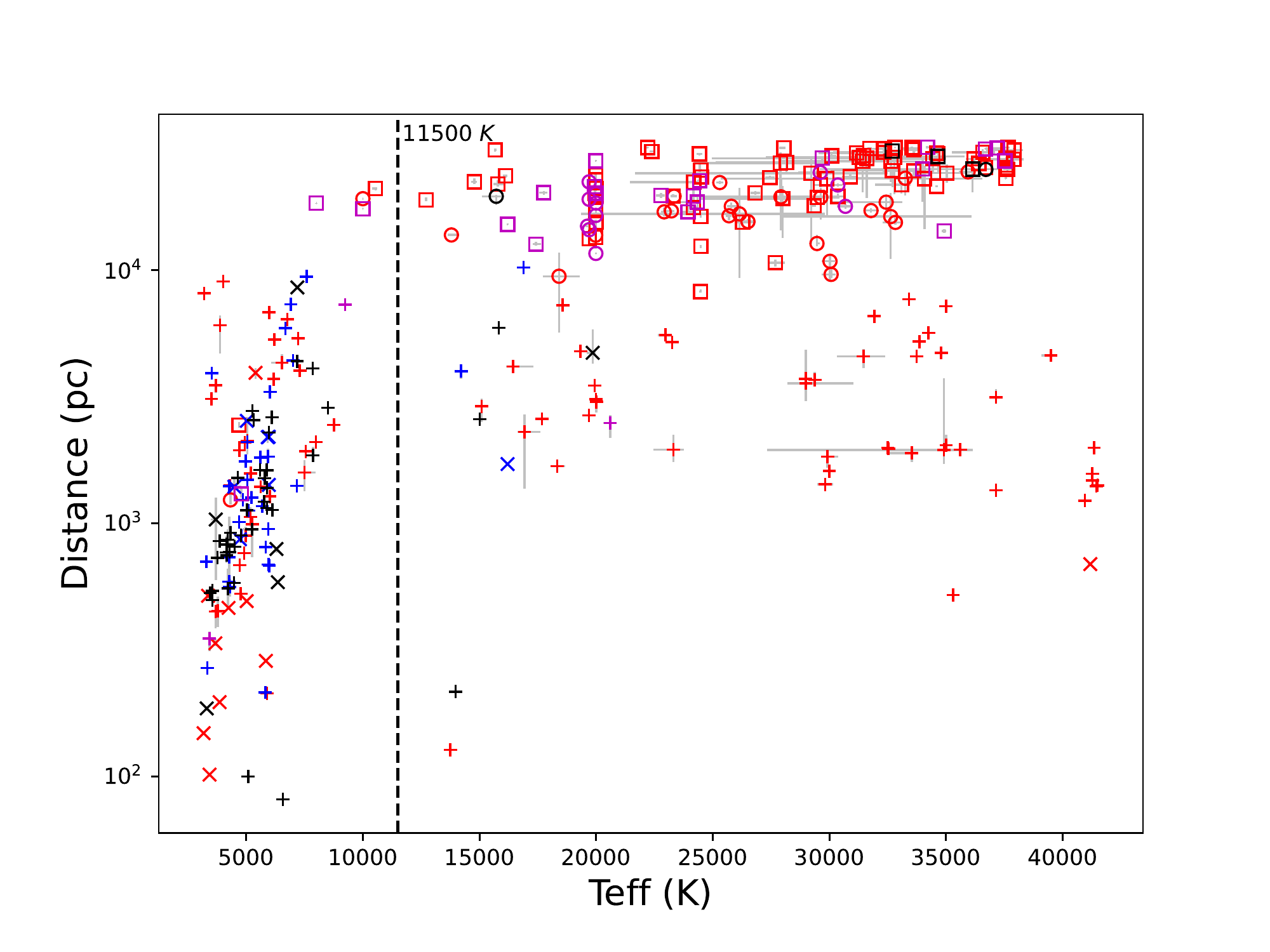}{0.3\textwidth}{(10)}
              \fig{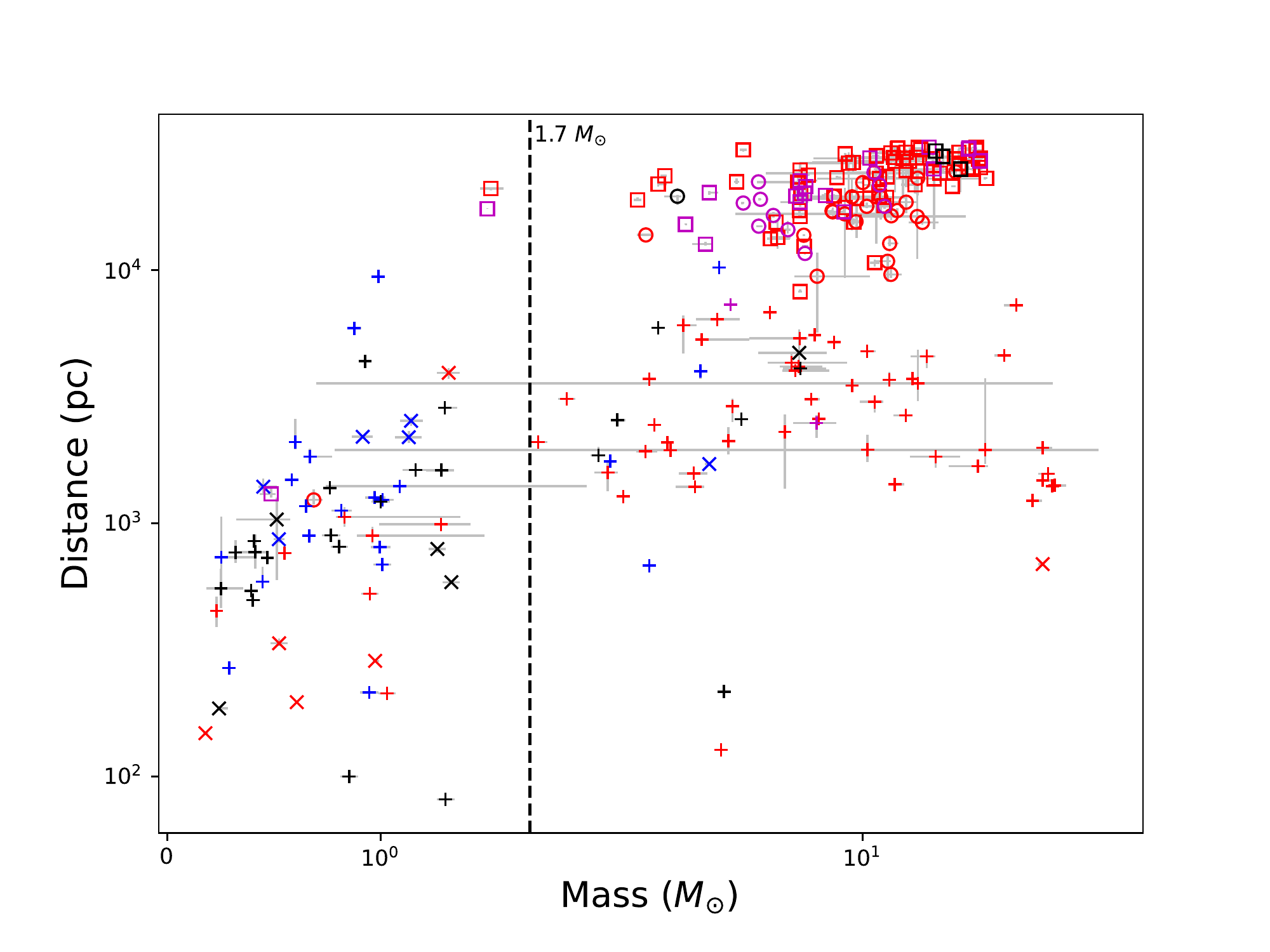}{0.3\textwidth}{(11)}
              \fig{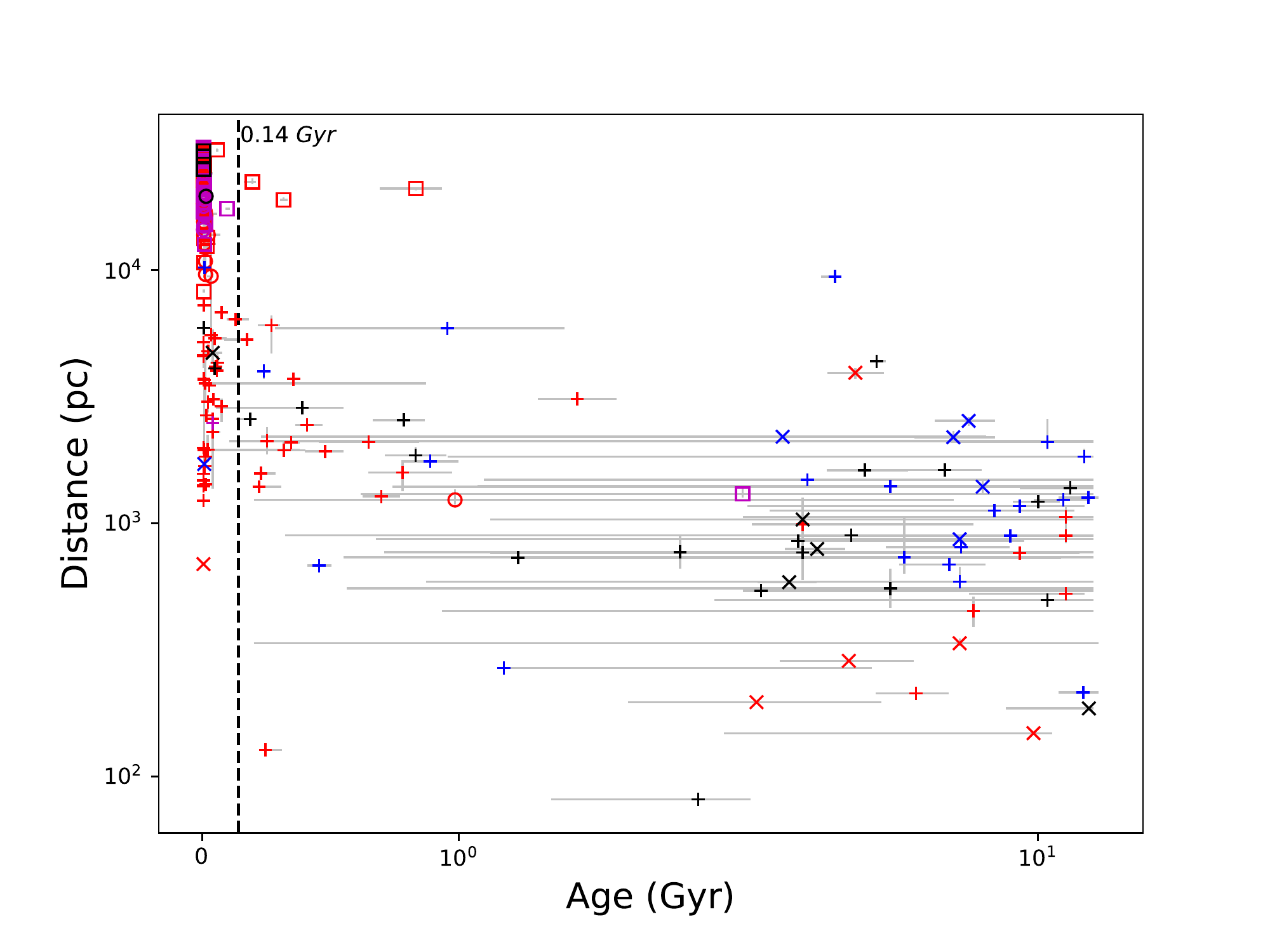}{0.3\textwidth}{(12)}
             }
\vspace{-0.6cm}
\gridline{ \fig{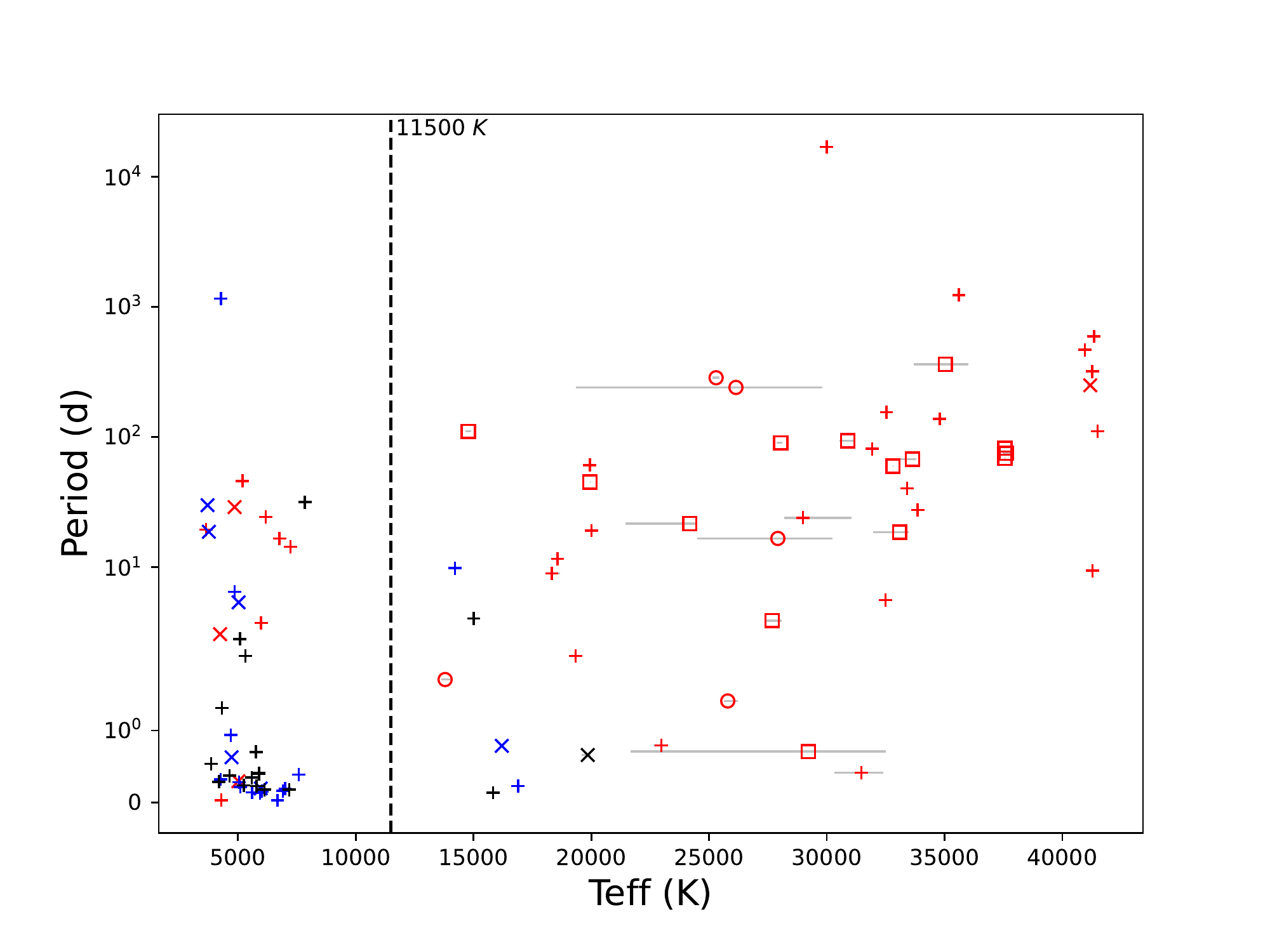}{0.3\textwidth}{(13)}
               \fig{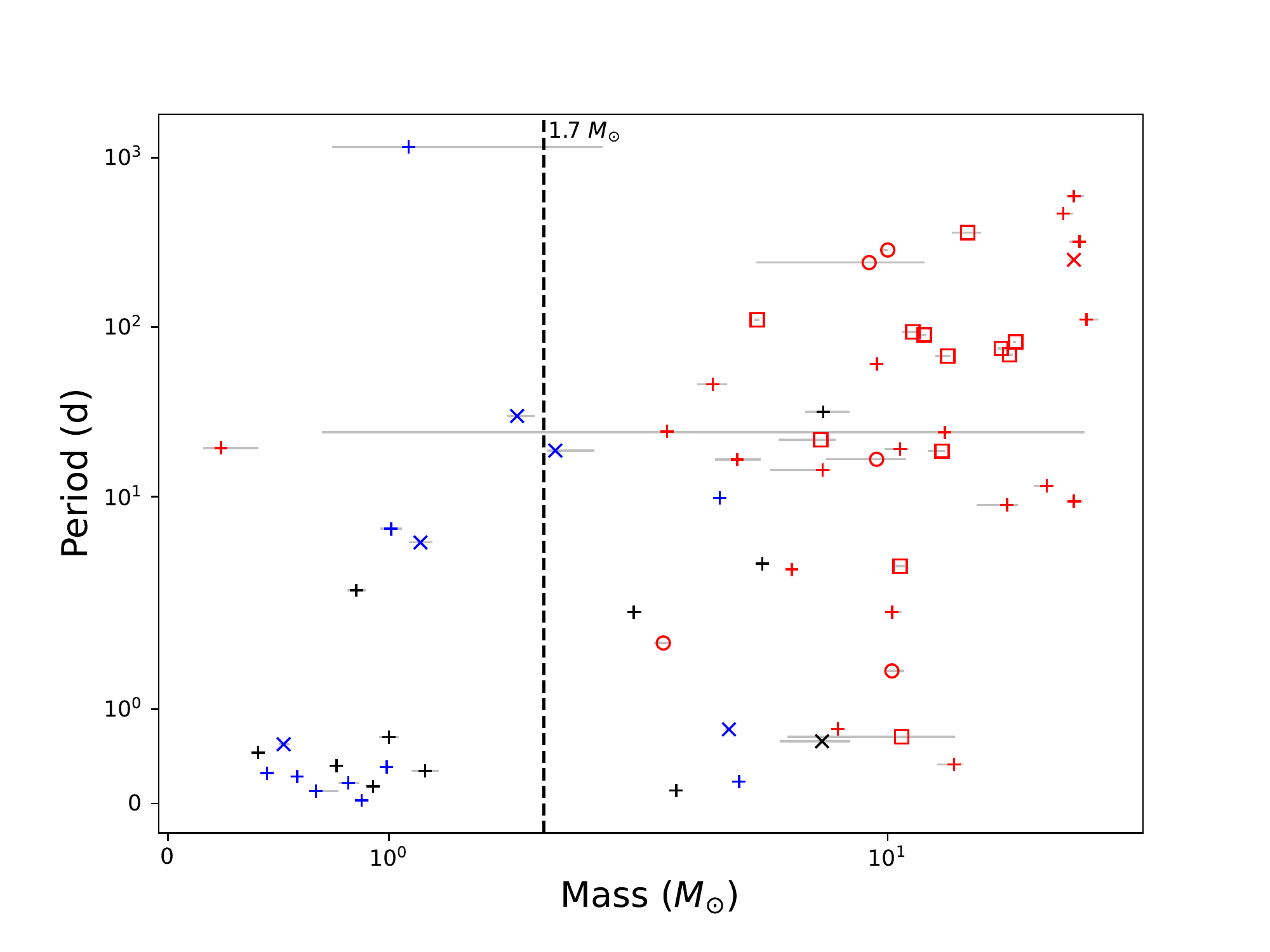}{0.3\textwidth}{(14)}
               \fig{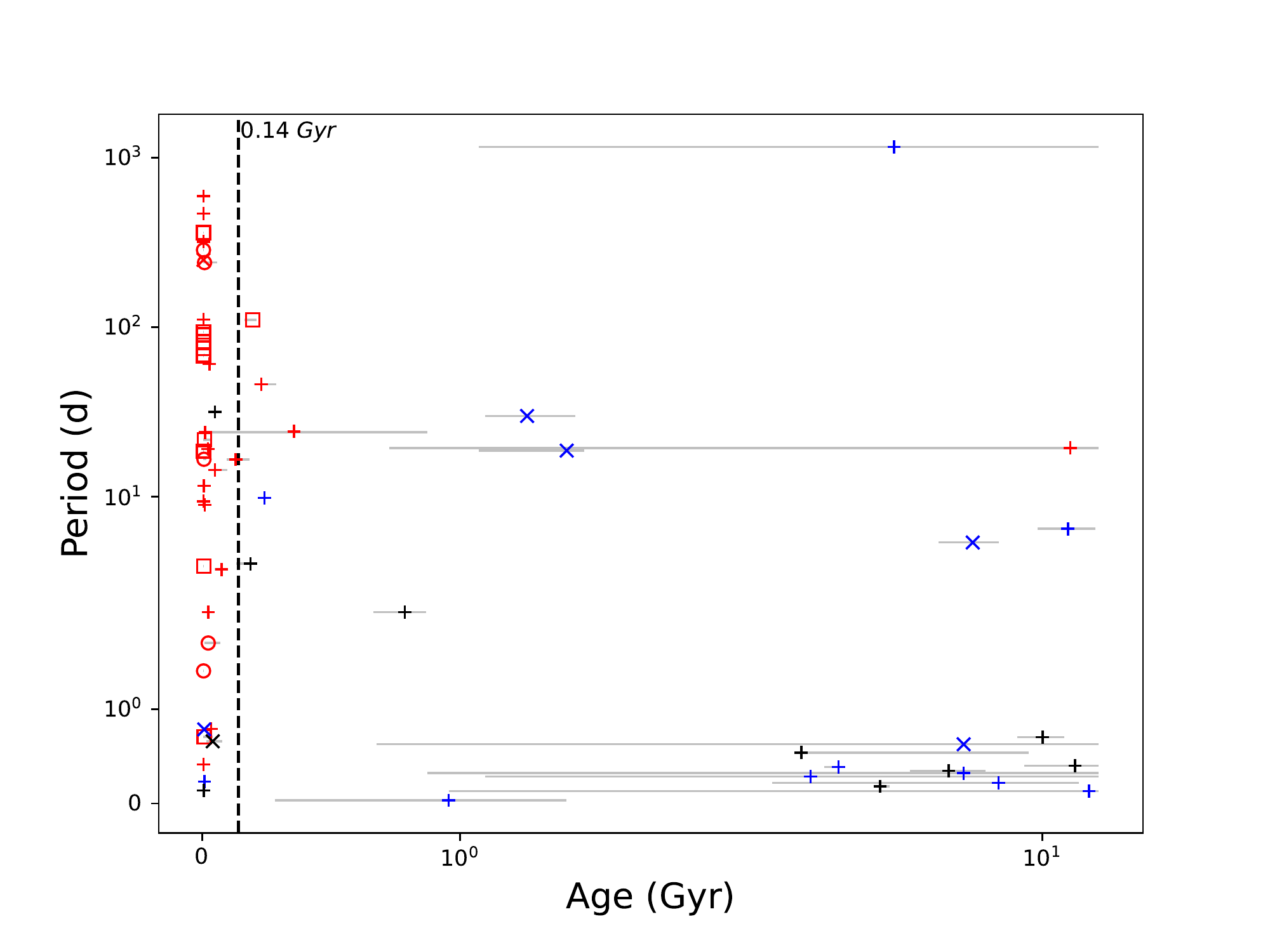}{0.3\textwidth}{(15)}
             }
\caption{Parameters relationship for the optical component of X-ray binary systems. Different colors represent different types of X-ray binary stars from literature. Red represents HMXB, blue represents LMXB, magenta represents candidates of HMXB, and black represents XB that have not been classified. The shape of the points represents the coordinate regions. Circles and squares represent the Large and Small Magellanic Clouds directions, + and $\times$ represent the galactic plane (-15 $<$ b $<$ 15) and other directions. 
\label{fig:relation}}
\end{figure*}

The panels 1-3 in Figure \ref{fig:relation} show the relationships between temperature, mass, and age. We marked the dividing lines from Figure \ref{fig:distribution} panels 1-3. It can be seen that most targets are distributed in the diagonal squares. In panel 1, most points are in the lower left and upper right boxes, which means that low-temperature X-ray binaries are mostly low-mass, and high-temperature X-ray binaries are mostly high-mass. Although some points also appear in the upper left square, their proportion is only 27/264=10.2\%, which means that about 90\% of X-ray binary stars have temperatures and masses that match each other.

In panels 2-3, this univariate distribution is more obvious, and old binary stars are almost low-temperature and low-mass, while young binary stars' temperature and mass are almost higher than the dividing lines. The proportions of points falling in other squares are 12/258=4.7\% and 25/258=9.7\%, respectively. Therefore, the consistency of the classification results given by the three classification parameters is 90\% or higher.

Here, we want to explain that the different colors of points in Figure \ref{fig:relation} represent different types of X-ray binaries. These types come from the original literature, not the new classification in this paper. The new classification no longer needs to be distinguished by color, as they have already been separated by the dividing line. It can be seen that blue points (LMXBs) are indeed more inclined to appear on the side of low-temperature, low-mass, and old.

The panels 1-3 of Figure \ref{fig:relation} confirm the compatibility and effectiveness of the new classification from panels 1-3 of Figure \ref{fig:distribution}. If only one parameter support a well-distributed classification, we can still classify the binary stars by that parameter, but the significance of the classification would be much weaker. Now that we have three parameters with good classification and compatibility, it is a highly significant classification.

The distribution pattern on two sides of the dividing lines are different, and the points around the dividing lines are sparse, as shown in panels 4-6 of Figure \ref{fig:relation}. Panel 4 shows that the metallicity of low-temperature X-ray binaries is widely distributed, but that of high-temperature has a lower limit of about -1. Panels 5 and 6 show that, for low-temperature binaries, there is no correlation can be seen, but for high-temperature binaries there is a clear linear relationship between the radius/luminosity (on a logarithmic scale) and the temperature.

The panels 7-9 in Figure \ref{fig:relation} show the relationship between mass and other parameters, and it can also be seen that the mass dividing line of 1.7 $M_{\odot}$ can divide the points into two parts very well, with different centers of aggregation and distribution characteristics on the left and right sides. Among them, the relationship between mass and luminosity is particularly notable. For mass less than 1.7 $M_{\odot}$, there is a clear curvature and branching; but for mass larger than 1.7 $M_{\odot}$, it is almost a straight line.

Panels 10-12 show the relationship between distance and three classification parameters. It can be seen that high-temperature, high-mass, and young X-ray binaries are much farther away. The X-ray binaries in Large and Small Magellanic Clouds are almost high-temperature, high-mass, and young. This indicates that classification and distance, or what galaxy the binaries are in, are related. 

In panel 7 of Figure \ref{fig:distribution}, binaries can be nicely divided into two groups by distance.
This suggests that the distance may also be a classification parameter. If we remove targets at the two Magellanic Clouds, it would be difficult to divide binary stars into two classes again. If we only look at the blue and orange colors (Disk and Other) in Panel 1-3, the significance of the bimodal structure will be significantly reduced; Similarly, if we only look at the + and $\times$ symbols in Figure \ref{fig:relation}, although the differences between two sides still exist, the significance has been greatly decreased.

In spite of that, we do not intend to add the distance as a classification parameter, due to that the distance is not a physical parameter, its essence is a certain region or a certain galaxy. The differences exhibited by different galaxies may be related to their formation history.

There is no obvious correlation between the period and the three classification parameters shown in panel 13-15. The type of X-ray binary is not closely related to its period, indicating that the initial period is unlikely to affect its type.

\section{Locations of X-ray binaries} \label{subsec:Location}

\begin{figure*}
\gridline{\fig{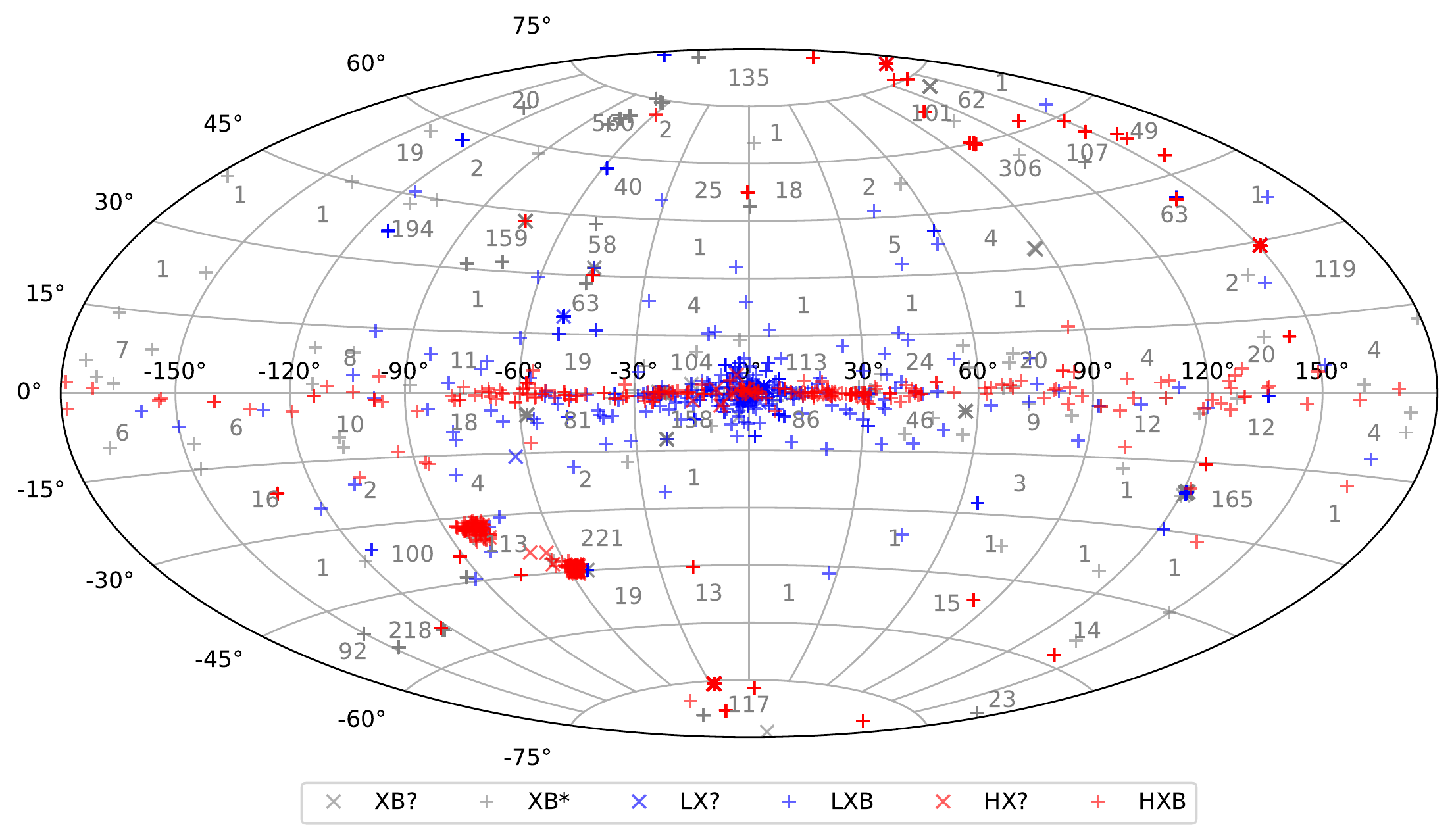}{0.5\textwidth}{(1)}
              \fig{fig/Coord_Teff344.pdf}{0.5\textwidth}{(2)}
              }
\caption{Panel 1: Galactic Coordinates distribution of all 4058 X-ray binaries, different colors and shapes represent different types from literature. Panel 2: Galactic Coordinates distribution of 344 X-ray binaries with temperatures, different colors represent types based on the new criterion of 11,500 K. The number of targets within each grid by latitude and longitude lines were also labeled. \label{fig:coord}}
\end{figure*}

The panel 1 of figure \ref{fig:coord} shows the Galactic Coordinates distribution of all 4058 X-ray binary systems, including both high-precision and low-precision coordinates. Different types are distinguished by different symbols and colors, and these types are from the original literature.

The panel 2 of figure \ref{fig:coord} is the position distribution of 344 X-ray binaries with temperatures. The red and blue points represent the temperature above and below 11,500 K. They are named High-Temperature X-ray Binaries (HTXBs) and Low-Temperature X-ray Binaries (LTXBs) based on our classification, and generally correspond to the HMXBs and LMXBs in the literature.

To begin with, it should be noted that in panel 1, it may seem that the majority of the points are concentrated in the Milky Way disk and the Magellanic Clouds, but this is not the case. Many points that appear as a single point actually represent tens to hundreds of targets that have very similar coordinates because they belong to the same galaxy. Therefore, the points appear close together on the all-sky map and look like a single point. This is why it is necessary to mark the number of stars on the plot. In fact, the number of X-ray binaries in directions outside the Milky Way is much greater than in the direction of the Milky Way disk. The number of stars in the direction of the disk (where l is within +/- 15 degrees) is 782, while the number in the direction of the Large and Small Magellanic Clouds is 83 and 183, and the number in other directions is 3010. At least 2748/3010=91\% of the X-ray binaries in other directions come from other galaxies outside the Milky Way.

For the extragalactic X-ray binaries collected in the paper, we list the works who observed and obtained the most targets. \citet{2008ApJ...689..983H} presented 1,132 low-mass X-ray binaries (LMXBs) from 24 early-type galaxies. The closest galaxy is NGC 3115, located 9 Mpc away, while the farthest is IC 4296, located 50.8 Mpc away. The criterion used to identify LMXBs is an X-ray luminosity of $L_{x} > 10^{37} erg/s$. \citet{2011A&A...533A..33Z} presented 185 LMXBs, 12 of which are from the Milky Way galaxy and the rest from seven other galaxies. \citet{2016ApJS..222...15T} presented 187 X-ray binaries from NGC 3115 with the assuming distance of 9.7 Mpc. \citet{2013A&A...555A..65H} presented 86 binaries, all from the central region of M31. \citet{2012MNRAS.419.2095M} provided 1,026 high-mass X-ray binaries (HMXBs) from 29 nearby galaxies, with the closest galaxy, NGC 5474, located 6.8 Mpc away, and the farthest, CARTWHEEL, located 122.8 Mpc away. \citet{2015AJ....150...94B} provided 132 binaries from three galaxies, NGC 55, NGC 2403 and NGC 4214.

According to our statistics, at least 3002 of the 4058 X-ray binaries in this paper are from outside of the Milky Way, accounting for 3002/4058=74\% of the total. However, few X-ray binaries targets from Extragalactic galaxies  (except for the Large and Small Magellanic Clouds) have accurate parameters for two reasons: X-ray surveys have not provided precise coordinates (with an error of less than 0.4 arcseconds), and the corresponding optical source are often too weak to obtain decent spectral data. That's why the vast majority of the targets in panel 2 come from the Milky Way and the two Magellanic Clouds.


Panel 2 shows the distribution of 344 X-ray binaries with temperature parameters. Among them, 164 are from the Milky Way, and 37+103=140 are from the Large and Small Magellanic Clouds. The distribution patterns obtained in this paper are based only on the binaries with parameters, and the conclusions are limited to the Milky Way and the Magellanic Clouds, and cannot be extended to other galaxies. The next two subsections will show that the types and occurrence probabilities are highly correlated with their host galaxies.

\subsection{X-ray binary types in different galaxies}

In the Milky Way, there are both HMXBs and LMXBs, and the numbers are comparable. In panel 1, we can see that in the direction of the Galactic disk, the red dots representing HMXBs are distributed horizontally along a narrow plane, while the blue LMXBs are more dispersed in the direction of the Galactic bulge. 

This is consistent with the age of both, as LMXBs are sufficiently old to have enough time to settle into the bulge region. On the contrary, the age of HMXBs are very short, so they should be located near their birthplace, i.e. on the disk.

In Large and Small Magellanic Clouds, almost all (252/253 = 99.6\%) the X-ray binaries are HMXBs, which is significantly different from that of the Milky Way. 

Why are there nearly no LMXBs in Large and Small Magellanic Clouds? Is it caused by the selection effect? Is it because the two Magellanic clouds are far away from us, so the faint LMXBs cannot be observed in the optical band? If the reason is the selection effect, it is difficult to explain why we were able to obtain the optical parameters of 14 and 5 LMXBs in M 31 and M 101, respectively. We should know that the distances of M 31 and M 101 are more than 10 and 100 times longer than that of LMC and SMC. Since LMXBs in much more distant places can be observed, they should be observed more easily if they are located in LMC and SMC. Therefore, the selection effect cannot effectively explain the reason for the scarcity of LMXBs in LMC and SMC. This indicates that in LMC and SMC, X-ray binaries may indeed be almost HMXBs.

\citet{2013A&A...555A..65H} found that ``Following the variability analysis, the majority of X-ray sources in M 31 seem to be LMXBs''.  Our statistics also strongly support this result. The rareness of HMXBs cannot be explained by the selection effect. We believe that the X-ray binary stars in M 31 are probably almost LMXBs.

Up to here, we have discovered that the types of X-ray binaries in different galaxies may vary greatly. The number of two types of X-ray binaries in the Milky Way is comparable. However, In LMC and SMC, HMXBs dominate, while in M 31, LMXBs dominate.

Further exploration of the reasons may require a detailed examination of the galactic environment, which is beyond the scope of this study.

\subsection{X-ray binary occurrence in different galaxies}

It is interesting to estimate the probability of X-ray binary occurrence in different galaxies. There are 83 X-ray binaries in the direction of the Large Magellanic Cloud, and 183 in the direction of the Small Magellanic Cloud. Combining the known number of stars in these two galaxies, the probability of X-ray binary occurrence in the Large Magellanic Cloud is 83/30 billion = 2.8 per billion, and in the Small Magellanic Cloud it is 183/3 billion = 61 per billion, a difference of more than 20 times. If we count the targets in the direction of the Galactic plane as within the Milky Way, the probability of X-ray binary occurrence in the Milky Way is 7.8 per billion out of 100 billion stars. This indicates that the probability of X-ray binary occurrence is highly correlated with the galactic environment.

Besides the types, occurrence rate of X-ray binary stars also vary greatly in different galaxies.

\section{Conclusion and Discussion} \label{subsec:Conclusion}

\subsection{X-ray binary catalog}

We present the largest and most parameter-rich catalog of X-ray binary stars to date, including 339 targets with atmospheric parameters and 264 targets with masses and radii of the optical components. Mass and radius were calculated using atmospheric parameters from Gaia and LAMOST spectroscopic data, and the PARSEC stellar models were used to perform these calculations. After comprehensive consideration and comparison, we estimate the typical relative error in mass and radius to be between 10-60\%.

\subsection{Bimodal distribution and a new classification proposal}

Based on these companion parameters of over 250 X-ray binaries, we performed statistical analyses on their distributions. We found that the temperature, mass, and age of the companions exhibit significant bimodal structures, with the troughs located at 11,500 K, 1.7 $M_{\odot}$, and 0.14 Gyr, respectively. If we consider an uncertainty in the location of the troughs or called dividing lines, the range is 10,529-12,499 K, 1.4-2.0 M, and 0.06-0.18 Gyr, which corresponds to the range of the lowest bar in the distribution plot. The number of targets at the trough is only 1/6 to 1/20 of the number at the peak.

Based on the prominent bimodal distribution, we propose to set the dividing line as a classification criterion for X-ray binaries. One class is characterized by high-temperature, high-mass, and young, while the other class is characterized by low-temperature, low-mass, and old. These two types of binaries exhibit significant differences in various parameters, showing distinct patterns or distributions, with few targets located near the dividing lines.

If we have to choose only one parameter as the classification criterion among the three, we recommend temperature. Firstly, because Figure \ref{fig:distribution} panel 1 shows that the number of points near the dividing line is close to 0, indicating that the intersection of the two types is very small. Secondly, Figure \ref{fig:relation} also repeatedly shows that the distribution on either side of the temperature dividing line is quite different. Finally, temperature is the easiest to measure and also the easiest to measure accurately, making it the most convenient to use.

Temperature is a better classification parameter than mass. Therefore, we suggest classifying X-ray binary stars into Low-Temperature X-ray Binaries (LTXBs) and High-Temperature X-ray Binaries (HTXBs), instead of LMXBs and HMXBs. If we adopt this classification, we no longer need to measure the mass of the companion star, which is very difficult and often has a large error. We only need to measure the temperature which is easy and accurate, and the classification result is good.

The results classified by temperature are basically consistent with those classified by mass. There are 344 binaries provided with temperature, 229 of them are classified the same as in the literature, 56 are classified conversely, and the remaining 59 are not classified in the literature. In the 229+56=285 binaries that are classified in the literature, 229/285=80\% of them are consistent with our new classification results. We have not provided a completely new classification, what we provided is a new quantitative classification criterion and a better classification parameter.

\subsection{Is there a selection effect?}

We should seriously discuss the reliability of the bimodal distribution. Is there a selection effect? Why are X-ray binary systems scarce in the intermediate region around the boundary lines? Is it because the actual number is small, or is it just difficult to detect them observationally? If it is due to selection effects, what are they?

Is it because the optical companions in the intermediate region are too faint? This is unlikely because they should be brighter than LMXBs and therefore easier to observe.

Is it because the X-ray emission from the X-ray binaries in the intermediate region is too faint or they rarely undergo X-ray outbursts? This is also unlikely because X-ray luminosity primarily depends on the mass and accretion rate of the compact star. There is no reason to believe that the compact stars in the intermediate region have lower mass or lower accretion rates. Relative to LMXBs, a heavier companion star should have higher wind loss and a larger radius, making it easier to fill the Roche lobe and undergo mass transfer.

Are the separations of X-ray binaries in the intermediate region larger? A larger separation would prevent the stars from filling the Roche lobe and undergoing efficient Roche lobe overflow accretion like in LMXBs. The relationship between the period and other parameters indicates that there is no notable difference between the periods of HMXBs and LMXBs, so there is no reason to believe that the periods of XBs in the intermediate region would be larger, and hence the separations should not be larger either.

The X-ray binaries in the intermediate region should be easier to detect than the LMXBs in any aspect. Therefore, the scarcity of X-ray binaries in the intermediate region should be due to the actual scarcity, rather than the selection effect.

The reason for the actual scarcity is beyond our research ability, and we speculate that it may be related to the different evolutionary modes of stars of different sizes and masses. Perhaps X-ray binaries in the intermediate region did exist, but they quickly evolved into LMXBs, resulting in a small number of observed X-ray binaries in the intermediate region. The 1.4-2.0 $M_{\odot}$ mass boundary is also close to the boundary between the convective and radiative envelopes at 1.5 $M_{\odot}$.

There is also a point that needs to be clarified, which is whether the presence of stripped He-rich stars would affect the bimodal distribution. Stripped He-rich stars can only exist in LMXBs, as the accretion of HMXBs is through stellar winds and companion star had not been stripped off. We previously assumed that the probability of the occurrence of this type of stripped star is very low (1/3000), so it can be ignored. Even if we assume that all companion stars of LMXBs are stripped He-rich stars, their masses would be overestimated by more than 5 times. In this case, the actual mass of LMXBs should be lower, and the gap of the bimodal distribution would be even larger. Therefore, stripped He-rich stars do not destroy the bimodal distribution, but enhance it.

In conclusion, we believe that the bimodal distribution is real.

\begin{acknowledgments}
This work is supported by the Chinese Natural Science Foundation (grant Nos. 11933008) and the Yunnan Natural Science Foundation (grant No. 202001AT070091).
This research has made use of the SIMBAD database, operated at CDS, Strasbourg, France.
This work has made use of data from the European Space Agency (ESA) mission {\it Gaia} (\url{https://www.cosmos.esa.int/gaia}), processed by the {\it Gaia} Data Processing and Analysis Consortium (DPAC, \url{https://www.cosmos.esa.int/web/gaia/dpac/consortium}). Funding for the DPAC has been provided by national institutions, in particular the institutions participating in the {\it Gaia} Multilateral Agreement.
Guoshoujing Telescope (the Large Sky Area Multi-Object Fiber Spectroscopic Telescope LAMOST) is a National Major Scientific Project built by the Chinese Academy of Sciences. Funding for the project has been provided by the National Development and Reform Commission. LAMOST is operated and managed by the National Astronomical Observatories, Chinese Academy of Sciences.
This research has made use of the International Variable Star Index (VSX) database, operated at AAVSO, Cambridge, Massachusetts, USA.

\end{acknowledgments}

\vspace{5mm}
\facilities{Gaia, LAMOST}

\bibliography{sample631}{}
\bibliographystyle{aasjournal}



\end{CJK}
\end{document}